\documentclass[11pt]{article}
\usepackage{hyperref}
\usepackage{amsmath,amssymb,amsthm,amscd}
\usepackage[all,cmtip]{xy}
\usepackage{xcolor}
\usepackage{blkarray}
\usepackage{float}

\usepackage{tikz,textcomp}
\usetikzlibrary{decorations.pathreplacing,calc,fadings,fit,shapes,arrows,positioning,chains,matrix}

\newcommand*{\ditto}{---\texttt{"}---}

\DeclareFontFamily{U}{wncy}{}
\DeclareFontShape{U}{wncy}{m}{n}{<->wncyr10}{}
\DeclareSymbolFont{mcy}{U}{wncy}{m}{n}
\DeclareMathSymbol{\Sh}{\mathord}{mcy}{"58} 

\numberwithin{equation}{section}

\begin{document}
\begin{titlepage}
{}~ \hfill\vbox{ \hbox{} }\break

\rightline{UWThPh-2021-13}

\vskip 3.5 cm

\centerline{\Large \bf Modular curves, the Tate-Shafarevich group} \vskip .4cm
\centerline{\Large \bf and Gopakumar-Vafa invariants with discrete charges}  \vskip 0.5 cm
\vskip 0.5 cm

\renewcommand{\thefootnote}{\fnsymbol{footnote}}
\vskip 5pt \centerline{ 
Thorsten Schimannek\footnote{thorsten.schimannek@univie.ac.at} } \vskip .5cm \vskip 20pt

\begin{center}

\textit{Faculty of Physics, University of Vienna}\\\textit{Boltzmanngasse 5, A-1090 Vienna, Austria}\\ [3 mm]

\end{center}

\setcounter{footnote}{0}
\renewcommand{\thefootnote}{\arabic{footnote}}
\vskip 60pt
\begin{abstract}
	We show that the stringy K\"ahler moduli space of a generic genus one curve of degree $N$, for $N\le 5$, is the $\Gamma_1(N)$ modular curve $X_1(N)$.
	This implies a correspondence between the cusps of the modular curves and certain large volume limits in the stringy K\"ahler moduli spaces of genus one fibered Calabi-Yau manifolds with $N$-sections.
	Using Higgs transitions in M-theory and F-theory as well as modular properties of the topological string partition function, we identify these large volume limits with elements of the Tate-Shafarevich group of the genus one fibration.
	Singular elements appear in the form of non-commutative resolutions with a torsional B-field at the singularity.
	The topological string amplitudes that arise at the various large volume limits are related by modular transformations.
	In particular, we find that the topological string partition function of a smooth genus one fibered Calabi-Yau threefold is transformed into that of a non-commutative resolution of the Jacobian by a Fricke involution.
	In the case of Calabi-Yau threefolds, we propose an expansion of the partition functions of a singular fibration and its non-commutative resolutions in terms of Gopakumar-Vafa invariants that are associated to BPS states with discrete charges.
	For genus one fibrations with 5-sections, this provides an enumerative interpretation for the partition functions that arise at certain irrational points of maximally unipotent monodromy.
\end{abstract}

\end{titlepage}
\vfill \eject

\newpage

\baselineskip=16pt    

\tableofcontents
\section{Introduction}
Compactifications on torus fibered Calabi-Yau manifolds play a pivotal role in the network of string dualities.
This could give reason to believe, that dualities which require such a fibration structure are restricted to small regions of the string landscape.
It is thus reassuring, that scans over the Kreuzer-Skarke list of Calabi-Yau hypersurfaces~\cite{Kreuzer:2000xy} and over complete intersections in products of projective spaces~\cite{CANDELAS1988493,Anderson:2015iia} suggest, that the vast majority of Calabi-Yau manifolds exhibit a torus fibration~\cite{Gray:2014fla,Anderson:2017aux,Huang:2018esr}.
In this way, torus fibrations probe the bulk of the string theory landscape.
Understanding generic properties of torus fibered Calabi-Yau manifolds and their moduli spaces is therefore of particular importance.

Efforts in this direction have largely been restricted to smooth fibrations, with exceptions being e.g.~\cite{Anderson:2013rka,Collinucci:2014taa,Arras:2016evy,Anderson:2017rpr,Grassi:2018wfy}, although the relevance of $\mathbb{Q}$-factorial terminal singularities in F-theory is well understood~\cite{Braun:2014nva,Braun:2014oya,Mayrhofer:2014laa,Arras:2016evy,Grassi:2018rva}.
These singularities can not be resolved with blowups that preserve the Calabi-Yau property.
However, at least in compactifications of Type II string theory, certain singularities of the compact space can also be regularized in terms of \textit{non-commutative resolutions}~\cite{Caldararu:2010ljp}.
The theory of non-commutative resolutions replaces the classical notion of a smooth space by the category of topological branes.
A typical example is a twisted derived category of B-branes, that probe the geometry with a non-trivial B-field ``along'' a vanishing cycle~\cite{Kapustin:1999di,Kapustin:2000aa,Berenstein:2001jr,caldararuThesis,Caldararu:2002ab,Caldararu:2003kt}.
The B-field stabilizes the singularity and obstructs the complex structure deformations that would remove it.
Another example are certain categories of matrix factorizations of Landau-Ginzburg models, that arise as homologically projective duals of certain smooth Calabi-Yau manifolds~\cite{Kuznetsov:2007ab,Caldararu:2010ljp}.
In general, the notion of a Calabi-Yau manifold is replaced by the concept of a Calabi-Yau category (see~\cite{Caldararu:2010ljp} for an accessible introduction).

Results by C\u{a}ld\u{a}raru on twisted derived equivalences~\cite{caldararuThesis,Caldararu:2002ab} suggest a particular non-commutative resolution of the $\mathbb{Q}$-factorial terminal singularities that arise in the fibers of torus fibrations.
To be precise, the singular Jacobian fibration $J$ of a sufficiently generic smooth genus one fibered Calabi-Yau threefold $X$ admits an analytic small resolution $\bar{J}$ which is a non-K\"ahler Calabi-Yau.
The genus one fibration determines an element of the Brauer group $\alpha\in\text{Br}(\bar{J})$, which can be thought of as corresponding to a non-trivial flat B-field, and there is a derived equivalence
\begin{align}
	D^b(X)\cong D^b(\bar{J},\alpha)\,.
\label{eqn:twderintro}
\end{align}
Here $D^b(X)$ is the bounded derived category of coherent sheaves on $X$, which is the category of topological B-branes, and $D^b(\bar{J},\alpha)$ is the bounded derived category of \textit{twisted sheaves}, that admits a corresponding interpretation as the category of B-branes in the presence of a B-field.
The derived equivalence~\eqref{eqn:twderintro} suggests, that the worldsheet theories associated to the Calabi-Yau $X$ and the small resolution $\bar{J}$ with a torsional B-field flow to the same topological sigma model~\cite{Caldararu:2002ab}. 
This naturally leads to the question, whether $D^b(\bar{J},\alpha)$ is a non-commutative resolution of the Jacobian fibration.

Before we continue this thought, let us consider a different line of research that at first appears to be independent.
If the Calabi-Yau is a smooth torus fibration with $N$-sections and $N\le 5$, one can show that the A-model topological string partition function admits an expansion in terms of $\Gamma_1(N)$-Jacobi forms~\cite{Huang:2015sta,Cota:2019cjx,Knapp:2021vkm}.
This can be seen as a consequence of the $\Gamma_1(N)$ monodromy group in the stringy K\"ahler moduli space of the generic fiber~\cite{Schimannek:2019ijf,Cota:2019cjx}.
As we discuss in Section~\ref{sec:strk}, the monodromy group also implies that the moduli space is isomorphic to the modular curve $X_1(N)=\overline{\mathbb{H}/\Gamma_1(N)}$.
A striking consequence is, that the cusps of the modular curve are in one-to-one correspondence with the large volume limits that appear in the large base limit of the fibrations.
The cusps of the modular curve are related by Fricke involutions and $\Gamma_0(M)$ transformations and these also transform the topological string partition functions into each other.
This generalizes a relation among homologically projective dual genus one fibered Calabi-Yau threefolds with $5$-sections that has been found in~\cite{Knapp:2021vkm}.
It also raises a new question, namely what is the geometric interpretation of the different large volume limits?

As the reader might have already guessed, we propose that the additional large volume limits correspond to non-commutative resolutions of fibrations with $\mathbb{Q}$-factorial terminal singularities.
Using Higgs transitions in F- and M-theory, we can interpret the limits as arising from an elliptic fibration via extremal transition.
Modular properties of the topological string partition function allow us to deduce the charges of the Higgs field and the presence of a non-trivial B-field along the singularity directly from the stringy K\"ahler moduli space of the generic fiber.
This identifies the singular geometries as particular elements of the Tate-Shafarevich group of the genus one fibration\footnote{In the presence of multiple fibers, one has to consider the Weil-Ch\^atelet group, which contains the Tate-Shafarevich group as a subgroup~\cite{dolgachev1992elliptic,Bhardwaj:2015oru,Anderson:2018heq,nadirThesis}. To avoid the associated technical issues, we restrict ourselves to fibrations without multiple fibers and also assume that the fibrations are flat. However, most of our results only depend on the properties of the generic fiber and we expect those to remain valid even when these assumptions are dropped.} and the presence of a B-field ``along'' the vanishing cycles at the singularities can be seen as the smoking gun of a non-commutative resolution.
In particular, we always find the Jacobian fibration with a torsional B-field in the limit where the volume of the generic fiber of the smooth genus one fibrations goes to zero.
Let us point out, that for smooth genus one fibered Calabi-Yaus with $N$-sections and $N\le 5$, the generic fiber can always be brought into a certain normal form of degree $N$ curves.
Our results only depend on the properties of the stringy K\"ahler moduli spaces of these normal forms and therefore apply to all of the corresponding fibrations.

Although we can demonstrate the presence of a torsional B-field, we will remain largely agnostic about the precise nature of the non-commutative resolution.
However, the fact that the large volume limits are realized in the same stringy K\"ahler moduli space implies that the associated brane categories are derived equivalent.
Interpreting the non-commutative resolutions as twisted derived categories on analytic small resolutions would then provide a stringy realization of the twisted derived equivalence~\eqref{eqn:twderintro}.
This supports the idea, that the non-commutative resolution arises as an infrared fixed point from the strings propagating on the analytic small resolution in the presence of a B-field.
Another interpretation in terms of large blowups that violate the Calabi-Yau condition will be discussed in Section~\ref{sec:derivedEquivalences}.
It could well be that both realizations are equally valid and flow to the same topological sigma model in the infrared.

We further support our interpretation of the large volume points by studying the enumerative information that is contained in the topological string partition functions.
Before we elaborate on this, recall that F-theory can not distinguish between different elements of the same Tate-Shafarevich group and the group itself is conjectured to be the discrete gauge symmetry of the compactifications~\cite{deBoer:2001wca,Braun:2014oya,Anderson:2014yva,Mayrhofer:2014laa,Cvetic:2015moa}.
This changes in the dual M-theory vacua, that are obtained from F-theory by compactifying on an additional circle.
The M-theory compactifications on the various elements of the Tate-Shafarevich group are conjectured to differ in the value of a discrete Wilson line along the circle.
This can break the discrete gauge symmetry and the full symmetry is only preserved in the compactification on the Jacobian fibration.
Assuming completeness of the charge lattice one can argue, that the compactifications with non-trivial discrete gauge group correspond to the fibrations with $\mathbb{Q}$-factorial terminal singularities.

On smooth Calabi-Yau threefolds, the topological string partition function is the generating function of Gopakumar-Vafa (GV) invariants.
The latter are defined as traces of multiplicities of BPS particles in the five-dimensional M-theory compactification that are invariant under complex structure deformations~\cite{Gopakumar:1998ii,Gopakumar:1998jq}.
We propose a generalized expansion of the partition functions of a singular fibration and its different non-commutative resolutions in terms of multiplicities of BPS particles with discrete charges. 
To extract these invariants, one has to combine the various partition functions.
Except for genus one fibrations with $4$-sections, we find that all of the relevant resolutions of singular elements of the Tate-Shafarevich group are realized as large volume limits in the moduli space of the smooth genus one fibration.
Together with the partition function of a smooth deformation this allows us to explicitly calculate the invariants.
In the case of fibrations with $4$-sections, the remaining singular geometries and resolutions can be found in the stringy K\"ahler moduli space of a genus one fibration with $2$-sections.
In the case of genus one fibrations with $5$-sections, the expansion in terms of Gopakumar-Vafa invariants with discrete charges naturally provides an enumerative interpretation of the partition functions with coefficients in $\mathbb{Q}[\sqrt{5}]$ that arise at certain irrational points of maximally unipotent monodromy.
Comparing the invariants to the multiplicities of the charged half-hypermultiplets in the M-theory compactification, together with the integrality of the invariants, provides a strong check of our proposal.

To avoid confusion, let us briefly comment on our conventions.
In this paper we will use the name torus fibration if we want to remain agnostic about the existence of a section and use genus one fibration to refer to a fibration without a section.
An elliptic fibration is a torus fibration with a section.
We will always assume that the fibration is flat and does not have multiple fibers.

The structure of the paper is as follows.
In Section~\ref{sec:ts} we review the relevant background on the Tate-Shafarevich group and compactifications of F-theory, M-theory and Type IIA string theory on torus fibered Calabi-Yau manifolds.
Of particular importance will be the interpretation of different elements of the Tate-Shafarevich group in terms of Higgs transitions in M-theory that is discussed in Section~\ref{sec:higgsMF}.
The expansion of the topological string partition function on non-commutative resolutions, and also Calabi-Yau manifolds with torsional cohomology, in terms of GV-invariants with discrete charges is proposed in Section~\ref{sec:gvtorsion}.
Then, in Section~\ref{sec:modularstructure}, we discuss the modular structure of the topological string partition function and illustrate our general results at the example of genus one fibrations with 2-sections.
This is followed by an in-depth study of the stringy K\"ahler moduli spaces of the normal forms of degree $N$ curves and analogous discussions for genus one fibrations with $N$-sections for $N\le 5$ in Section~\ref{sec:strk}.
Detailed examples and calculations of GV-invariants with discrete charges are provided in Section~\ref{sec:ex2sec} and~\ref{sec:ex4sec}.
A third example in Section~\ref{sec:ex5sec} illustrates the relation of the Gopakumar-Vafa invariants to irrational MUM-points in the case of genus one fibrations with $5$-sections.
In Section~\ref{sec:outlook} we discuss some relations to previous work and give an outlook on future research directions.

\section*{Acknowledgments}
We would like to thank Cesar Fierro Cota, Albrecht Klemm, Johanna Knapp, Paul Oehlmann and Emanuel Scheidegger for many fruitful discussions and collaborations on related projects.
We are particularly grateful to Emanuel Scheidegger, Johanna Knapp, Albrecht Klemm and Paul Oehlmann for valuable comments on the draft.
We further want to thank Jan Manschot for an inspiring question regarding the cusps of the modular curve, that we hope to have answered conclusively in this paper.
The work of the author is supported by the Austrian Science Fund (FWF): P30904-N27.

\section{String Theory and the Tate-Shafarevich group}
\label{sec:ts}
The Tate-Shafarevich group associated to a Jacobian fibration is closely connected to various aspects of string theories on torus fibrations.
Yet, perhaps due to the daunting mathematical formalism surrounding it, some of the details of that connection still remain to be understood.
A mathematical analysis of the Tate-Shafarevich group for threefolds has been performed in~\cite{dolgachev1992elliptic} and for an accessible introduction we refer to~\cite{caldararuThesis}.
Physical discussions can be found in~\cite{Braun:2014oya,Cvetic:2015moa,Cvetic:2018bni}.
Already to define it as a mathematical object requires some work and we will brush over most of the technical details.

\subsection{The Tate-Shafarevich group}
\label{sec:tsgroup}
Every torus fibration $X$ has an associated Jacobian fibration $J$ that is, roughly speaking, the fibration of moduli spaces of degree zero line bundles on the fibers of $X$.
The trivial line bundle identifies a unique point on the fibers of $J$ and they combine into a section, making the Jacobian fibration elliptic.
Similarly, the torus fibration $X$ itself can be interpreted, again ignoring various subtleties related to the singular fibers, as the fibration of moduli spaces of degree one line bundles on its fibers.

By tensoring the associated bundles on the fibers, the Jacobian fibration $J$ acts on the torus fibration $X$ which in that way becomes a $J$-torsor.
In a non-trivial manner, the $J$-torsors for a given elliptic fibration $J$ can be equipped with a group structure and form the Weil-Ch\^{a}telet group $\text{WC}(J)$.
Of course, this group depends on the choice of a fibration structure which we keep implicit in our notation.
Details about the group structure can be found in Appendix~A of~\cite{Cvetic:2015moa} but will not be relevant for our discussion.

The subgroup of the Weil-Ch\^{a}telet group of torus fibrations that are locally elliptic is called the Tate-Shafarevich group $\Sh(J)$.
The elements are essentially the fibrations that do not have multiple fibers.
In the case of hypersurfaces, multiple fibers arise when the defining polynomial over a point in the base becomes a product of factors that have exponents with a common divisor greater than one.
A more general definition can be found in~\cite{griffiths1978principles} and from now on we will only consider fibrations without multiple fibers.
In a minor abuse of notation, we will also write $\Sh(X)$ to denote the Tate-Shafarevich group associated to the Jacobian of a genus one fibration $X$.

The Tate-Shafarevich group of torus fibered threefolds has been studied in~\cite{dolgachev1992elliptic}.
The simplest situation arises, when a threefold $X$ is a smooth elliptic fibration without reducible fibers.
If we denote the base of the fibration by $B$, one can then show that
\begin{align}
	\Sh(X)=\text{coker}\left(\text{Br}^\prime(B)\rightarrow\text{Br}^\prime(X)\right)\,,
	\label{eqn:smoothTS}
\end{align}
where $\text{Br}^\prime(X)$ is the cohomological Brauer group.
On Calabi-Yau threefolds and smooth surfaces, the cohomological Brauer group is identical to the ordinary Brauer group $\text{Br}(X)$ and both are torsion groups.
In particular, if $X$ is a Calabi-Yau threefold one has the isomorphism~\cite{Batyrev:2005jc}
\begin{align}
	\text{Br}(X)\cong \text{Tors}\left(H^3(X,\mathbb{Z})\right)\,.
	\label{eqn:brauerisomorphism}
\end{align}
We will talk more about the Brauer group in Section~\ref{sec:derivedEquivalences}.

Unfortunately, the relation~\eqref{eqn:smoothTS} provides a useful intuition but is not directly applicable to genus one fibered Calabi-Yau threefolds.
Responsible are the $I_2$ fibers which lead to singularities in the Jacobian fibration.
On a smooth genus one fibration, they are resolved and consist of two $\mathbb{P}^1$'s that intersect transversely in two points.
However, the same resolution is non-crepant on the Jacobian fibration and would violate the Calabi-Yau condition.
More precisely, the singularities are $\mathbb{Q}$-factorial terminal singularities and physically correspond to localized uncharged matter~\cite{Arras:2016evy}.
We will elaborate on this in Section~\ref{sec:higgsMF}.
In general, the Tate-Shafarevich group also contains genus one fibrations with $\mathbb{Q}$-factorial terminal singularities.

To be able to calculate the Tate-Shafarevich group for torus fibered Calabi-Yau threefolds, we therefore have to restrict ourselves to fairly simple fibrations.
A torus fibration $X$ over $B$ is considered \textit{generic}~\cite{Caldararu:2002ab}, if both $X$ and $B$ are smooth, the fibration does not have non-flat fibers, multiple fibers or fibral divisors and the fibers over the discriminant locus have Kodaira type $I_1,\,I_2$ or $II$.
Note that this excludes F-theory compactifications with non-Abelian gauge symmetry.
Further assuming that the base is $\mathbb{P}^2$, one can then use an argument made in~\cite{dolgachev1992elliptic}, there applied to fibrations of cubics, and calculate the Tate-Shafarevich group explicitly.
It follows that if $X$ is a generic torus fibered Calabi-Yau threefold over $\mathbb{P}^2$ that has an $N$-section, with $N$ minimal, and torsion free cohomology,
\begin{align}
	\Sh(X)=\mathbb{Z}_N\,. 
	\label{eqn:ZNTS}
\end{align}
Moreover, $X$ will be an element of order $N$ in this group.
In all of our examples, the Tate-Shafarevich group will contain at least one element that satisfies these requirements.
However, many of the results in this paper follow from physical arguments and properties of the generic fiber.
We expect those to carry over also to non-generic Calabi-Yau threefolds, fibrations over different bases and to $d$-folds with $d>3$.

There is only one generic elliptically fibered Calabi-Yau threefold over $\mathbb{P}^2$ with $h^{1,1}=2$.
This is the degree $18$ hypersurface $X^{(1)}_0$ in the weighted projective space $\mathbb{P}_{11169}$, which has been studied e.g. in~\cite{Candelas:1994hw,Huang:2015sta}.
However, the generic smooth genus one fibered Calabi-Yau threefolds that have $N$-sections with $N>1$ and two K\"ahler parameters are not unique.
They differ in the number of the $I_2$ singular fibers that have a component which intersects the $N$-section once.
Nevertheless, the structures that we are studying in this paper are essentially the same for all of these geometries and we will consider one representative case $X^{(N)}_1$ for each value $1<N\le 5$.
The corresponding Jacobian fibrations will be denoted by $X^{(N)}_0$.
The geometries that are associated to the elements of the corresponding $\mathbb{Z}_N$ Tate-Shafarevich group are listed in Table~\ref{tab:tselements}.
\begin{table}[h!]
	\begin{align*}
		{\def\arraystretch{1.4}
		\begin{array}{c|cccccc}
			&n\in\mathbb{Z}_N=0&1&2&3&4&5\\\hline
			N=1&{J^{(1)}=X^{(1)}_0}&&&&&\\
			2&{\color{blue}J^{(2)}=X^{(2)}_0}&X^{(2)}_1&&&&\\
			3&{\color{blue}J^{(3)}=X^{(3)}_0}&X^{(3)}_1&X^{(3)}_1&&&\\
			4&{\color{blue}J^{(4)}=X^{(4)}_0}&X^{(4)}_1&{\color{blue}X^{(4)}_2}&X^{(4)}_1&&\\
			5&{\color{blue}J^{(5)}=X^{(5)}_0}&X^{(5)}_1&X^{(5)}_2&X^{(5)}_2&X^{(5)}_1&
		\end{array}
		}
	\end{align*}
	\caption{The geometries that are associated to the elements of the $\mathbb{Z}_N$ Tate-Shafarevich group of representative generic torus fibered Calabi-Yau threefolds over $\mathbb{P}^2$ with $h^{1,1}=2$.
	Singular geometries are highlighted in blue.}
	\label{tab:tselements}
\end{table}

\subsection{M- and F-theory on torus fibered Calabi-Yau threefolds}
\label{sec:mftheory}
In F-theory compactifications on torus fibered Calabi-Yau manifolds, the discrete symmetry is conjectured to be isomorphic to the Tate-Shafarevich group of the fibration~\cite[p.~63]{deBoer:2001wca}\cite{Braun:2014oya}.
The connection can be understood, using the duality with M-theory and Higgs transitions from continuous to discrete gauge symmetries.
This perspective provides a valuable tool to study the group also for general torus fibered Calabi-Yau manifolds.
Before we elaborate on this, let us first review the relevant general aspects of the physics of F-theory and the dual M-theory compactifications.
This will serve to fix our notation and to keep the presentation somewhat self-contained. For a thorough review of F-theory we refer the reader to~\cite{Weigand:2018rez,Cvetic:2018bni}.

Considering F-theory on an elliptically fibered Calabi-Yau $3$-fold $X$, the rank $k$ of the continuous Abelian part in the gauge group of the six-dimensional effective theory geometrically corresponds to the rank of the free part of the Mordell-Weil group $\text{MW}(X)$~\cite{Morrison:1996pp}.
The Mordell-Weil group is formed by the sections of the fibration.
On generic fibers, concretely those that are irreducible and reduced, the addition of sections follows from the ordinary group law on elliptic curves.
Explanations of the Mordell-Weil group can be found for example in~\cite{silverman2009arithmetic,Weigand:2018rez}.

The relation of this group with Abelian gauge symmetries can be seen by compactifying on an additional circle, that is considering F-theory on $X\times S^1$.
One can then use a dual description in terms of M-theory on $X$.
M-theory contains a 3-form field $C_3$, that can be expanded along the harmonic $2$-forms $\omega^i,\,i=0,\ldots,k$, which are associated to the sections in cohomology, as
\begin{align}
	C_3=\ldots+\sum A_i\wedge\omega^i\,.
\end{align}
This leads to $\text{MW}(X)+1$ Abelian gauge fields $A_i$ in five dimensions. 
However, the effective supergravity has obtained an additional Kaluza-Klein $U(1)$ gauge symmetry due to the circle compactification.
Therefore only $\text{MW}(X)$ of the $U(1)$ factors are preserved after decompactifying to six dimensions.

The cohomology classes that lead to the Kaluza-Klein gauge symmetry and to the symmetries that have a six-dimensional origin can be constructed explicitly.
To this end, we first choose a $0$-section $s_0$ on the elliptic fibration.
This can be used to construct the Shioda map $\sigma:\,\text{MW}(X)\rightarrow \text{NS}(X)$, which is a homomorphism from the Mordell-Weil group to the N\'eron-Severi group $\text{NS}(X)$ of the fibration~\cite{shioda1}.
Recall that the N\'eron-Severi group is formed by the Cartier divisors modulo algebraic equivalence.
The homomorphism maps a section $s_1\in\text{MW}(X)$ to the divisor
\begin{align}
	\sigma(s_1)=S_1-S_0+[K_B]-\pi^*\pi(S_1\cdot S_0)+\ldots\,.
	\label{eqn:shioda}
\end{align}
Here we use $S_i$ to denote the divisor class associated to a section $s_i$ and $K_B$ is the canonical divisor of the base of the fibration.
We have omitted the contributions from fibral divisors.
The images of the generators of the Mordell-Weil group under the Shioda map are precisely dual to those elements in cohomology that lead to $U(1)$ gauge symmetries which lift to six dimensions~\cite{Park:2011ji,Morrison:2012ei}.
On the other hand, the element that is associated to the Kaluza-Klein $U(1)$ is given by
\begin{align}
	\tilde{S}_0=S_0-\frac12[K_B]\,.
	\label{eqn:kksection}
\end{align}
Abelian gauge symmetry can also arise in compactifications on genus one fibrations with multiple $N$-sections that are independent in cohomology.
This will not be relevant for our discussion and we refer the interested reader to~\cite{Klevers:2014bqa,Cvetic:2018bni}.

A $U(1)$ gauge symmetry can be broken to $\mathbb{Z}_n$, by giving a non-vanishing vacuum expectation value to a massless scalar field of charge $n$.
In F-theory compactifications on Calabi-Yau threefolds, the charged scalar fields are contained in six-dimensional hypermultiplets.
The hypermultiplets are in turn composed of two five-dimensional half-hypermultiplets which transform in conjugate representations of the gauge group.
To avoid unnecessary details, we will now assume that the Calabi-Yau threefold $X$ does not contain any fibral divisors, multiple fibers or non-flat fibers.
The six-dimensional hypermultiplets and the corresponding five-dimensional half-hypermultiplets then fall into three categories with different geometric origins:
\begin{itemize}
	\item \textbf{Charged matter}
		In the dual five-dimensional M-theory compactification on $X$, the charged half-hypermultiplets arise from M2-branes that wrap components of resolved $I_2$ fibers.
		Denoting the component by $C$, the charge under the six-dimensional $U(1)$ that corresponds to a section $s_1$ is given by $\sigma(s_1)\cdot C$.
		On the other hand, the Kaluza-Klein charge is given by $\tilde{S}_0\cdot C$.
		The mass of a half-hypermultiplet with $U(1)$ charge $q$ and KK-charge $n$ is given by
		\begin{align}
			m_{n,q}=|q\xi+n|\,,
			\label{eqn:hhmass}
		\end{align}
		where $\xi$ is the expectation value of the $U(1)$ Wilson line along the circle and we have normalized the Kaluza-Klein scale to one.
		Geometrically, the mass can be identified with the volume of $C$ and the half-hypermultiplet becomes massless on the boundary of the K\"ahler cone where $C$ shrinks to a point.
		Only states with Kaluza-Klein charge $0$ lift to six-dimensions and every isolated $I_2$ fiber leads to a pair of half-hypermultiplets with opposite charges that combine to one six-dimensional massless hypermultiplet.

		It is important to note that, starting from a six-dimensional theory, every hypermultiplet of a given charge $\pm q$ leads to an infinite tower of five-dimensional states with Kaluza-Klein charges $n\in\mathbb{Z}$.
		Consider a rational irreducible component $C$ of a resolved $I_2$ fiber that leads to a half-hypermultiplet with $U(1)_{KK}\times U(1)$ charge $(n,q)$.
		The M2-brane that wraps $C\pm T^2$, with $T^2$ being the class of the generic fiber, leads to a state with charge $(n\pm1,q)$.

	\item \textbf{Uncharged localized matter}
		As we discussed in the previous section, $X$ can contain $I_2$ singular fibers that can not be resolved without violating the Calabi-Yau condition.
		These give rise to $\mathbb{Q}$-factorial terminal singularities of the fibration.
		In the case of charged half-hypermultiplets, the volume of an exceptional curve that resolves the fiber was related to the expectation value of a $U(1)$ Wilson line.
		This generates a mass term for charged particles.
		Correspondingly, the singular fibers lead to half-hypermultiplets that are not charged under any gauge symmetry except for the Kaluza-Klein $U(1)$~\cite{Arras:2016evy}.
		From this perspective, the origin of the singularities in the Jacobian fibration of a genus one fibered Calabi-Yau threefold will become clear when we consider Higgs transitions.
		Again, two uncharged localized half-hypermultiplets of Kaluza-Klein charge $0$ combine into one hypermultiplet that lifts to six dimensions.

	\item \textbf{Uncharged unlocalized matter}
		Uncharged matter is also present in compactifications on smooth fibrations.
		In M-theory, the elementary unlocalized half-hypermultiplets arise from expanding $C_3$ along the harmonic forms associated to generators of $H^3(X,\mathbb{Z})$.
		All of those are massless and can be lifted to six dimensions.
		This leads to $h^{2,1}(X)+1$ hypermultiplets in F-theory.
\end{itemize}
Finally, let us note that the vertical divisors that arise from the base $B$ of the fibration lead to $b_2(B)-1$ tensor multiplets in six dimensions, which in five dimensions can also be dualized into vector mulitiplets.

Having discussed the geometric origin of $U(1)$ gauge symmetries and hypermultiplets in F-theory and the dual M-theory compactifications, let us now come to the discrete symmetries.

\subsection{Discrete symmetries and Higgs transitions in M- and F-theory}
\label{sec:higgsMF}
The physics of F-theory compactified on a torus fibered Calabi-Yau $X$ only depends on the associated Jacobian fibration.
In that way, every element of a Tate-Shafarevich group leads to the same effective theory.
On the other hand, as mentioned before, the discrete symmetry of an F-theory compactification is conjecturally isomorphic to the Tate-Shafarevich group itself.

In five dimensions, the discrete symmetries of M-theory on smooth Calabi-Yau threefolds arise from torsional 3-forms~\cite{Camara:2011jg,Mayrhofer:2014laa}.
A $k$-torsional 3-form $\alpha$ implies the existence of a non-exact 2-form $w$ such that
\begin{align}
	k\cdot\alpha =dw\,.
\end{align}
One can then expand the M-theory 3-form $C_3$ along $w$ and obtain the gauge potential of a massive $U(1)$ symmetry that is broken to $\mathbb{Z}_k$.
Moreover, on Calabi-Yau threefolds there is an isomorphism
\begin{align}
\text{Tors}\left(H^3(X,\mathbb{Z})\right)\cong\text{Tors}\left(H_2(X,\mathbb{Z})\right)\,.
\end{align}
The right-hand side can be identified with the discrete charge lattice, and the corresponding states are realized by M2-branes that wrap torsional 2-cycles.

This seemingly poses a puzzle, because a smooth genus one fibered Calabi-Yau leads to a discrete symmetry in F-theory but not in the dual M-theory vacuum, that one obtains after circle compactification.
The resolution is provided by the additional degrees of freedom to turn on a discrete Wilson line along the compactification circle~\cite{deBoer:2001wca,Braun:2014oya,Anderson:2014yva,Mayrhofer:2014laa,Cvetic:2015moa}.
Taking this into account, the M-theory compactifications on the different elements of the Tate-Shafarevich group are dual to F-theory compactified on an additional circle with different choices for this ``flux''.
A non-vanishing Wilson line breaks parts of the discrete symmetry.
As a consequence, M-theory on a smooth genus one fibered Calabi-Yau $X$ should be dual to the F-theory compactification on $X\times S^1$ with a choice of discrete Wilson line that breaks the discrete symmetry completely.
On the other hand, M-theory on the associated Jacobian fibration $J(X)$ is dual to F-theory on $X\times S^1$ with a vanishing discrete Wilson line.
If the Jacobian fibration of a genus one fibered Calabi-Yau threefold were smooth, this could be seen as a physical interpretation of the relation~\eqref{eqn:smoothTS}.

However, the singularities force us to take a more indirect approach.
It is widely believed, that consistent theories of quantum gravity can not exhibit any global symmetries~\cite{Banks:2010zn,Hellerman:2010fv}.~\footnote{A proof in the context of AdS/CFT duality has been proposed in~\cite{Harlow:2018tng}.}
As a consequence, discrete symmetries should also be gauged.
They are then assumed to be associated to massive vector fields, that acquired a mass from the breaking of continuous gauge symmetries due to the vacuum expectation value of a Higgs field.
Every F-theory compactification with a discrete symmetry should therefore be connected via ``un-Higgsing'' to another compactification, where the discrete symmetry is embedded into an unbroken continuous gauge symmetry.
In the cases that we study in this paper, the theory with a continuous gauge group arises from an F-theory compactification on a smooth elliptically fibered Calabi-Yau threefold.
The Higgs transition geometrically corresponds to an extremal transition from this elliptic fibration to a genus one fibration.

Let us assume that $X$ is a smooth elliptically fibered Calabi-Yau with $\text{rk}\,\text{MW}(X)=1$ and no fibral divisors, multiple fibers or non-flat fibers.
We denote the F-theory gauge group by $U(1)$ and the additional Kaluza-Klein gauge symmetry in the dual M-theory compactification by $U(1)_{KK}$.
This necessarily implies that the fibration contains $I_2$ fibers, otherwise there would be no 2-cycle that is dual to the additional section, and we assume that some of them lead to half-hypermultiplets of $U(1)$ charge $\pm q$.
As we discussed in the previous section, the half-hypermultiplets are part of an infinite tower of half-hypermultiplets that populate the Kaluza-Klein charge lattice $\mathbb{Z}$.
Moreover, the mass of a half-hypermultiplet with $U(1)_{KK}\times U(1)$ charge $(n,q)$ is given by
\begin{align}
	m_{n,q}=|q\xi + n\tau|\,.
	\label{eqn:mass}
\end{align}
with $\xi$ being the vacuum expectation value of the $U(1)$ Wilson line along the circle and $\tau$ the Kaluza-Klein scale.

To be able to give the scalar fields in the half-hypermultiplets with charge $(n,q)$ a non-zero vacuum expectation value, we first have to fix the value of the Wilson line
\begin{align}
	\xi =-\frac{n}{q}\tau\,,
	\label{eqn:wilsonline}
\end{align}
such that the scalar particles become massless.
This corresponds to a K\"ahler deformation of the Calabi-Yau metric which shrinks the corresponding curve to a point.
Turning on a vacuum expectation value for the massless scalar field then breaks the gauge group to the subgroup of transformations that leave the field invariant.
Geometrically, the vacuum expectation value for the scalar fields in half-hypermultiplets corresponds to a complex structure deformation.
This removes the $I_2$ singular fibers associated to the charge $q$ states and completes the extremal transition.

To determine the subgroup of the gauge group that remains unbroken after the Higgs transition, one first needs to use the freedom of performing a unimodular transformation
\begin{align}
	\left(\begin{array}{c}
		U(1)_{KK}\\
		U(1)
	\end{array}\right)\mapsto\left(\begin{array}{c}
		U(1)_1'\\
		U(1)_2'
	\end{array}\right)=\left(\begin{array}{cc}
		a&b\\
	c&d\end{array}\right)\left(\begin{array}{c}
		U(1)_{KK}\\
		U(1)
	\end{array}\right)\,,
	\label{eqn:unimodU1trafo}
\end{align}
with $a,b,c,d\in\mathbb{Z}$ and $ad-bc=1$, such that the Higgs field is only charged under $U(1)_2'$.
Using $k=\text{gcd}(n,q)$, the unbroken subgroup is then given by
\begin{align}
	U(1)_1'\times\mathbb{Z}_k\,,\quad\text{with}\quad U(1)_1'=\frac{q}{k}\cdot U(1)_{KK}-\frac{n}{k}\cdot U(1)\,.
\end{align}

A change of basis~\eqref{eqn:unimodU1trafo} can also be used to assume that, without loss of generality, the KK-charge takes values in the range $0\le n\le q$.
This leads to $n$ physically inequivalent choices for the Wilson line~\eqref{eqn:wilsonline} and the corresponding massless Higgs fields.
The M-theory Higgs transition with $n=1$ breaks the discrete gauge symmetry completely and conjecturally corresponds to an extremal transition from the elliptic fibration $X$ to a genus one fibration $X'$ with $q$-sections.
On the other hand, the transition associated to $n=0$ connects it to the Jacobian fibration $J(X')$ and the remaining $n-2$ values lead to the other elements of the Tate-Shafarevich group. 

Note that multiplying all $U(1)_{KK}$ charges with $-1$ does not change the physics, as it can be compensated by choosing a different generator for the gauge symmetry.
This reflects the fact that an element of the Tate-Shafarevich group and its inverse correspond to the same torus fibration that is equipped with opposite actions of the Jacobian~\cite{Cvetic:2015moa}.

The Higgs transitions also provide a physical explanation why the Jacobian fibration of a smooth genus one fibered Calabi-Yau threefold is singular.
Closely related to the absence of global symmetries, a consistent theory of quantum gravity is conjecture to contain states that populate every point in the charge lattice~\cite{Banks:2010zn}.
This implies that if the Higgs field carries charges $(n,q)$ with $k=\text{gcd}(n,q)>1$, there should also be states with charges
\begin{align}
	(n',q')=\frac{m}{k}\left(n,\,q\right)\,,\quad m=1,\ldots,k-1\,.
	\label{eqn:lcharges}
\end{align}
These also have to arise from isolated $I_2$ fibers in the elliptic fibration and, as can be seen from~\eqref{eqn:mass}, become massless for the same choice of Wilson line~\eqref{eqn:wilsonline}.
After performing the Higgs transition, those states are not charged under the remaining six-dimensional gauge symmetry and thus lead to uncharged localized matter.
Geometrically, the extremal transition only removes the $I_2$ fibers associated to the $(n,q)$ states.
The fibers corresponding to the states~\eqref{eqn:lcharges} end up with $\mathbb{Q}$-factorial terminal singularities.

\subsection{Type II strings and (twisted) derived equivalences}
\label{sec:derivedEquivalences}

Compactifying further down to four dimensions, M-theory on $X\times S^1$ and F-theory on $X\times T^2$ are dual to Type IIA string theory on $X$.
In Type IIA string theory on smooth Calabi-Yau threefolds, discrete symmetries arise, just as in M-theory, from non-trivial torsion cohomology
\begin{align}
\text{Tors}\left(H^3(X,\mathbb{Z})\right)\cong\text{Tors}\left(H_2(X,\mathbb{Z})\right)\,.
\end{align}
The discussion from the previous section still applies, with the M-theory 3-form field and the M2-branes being replaced by the Type IIA Ramond-Ramond 3-form field and D2-branes.

However, in Type IIA string theory, the torsional 2-cycles open up yet another degree of freedom, namely the choice of a non-trivial B-field along those cycles.
As we are going to discuss in more detail in Section~\ref{sec:gvtorsion}, such a choice is best understood as a homomorphism
\begin{align}
	b:\,H_2(X,\mathbb{Z})_{\text{tors.}}\rightarrow\mathbb{C}^\times\,.
\end{align}
The group of such homomorphisms is itself isomorphic to $H_2(X,\mathbb{Z})_{\text{tors.}}$.
Again, ignoring the singularities and assuming the absence of torsional cycles in the base of the fibration, the torsional B-field choices in a Type IIA compactification on the Jacobian fibration $J(X)$ of a smooth genus one fibered Calabi-Yau threefold $X$ appear to be captured by $\Sh(J)$.

The physical effect of such a discrete B-field is closely connected to the relation~\eqref{eqn:smoothTS}, of the Tate-Shafarevich group to the Brauer groups of the fibration and the base.
The elements of the Brauer group are characteristic classes of \textit{1-gerbes}.
Gerbes are generalizations of line bundles, with line bundles being 0-gerbes, and a 1-gerbe can be interpreted as a twist of a line bundle, such that the composition of the transition functions on triple overlaps is no longer the identity.
From a string theory perspective, a flat B-field is a connection on a 1-gerbe and the bundles on D-branes, in the presence of a non-trivial B-field, are twisted by the gerbe~\cite{Kapustin:1999di,Kapustin:2000aa,Hitchin:1999fh,Berenstein:2001jr,Caldararu:2002ab,Caldararu:2003kt}.
An introduction to gerbes for string theorists can be found for example in~\cite{Hitchin:1999fh,Sharpe:1999pv,Sharpe:1999xw,Sharpe:2003dr}.

Restricting attention to the subsector that is captured by the topological string, the (untwisted) topological B-branes can be described as elements of the bounded derived category of coherent sheaves $D^b(X)$~\cite{Kontsevich:1994dn,Douglas:2000gi,Aspinwall:2004jr}.
The latter is constructed from the category of bounded complexes of coherent sheaves by identifying complexes that are quasi-isomorphic.
Two complexes are quasi-isomorphic if there exists a map between the complexes that induces an isomorphism on the cohomologies.
In this way, every element $\mathcal{F}^\bullet\in D^b(X)$ can be represented by a bounded complex of locally free sheaves,
\begin{align}
	\mathcal{F}^\bullet=0\rightarrow\ldots\rightarrow\mathcal{F}_{i-1}\rightarrow\mathcal{F}_i\rightarrow\mathcal{F}_{i+1}\rightarrow\ldots\rightarrow 0\,.
	\label{eqn:lfcomplex}
\end{align}
Recall that locally free sheaves are just vector bundles on $X$.
Physically speaking, a vector bundle on $X$ can be thought of as a 6-brane, potentially with some lower dimensional brane charges dissolved on the worldvolume.
Vector bundles at odd-positions in the complex~\eqref{eqn:lfcomplex} are identified with the corresponding anti-branes.\footnote{The $\mathbb{Z}_2$ grading of branes and anti-branes is refined into a ghost number grading taking values in $\mathbb{Z}$.
This can be derived from the open string worldsheet theory and is necessary to reproduce the spectrum of boundary operators as morphisms in the category of branes.}
Quasi-isomorphic complexes can be deformed into each other by D-term deformations on the boundary of the open string worldsheet theory and those are washed out after renormalization group flow to the infrared fixed point~\cite{Herbst:2008jq}.
In this way the derived category captures the infrared limit of branes after tachyon condensation.

A coherent sheaf $\mathcal{E}$ on $X$ is an $\mathcal{O}_X$-module, with $\mathcal{O}_X$ being the structure sheaf on $X$.
This just means, that regular functions on open subsets $U\subseteq X$ act by multiplication on the spaces of sections $\Gamma(U,\mathcal{E}|_U)$.
To obtain twisted sheaves, let us first note that the elements of the Brauer group $\text{Br}(X)$ admit a natural interpretation as equivalence classes of sheaves of Azumaya algebras on $X$~\cite{milne}.
Roughly speaking, those are coherent sheaves of which the sections form an algebra that is locally isomorphic to a matrix algebra.
An introduction can be found e.g. in~\cite{poonen2017rational} but the details won't be relevant for our discussion.
Given a sheaf of Azumaya algebras $\mathcal{A}$ on $X$, the corresponding twisted sheaves are $\mathcal{A}$-modules that are locally coherent as $\mathcal{O}_X$-modules.
With this definition, it is possible to construct a derived category of twisted sheaves $D^b(X,[\mathcal{A}])$, which only depends on the equivalence class of $\mathcal{A}$ in $\text{Br}(X)$~\cite{Kapustin:2000aa,caldararuThesis,Caldararu:2002ab}.
As mentioned before, this should be thought of as the category of topological B-branes in the presence of a non-vanishing flat B-field.

In many cases, the complexified K\"ahler moduli space of the topological string A-model compactified on a Calabi-Yau $X$ contains another large volume limit that is associated to a Calabi-Yau $Y$.
We can then continuously deform the open topological string along a path that connects the two limits and in this way obtain a functor that identifies the B-branes on both spaces.
This implies that the Calabi-Yaus are derived equivalent, i.e.
\begin{align}
	D^b(X)\cong D^b(Y)\,.
\end{align}
The transport of B-branes can often be carried out explicitly in an associated gauged linear sigma model (GLSM)~\cite{Herbst:2008jq}.
A simple example of different large volume limits in the same stringy K\"ahler moduli space are the phases of the extended K\"ahler cone of a Calabi-Yau threefold.
They correspond to birational Calabi-Yaus, which are known to be derived equivalent~\cite{Bridgeland1}.
On the other hand, a highly non-trivial example is the R{\o}dland pair of a complete intersection Calabi-Yau threefold in the Grassmannian $G(2,7)$ and a Pfaffian Calabi-Yau threefold in $\mathbb{P}^{20}$~\cite{Rodland:2000ab}.
Both are realized as phases in the FI-parameter space of the same non-Abelian GLSM und thus exist in the same stringy K\"ahler moduli space~\cite{Hori:2011pd}.
They are not birational and in this case the derived equivalence is an example of homological projective duality~\cite{Kuznetsov:2007ab}.

A construction analogous to that by R{\o}dland can be used to realize genus one curves of degree $5$ as Pfaffian curves in $\mathbb{P}^4$ and complete intersection curves in $G(2,5)$~\cite{Fisher:2010ab,Hori:2013gga}.
In~\cite{Knapp:2021vkm}, those curves were fibered over surfaces to obtain genus one fibered Calabi-Yau threefolds with $5$-sections.
Again the fibrations come in pairs which are realized in different phases of the same GLSM and are therefore derived equivalence.
This can be interpreted as a relative version of homological projective duality.
It was further argued, that the dual fibrations of Pfaffian curves and of curves in Grassmannians are elements of the same Tate-Shafarevich group.
Over the base $\mathbb{P}^2$, the Tate-Shafarevich group is given by $\mathbb{Z}_5$ and the dual smooth genus one fibrations, each equipped with the two actions of the Jacobian fibration, provide all of the non-trivial elements.
As we discuss now, the derived equivalence can then be seen as a special case of a more general class of twisted derived equivalences among elements of the same Tate-Shafarevich.

Some of the technical issues, that arise due to the $\mathbb{Q}$-factorial terminal singularities of the Jacobian fibration, can be circumvented by performing an analytic small resolution~\cite{Caldararu:2002ab}\footnote{An analytic variety is locally the common vanishing locus of holomorphic functions on $\mathbb{C}^n$. In this sense it is a generalization of the concept of an algebraic variety over $\mathbb{C}$.}.
This does not violate the Calabi-Yau condition but the geometry is no longer K\"ahler.
One can then show that a generic genus one fibered Calabi-Yau threefold $X$ determines an element $\alpha\in\text{Br}^\prime(\bar{J})$ in the Brauer group of the small resolution $\bar{J}$ of the Jacobian fibration.
This in turn can be used to construct the derived category of twisted sheaves $D^b(\bar{J},\alpha)$ and to find a derived equivalence~\cite{Caldararu:2002ab}
\begin{align}
	D^b(X)\cong D^b(\bar{J},\alpha)\,.
	\label{eqn:twistedDerivedEquivalence}
\end{align}
If $n$ is the order of $\alpha$ in $\text{Br}(\bar{J})$ and $k$ is coprime to $n$, one can also show that
\begin{align}
	D^b(\bar{J},\alpha)\cong D^b(\bar{J},\alpha^k)\,.
\end{align}
Recall, that the derived category naturally identifies branes that flow to the same configuration in the infrared.
As pointed out in~\cite{Caldararu:2002ab}, the fact that the twisted derived category $D^b(\bar{J},\alpha)$ is derived equivalent to the ordinary category of topological B-branes on $X$ suggests that they correspond to the same topological sigma model.
This would imply that the worldsheet theory on $\bar{J}$ with a torsional B-field flows to a non-commutative resolution of the singular fibration $J$.

Another resolution of the singularities proceeds via large blowups, that destroy the Calabi-Yau property and introduce exceptional divisors.
This has been performed explicitly for genus one fibrations with $N$-sections for $N\in\{2,3\}$ in~\cite{dolgachev1992elliptic,Mayrhofer:2014laa} and it was shown that the resolved geometry contains $N$-torsional 2-cycles.
For the fibration with $3$-sections over $\mathbb{P}^2$ that was studied in~\cite{dolgachev1992elliptic}, it was explicitly shown that the Tate-Shafarevich group of the large resolution is again $\mathbb{Z}_3$ and can be identified with the torsional 3-cohomology.
Similar large blowups have been used in~\cite{Aspinwall:1995rb} and were argued to correspond to irrelevant deformations of the worldsheet theory.
This suggests a second interpretation of the non-commutative resolution as the infrared fixed point of the sigma model associated to  a non-crepant resolution with a flat B-field along torsional 2-cycles.

In the following we will remain largely agnostic about the precise nature of the non-commutative resolution.
However, both of the potential realizations imply, that for the purpose of choosing a B-field the singularity looks effectively like a torsional 2-cycle.
We will verify in particular examples that the effective torsion 2-homology is precisely the unbroken discrete gauge symmetry in the M-theory and Type IIA compactification.
As we discuss in more detail in the next section, the choice of a B-field along an $N$-torsional 2-cycle corresponds to a homomorphism
\begin{align}
	b:\,\mathbb{Z}_N\rightarrow \mathbb{C}^\times\,.
\end{align}
This is completely determined by the image of the generator
\begin{align}
	b(1)=\exp\left(2\pi i\frac{k}{N}\right)\,,
\end{align}
with $k=0,\ldots,N-1$. We will denote the corresponding non-commutative resolution of a singular fibration $X^{(n)}_{i}$ by $X^{(n)}_{i,\text{nc.} k}$.

\section{Gopakumar-Vafa invariants in the presence of torsion}
\label{sec:gvtorsion}
On a smooth Calabi-Yau manifold, the genus $g$ amplitudes of the topological string A-model are encoded in the free energies $F_g(\underline{t},\bar{\underline{t}})$, which depend only on the complexified K\"ahler parameters $\underline{t}$.
In the holomorphic limit $\bar{t}\rightarrow \infty$~\cite{Bershadsky:1993cx}, the free energies $F_g(\underline{t})$, up to classical terms that arise for $g=0,1$, can also be defined as the generating functions of genus $g$ Gromov-Witten invariants~\cite{Cox:2000vi}.
The latter are integrals over virtual fundamental classes on moduli stacks of stable maps from genus $g$ curves into the Calabi-Yau.
For Calabi-Yau threefolds, the virtual dimensions of the relevant moduli stacks are exactly zero and non-trivial invariants exist for all genera.
However, the moduli stacks are orbifolds.
This makes the Gromov-Witten invariants $\mathbb{Q}$-valued and in general they do not have a direct enumerative interpretation.

Again in the case of Calabi-Yau threefolds, the topological string partition function
\begin{align}
	Z_{\text{top.}}(\underline{t},\lambda)=\exp\left(\sum\limits_{g=0}^\infty\lambda^{2g-2}F_g(\underline{t})\right)\,,
\end{align}
where $\lambda$ is the topological string coupling, admits an alternative interpretation as the generating function of Gopakumar-Vafa invariants~\cite{Gopakumar:1998ii,Gopakumar:1998jq}.
These invariants correspond to traces of multiplicities of BPS states in the five-dimensional M-theory compactification on the Calabi-Yau threefold $X$, which makes them manifestly integral.
From the field theoretic perspective, the invariants depend on the charges of the states and their representations under half of the little group
\begin{align}
	SO(4)=SU(2)_1\times SU(2)_2\,.
\end{align}
To define the trace, it is necessary to compactify on an additional circle which leads to a gauge group $U(1)^{b_2(X)}$. 
Assuming that the Calabi-Yau is smooth and does not contain torsional cycles, one can identify the $U(1)$ charge lattice with $H_2(X,\mathbb{Z})$.
Let $N^\beta_{j_1,j_2}$ denote the multiplicity of BPS particles with charge $\beta\in H_2(X,\mathbb{Z})$ that transform in the representation
\begin{align}
	\left[\left(\frac12,0\right)\oplus2(0,0)\right]\otimes(j_1,j_2)\,,
	\label{eqn:gvrep}
\end{align}
of the little group.
The Gopakumar-Vafa invariants $n^{(g)}_\beta$ are then defined as
\begin{align}
	\sum\limits_{g=0}^\infty n^{(g)}_{\beta}I_g=\sum\limits_{j_1,j_2}(-1)^{2j_2}(2j_2+1)N_{j_1,j_2}^{\beta}\cdot [j_1]\,,
	\label{eqn:gvtrace}
\end{align}
where $I_g=([\frac12]+2[0])^g$.
They are conjecturally encoded in the topological string partition function via the expansion
\begin{align}
	\log\left[Z_{\text{top}.}\left(\underline{t},\lambda\right)\right]=\sum\limits_{g=0}^\infty\sum\limits_{\beta\in H_2(X,\mathbb{Z})}\sum\limits_{m=1}^\infty n^{(g)}_{\beta}\cdot\frac{1}{m}\left(2\sin\frac{m\lambda}{2}\right)^{2g-2}Q^{m\beta}\,.
\end{align}
Here $Q^\beta$ is the exponentiated complexified volume of the curve $\beta$.

As can be seen in~\eqref{eqn:gvtrace}, the trace that is used to define the invariants does not involve the gauge charges.
It is therefore straightforward to extend the definition of the Gopakumar-Vafa invariants to incorporate also the charges under discrete factors of the five-dimensional gauge symmetry.
This leads to the question, how the invariants with discrete charges are encoded in the topological string partition function of the Calabi-Yau that is potentially singular.
As we have discussed in the previous section, discrete gauge symmetry arises in M-theory from torsional 2-cycles or from $\mathbb{Q}$-factorial terminal singularities that, at least from the Type IIA perspective, effectively behave like those cycles.
The answer to our question turns out to be closely connected to the new degree of freedom of turning on a B-field along those cycles.

Let us first assume that $X$ is a smooth Calabi-Yau manifold without torsional curves.
In this case, the contribution of a stable map into a curve of class $\beta\in H_2(X,\mathbb{Z})$ to the A-model topological string partition function is weighted by the integral of the complexified K\"ahler class $\omega=B+iJ$ over that curve
\begin{align}
	Q^\beta=e^{iS}=\exp\left(\int_CB+ iJ\right)\,.
\end{align}
This would at first suggest that the contribution of torsional curves is not visible to the topological string.
However, it was pointed out in~\cite{Aspinwall:1995rb,Braun:2007xh,Braun:2007vy} that if the manifold has a non-trivial Brauer group and thus contains torsional elements in $H_2(X,\mathbb{Z})$ it is better to think about the B-field as a homomorphism
\begin{align}
	e^{iS}:\,H_2(X,\mathbb{Z})\rightarrow \mathbb{C}^\times\,.
	\label{eqn:bfieldhom}
\end{align}
Let us now assume that $X$ does contain torsional $2$-cycles and decompose the homology as
\begin{align}
	H_2(X,\mathbb{Z})=H_2^{\text{free}}(X,\mathbb{Z})\oplus H_2^{\text{tors.}}(X,\mathbb{Z})=\mathbb{Z}^r\oplus\mathbb{Z}_{m_1}\oplus\ldots\oplus\mathbb{Z}_{m_k}\,.
	\label{eqn:xtorsh2}
\end{align}
The definition~\eqref{eqn:bfieldhom} now implies that, to measure the weight of the contribution of a curve in $H_2(X,\mathbb{Z})$ to the A-model partition function, we need to specify a complexified K\"ahler class $\omega\in H^2(X,\mathbb{C})$ as well as a set of homomorphisms
\begin{align}
	b_i:\,\mathbb{Z}_{m_i}\rightarrow\mathbb{C}^\times\,,
	\label{eqn:bihom}
\end{align}
that together assign a phase to each torsional curve.
This was used in~\cite{Braun:2007xh} for a Calabi-Yau threefold $Y$ with $H_2(Y,\mathbb{Z})=\mathbb{Z}^3\oplus\mathbb{Z}_3\oplus\mathbb{Z}_3$ to define genus zero instanton numbers with degrees in $\mathbb{Z}^3\oplus\mathbb{Z}_3\oplus\mathbb{Z}_3$ that are encoded in the topological string partition functions on $Y$ with different choices for $b_1,b_2$.

We now generalize this to an expansion of the topological string partition function on Calabi-Yau threefolds with torsional curves in terms of Gopakumar-Vafa invariants with discrete charges.
Assume again that $X$ is a Calabi-Yau threefold with $H_2(X,\mathbb{Z})$ as in~\eqref{eqn:xtorsh2}.
The corresponding M-theory compactification on $X\times S^1$ has a gauge symmetry
\begin{align}
	G=U(1)^r\times\mathbb{Z}_{m_1}\times\ldots\times\mathbb{Z}_{m_k}\,.
\end{align}
Let $N^{\beta,\tilde{\beta}}_{j_1,j_2}$ denote the multiplicity of five-dimensional BPS particles with $U(1)$ charge determined by $\beta\in H_2^{\text{free}}(X,\mathbb{Z})$ and discrete charge $\tilde{\beta}\in H^{\text{tors.}}_2(X,\mathbb{Z})$, that transform in the representation~\eqref{eqn:gvrep}.
We can then define Gopakumar-Vafa invariants $n^{(g)}_{\beta,\tilde{\beta}}$ via the usual trace~\eqref{eqn:gvtrace}.
We denote the A-model topological string partition function by $Z_{\text{top.}}(\underline{t},b,\lambda)$ with $\lambda$ being the string coupling, $b=\prod_{i=1}^kb_i$ for a particular choice of the homomorphisms $b_i$ in~\eqref{eqn:bihom} and $\underline{t}$ parametrizing the complexified K\"ahler form $\omega$ on $X$ as
\begin{align}
	\omega =B+iJ=\sum_{i=1}^kt^iJ_i\,.
\end{align}
Our conjecture is then that the invariants $n^{(g)}_{\beta,\tilde{\beta}}$ are encoded in $Z_{\text{top.}}(\underline{t},b,\lambda)$ via
\begin{align}
	\log\left[Z_{\text{top}.}\left(\underline{t},b,\lambda\right)\right]=\sum\limits_{g=0}^\infty\sum\limits_{\substack{(\beta,\tilde{\beta})\in\\H_2(X,\mathbb{Z})}}\sum\limits_{m=1}^\infty n^{(g)}_{\beta,\tilde{\beta}}\cdot\frac{1}{m}\left(2\sin\frac{m\lambda}{2}\right)^{2g-2}b(\tilde{\beta})^mQ^{m\beta}\,.
	\label{eqn:torsiongvex}
\end{align}

We expect an analogous relation to hold for topological strings on non-commutative resolutions that are associated to a torsional flat B-field.
For our generic examples over $\mathbb{P}^2$, let us assume that $X\equiv X^{(N)}_{i}$ has $\mathbb{Q}$-factorial terminal singularities and leads to an unbroken $\mathbb{Z}_M$ symmetry in five dimensions.
As discussed in the previous section, one then obtains $M$ different non-commutative resolutions $X_{\text{n.c.}k=1,\ldots,M}$, where $X=X_{\text{n.c.}0}$ corresponds to the singular geometry with a vanishing B-field.
We denote the partition function on $X_{\text{n.c.}k}$ by $Z_k$ and propose the expansion
\begin{align}
	\begin{split}
		\log\left[Z_k\left(\underline{t},\lambda\right)\right] =\sum\limits_{g=0}^\infty\sum\limits_{\beta\in H_2(X,\mathbb{Z})}\sum\limits_{q=0}^{M-1}\sum\limits_{m=1}^\infty n^{(g)}_{\beta,q}\cdot\frac{1}{m}\left(2\sin\frac{m\lambda}{2}\right)^{2g-2}e^{mq\frac{2\pi ik}{M}}Q^\beta\,,
	\end{split}
	\label{eqn:torsiongvexs}
\end{align}
where $n^{(g)}_{\beta,q}$ are the Gopakumar-Vafa invariants with $U(1)$ charges $\beta$ and $\mathbb{Z}_M$ charge $q$.

Note that knowledge of the topological string partition function for some choice of $b$ in~\eqref{eqn:torsiongvex} or $k$ in~\eqref{eqn:torsiongvexs} is not sufficient to determine the invariants $n^{(g)}_{\beta,\tilde{\beta}}$.
Instead, the information that one obtains from all possible choices has to be combined.
We demonstrate this for torus fibration with $\mathbb{Q}$-factorial terminal singularities in Sections~\ref{sec:ex2sec},~\ref{sec:ex4sec} and~\ref{sec:ex5sec}.
Using our knowledge of the associated M-theory spectrum, we can perform a highly non-trivial check that our interpretation of the invariants as traces of multiplicities of BPS states with discrete charges is indeed correct.

\section{Modularity of the topological string partition function}
\label{sec:modularstructure}
The topological string partition function on a smooth torus fibered Calabi-Yau threefold $X$ with $N$-sections can be expanded as
\begin{align}
	Z_{\text{top.}}(\tau,\underline{m},t,\lambda)=Z_0(\tau,\underline{m},\lambda)\left(1+\sum\limits_{\beta\in H_2(B)}Z_\beta(\tau,\underline{m},\lambda)Q^\beta\right)\,,
	\label{eqn:ztop}
\end{align}
in terms of $\Gamma_1(N)$ lattice Jacobi forms $Z_\beta(\tau,\underline{m},\lambda)$.
The expansion parameter is $Q^\beta=\exp(2\pi t^i\beta_i)$, with $t^i$ being (shifted) complexified volumes of curves in the base, 
and the modular parameter $\tau$ corresponds to $1/N$-times the complexified volume of the generic fiber.
The elliptic arguments $\underline{m}$ are Coulomb branch parameters that correspond to complexified volumes of components of reducible fibers. 
The topological string coupling $\lambda$ also transforms as an elliptic parameter.
As has been worked out over the last three decades~\cite{Candelas:1994hw,Klemm:2012sx,Alim:2012ss,Huang:2015sta,Schimannek:2019ijf,Cota:2019cjx,Knapp:2021vkm}, the modular properties can be seen as a consequence of the monodromies in the stringy K\"ahler moduli space.

More recently in~\cite{Knapp:2021vkm}, it was found that the topological string partition function also transforms as a vector valued Jacobi form under the $\Gamma_0(5)$-transfer matrix that connects two large volume limits in the moduli space of genus one fibrations with five sections.
The two genus one fibrations have been identified as corresponding to different elements of the same Tate-Shafarevich group.
In this section we will show, that this is part of a more general network of relations between the topological string partition functions on all elements of the Tate-Shafarevich group.
It turns out that the modular structure of the topological string partition function also provides a powerful tool to identify the Kaluza-Klein charge of the Higgs field that has been used in the transition to arrive at a particular geometry.
In particular, it can be used to deduce the presence of a non-vanishing ``torsional'' B-field, or, in other words, determine that the target space is a non-commutative resolution of a singular fibration.
We will illustrate this in detail for generic genus one fibered Calabi-Yau threefolds with $N$-sections for $N\le 5$.

Before that, let us discuss the precise form of the base degree $\beta$ partition function $Z_\beta(\tau,\underline{m},\lambda)$ in~\eqref{eqn:ztop}.
We will only need the special cases of an elliptic fibration with two-sections and a genus one fibration with one $N$-section.

\subsection{Modular structure of the topological string partition function}
The modular structure of the topological string partition function is only manifest in an appropriate parametrization of the K\"ahler form.
To simplify the exposition we will fix the base to be $B=\mathbb{P}^2$ but the discussion in this section only relies on properties of the generic fiber and applies to other bases as well.
We denote the vertical divisor that is associated to the hyperplane class in $\mathbb{P}^2$ by $D_b$.
Details on the relevant Jacobi forms and rings of $\Gamma_1(N)$ modular forms can be found in Appendix~\ref{app:modforms}.

\paragraph{Elliptic fibrations over $\mathbb{P}^2$ with 2 sections}
If the fibration is elliptic, we assume that it has two sections $s_0,\,s_1$ and no fibral divisors.
Choosing $s_0$ to be the zero section and denoting the divisor classes associated to $s_i$ by $S_i$, the image of $s_1$ under the Shioda map is
\begin{align}
	\sigma(s_1)=S_1-S_0-\pi^*\left[c_1(B)+\pi(S_1\cdot S_0)\right]\,,
	\label{eqn:shiodagen}
\end{align}
with the height pairing
\begin{align}
	b=\pi^*\pi\left(\sigma(s_1)\cdot\sigma(s_1)\right)\,,
\end{align}
and we also need the height pairing of the zero section with itself
\begin{align}
	D=-\pi^*\pi(S_0\cdot S_0)=\pi^*c_1(B)\,.
\end{align}
Having fixed the base to be $\mathbb{P}^2$, we could immediately insert $\pi^*c_1(B)=3D_b$.
However, to highlight the general structure we will often keep the dependence on $c_1(B)$ explicit.
The modular parametrization of the K\"ahler form then reads
\begin{align}
	\omega=\tau\cdot\left(S_0+\frac12 D\right)+m\cdot \sigma(s_1)+t\cdot D_b\,.
\end{align}
One can then expand the topological string partition function as in~\eqref{eqn:ztop}, with $\beta\in H_2(\mathbb{P}^2,\mathbb{Z})\cong\mathbb{Z}$, and the coefficients take the form~\cite{Lee:2018urn,Lee:2018spm}
\begin{align}
	Z_\beta(\tau,m,\lambda)=\frac{1}{\eta(\tau)^{12\beta\cdot c_1(B)}}\frac{\phi_\beta(\tau,m,\lambda)}{\prod_{s=1}^\beta\phi_{-2,1}(\tau,s\lambda)}\,.
\end{align}
Here $\phi_{-2,1}(\tau,z)$ is the Jacobi form of weight $-2$ and index $1$ that together with $\phi_{0,1}(\tau,z)$ generates the ring of weak Jacobi forms.
For a reference on Jacobi forms we refer the reader to the original work~\cite{eichler1985theory} or to the summaries of the relevant properties in~\cite{Huang:2015sta,Cota:2019cjx}.
The numerator $\phi_\beta(\tau,m,\lambda)$ is a lattice Jacobi form with two elliptic parameters $m$ and $\lambda$ that is an element of the ring
\begin{align}
	\phi_\beta(\tau,m,\lambda)\in M_\bullet(1)\left[\phi_{-2,1}(\tau,m),\phi_{0,1}(\tau,m),\phi_{-2,1}(\tau,\lambda),\phi_{0,1}(\tau,\lambda)\right]\,,
\end{align}
where $M_\bullet(1)$ is the ring of $\Gamma_1(1)=SL(2,\mathbb{Z})$ modular forms.
The weight and index of $\phi_\beta$ is determined by the requirement that $Z_\beta(\tau,m,\lambda)$ is a lattice Jacobi form of weight zero and index matrix $M=\text{diag}(r_m,r_\lambda)$, with
\begin{align}
	r_m=-\frac12 \beta\cdot b\,,\quad r_\lambda=\frac12 \beta\cdot(\beta-c_1(B))\,.
\end{align}

\paragraph{Genus one fibrations over $\mathbb{P}^2$ with one $N$-section}
Let us now consider the topological string partition function on a smooth genus one fibered Calabi-Yau threefold over $\mathbb{P}^2$
 with one $N$-section $s_0^{(N)}$ and no fibral divisors.
We denote the class of the $N$-section again by $S_0$ and the height pairing by
\begin{align}
	D=-\pi^*\pi\left(S_0\cdot S_0\right)=\tilde{a}D_b\,,\quad \tilde{a}=-S_0\cdot C\,,
\end{align}
where $C=S_0\cdot D_b$.
The modular parametrization of the K\"ahler form now reads
\begin{align}
	\omega=\tau\cdot\left(S_0+\frac12 D\right)+t\cdot D_b\,.
\end{align}
The $Z_\beta(\tau,\lambda)$ in the expansion~\eqref{eqn:ztop} takes the form
\begin{align}
	Z_\beta(\tau,\lambda)=\frac{1}{\eta(N\tau)^{12\beta\cdot c_1(B)}}\frac{\phi_\beta(\tau,\lambda)}{\prod_{s=1}^\beta\phi_{-2,1}(N\tau,s\lambda)}\,,
	\label{eqn:nsecZtop}
\end{align}
with the numerator $\phi_\beta(\tau,\lambda)$ being a $\Gamma_1(N)$ weak Jacobi form
\begin{align}
	\phi_\beta(\tau,\lambda)=M_\bullet(N)\left[\phi_{-2,1}(\tau,\lambda),\phi_{0,1}(\tau,\lambda)\right]\cdot \Delta_{2N}(\tau)^{k_\beta}\,,
	\label{eqn:nsecZtopnum}
\end{align}
where $M_\bullet(N)$ is the ring of $\Gamma_1(N)$ modular forms.
The overall coefficient $\Delta_{2N}(\tau)$ is a $\Gamma_1(N)$ modular form of weight $2N$ and the fractional part of the exponent is determined by the congruence relation
\begin{align}
	k_\beta\equiv\frac12\left[Nc_1(B)-\frac{1}{N}D\right]\cdot\beta\text{ mod }1\,.
	\label{eqn:deltaexcong}
\end{align}
Its occurence can be understood from the Higgs transitions that we discussed in Section~\eqref{sec:higgsMF} and we will elaborate on this in the next section.
The weight of $Z_\beta(\tau,\lambda)$ is again zero but the index is
\begin{align}
	r=\frac{1}{2N} \beta\cdot(\beta-c_1(B))\,,
\end{align}
which just reflects that $\phi_{-2,1}(N\tau,z)$ is formally an index $1/N$ Jacobi form.

\subsection{Relating partition functions via Higgs transitions}
\label{sec:higgs2}
In~\cite{Cota:2019cjx}, the Higgs transitions in M- and F-theory that we discussed in Section~\ref{sec:higgsMF} have been used to determine the ansatz~\eqref{eqn:nsecZtop}.
The idea is, that the base degree zero part of the topological string partition function on a smooth elliptic fibration with two sections and matter with $U(1)$ charges up to $q_{\text{max}}=N$ takes the form
\begin{align}
	Z_0(\tau,m,\lambda)\approx 1+\frac{1}{\lambda^2}\sum\limits_{n=0}^\infty\sum\limits_{k=-N}^Nn_{n,q}\text{Li}_3\left(q^{n}\zeta^{k}\right)+\mathcal{O}(\lambda^0)\,,
	\label{eqn:z0exp}
\end{align}
with $q=e^{2\pi i\tau}$ and $\zeta=e^{2\pi i m}$.
In the five-dimensional M-theory compactification, $\tau$ sets the Kaluza-Klein scale associated to the F-theory circle, while $m$ is the vacuum expectation value of the $U(1)_{6d}$ Wilson line along the circle. 
For $q\ne0$, the coefficients $n_{n,q}$ are the numbers of complex scalar fields with Kaluza-Klein charge $n$ and $U(1)_{6d}$ charge $k$ in five dimensions.
All coefficients $n_{n,0}$ are equal to minus the Euler characteristic of the Calabi-Yau.
Let us point out, that for a given elliptically fibered Calabi-Yau threefold not all of the terms in~\eqref{eqn:z0exp} are encoded in the genus zero Gopakumar-Vafa invariants.
This is because some of the fiber components that lead to fields with small KK-charge are only realized in different birational phases of the Calabi-Yau or, physically speaking, chambers of the five-dimensional Coulomb branch.
However, the expansion~\eqref{eqn:z0exp} can be seen as the modular completion of the base degree zero topological string partition function on the Calabi-Yau and it will allow us to study Higgs transitions independent of an explicit realization of the phase.

Let us assume, that the elliptic Calabi-Yau $X_{\text{el.}}^{(2)}$ describes an unhiggsing of the theory associated to a generic genus one fibration $X^{(2)}_1$ with $2$-sections over $\mathbb{P}^2$ that has $h^{1,1}=2$.
We denote the Jacobian fibration by $J^{(2)}$ and the Tate-Shafarevich group is then
\begin{align}
	\Sh\left(J^{(2)}\right)=\mathbb{Z}_2\,,
\end{align}
with the elements $\{J^{(2)},\,X_1^{(2)}\}$, as listed in Table~\ref{tab:tselements}.
To increase readability, we will use $X_1\equiv X^{(2)}_1,\,J\equiv J^{(2)}$ and $X_{\text{el.}}=X^{(2)}_{\text{el.}}$ for the remainder of this section.
Following the discussion in Section~\ref{sec:higgsMF}, the fibrations $J$ and $X_1$ are related to $X_{\text{el.}}$ via extremal transitions that physically correspond to Higgs transitions with different values for the $U(1)_{6d}$ Wilson line along the circle.
The corresponding specializations of the K\"ahler parameters $m$ and $\tau$ relate the topological string partition functions on the fibrations and, as we will now discuss, enable us to deduce the charge of the Higgs field from the partition function associated to the elements of the Tate-Shafarevich group of the genus one fibration.

To perform the Higgs transition that corresponds to an extremal transition $X_{\text{el.}}\rightsquigarrow X_1$, we can give a vacuum expectation value to the scalar fields with $U(1)_{KK}\times U(1)_{6d}$ charge $(1,-2)$.
Using the mass formula~\eqref{eqn:mass} one finds that these fields become massless if the Kaluza-Klein scale and the Coulomb branch parameter are related as $2m=\tau$.
As has been observed in~\cite{Cota:2019cjx}, this can be achieved by setting the K\"ahler parameters to
\begin{align}
	\tau\rightarrow2\tau\,,\quad m\rightarrow\tau\,
	\label{eqn:tau2res}
\end{align}
which puts us on the boundary of the K\"ahler cone.
One might ask why the correct choice is~\eqref{eqn:tau2res} and not $\tau\rightarrow\tau,\,m\rightarrow \tau/2$.
This is a consequence of the fact that we want the restricted Coulomb branch parameter to correspond to a normalization of the unbroken $U(1)$ with integral charges.
The replacement~\eqref{eqn:tau2res} relates the generators $\phi_{-2,1}(\tau,m),\,\phi_{0,1}(\tau,m)$ of the ring of weak Jacobi forms to $\Gamma_1(2)$ modular forms via
\begin{align}
	q^{\frac{1}{2}}\phi_{-2,1}(2\tau,\tau)=&(\Delta_4)^{-\frac12}\,,\quad q^{\frac{1}{2}}\phi_{0,1}(2\tau,\tau)=E_{2,2}(\Delta_4)^{-\frac12}\,,
	\label{eqn:2secG1higgs}
\end{align}
where $\Delta_4,\,E_{2,2}\in M_\bullet(2)$ are defined in Appendix~\ref{app:modforms} and have respective weights four and two.
This is the origin of the overall factor $\Delta_4^{k_\beta}$ in the ansatz for the numerators of the base degree $\beta$ topological string partition function~\eqref{eqn:nsecZtopnum}.
The congruence relation~\eqref{eqn:deltaexcong} that determines the fractional part of $k_\beta$ has been determined in~\cite{Cota:2019cjx} by a careful consideration of the cancellation of poles.

The analogous Higgs transition associated to $X_{\text{el.}}\rightsquigarrow J$ can be performed with scalar fields of charge $(0,2)$, which become massless in the limit $m=0$.
Since the generators of weak Jacobi forms restrict to constants
\begin{align}
	\phi_{-2,1}(\tau,0)=0\,,\quad\phi_{0,1}(\tau,0)=12\,,
\end{align}
there is no fractional power of a modular form appearing for elliptic fibrations.
Note that although $J$ is singular, it is related via complex structure deformation to the generic elliptic fibration over $\mathbb{P}^2$ which is the degree $18$ hypersurface $X^{(1)}_0$ in $\mathbb{P}_{11169}$.
The topological A-model partition function is invariant under complex structure deformations and thus cannot distinguish between $J$ and $X^{(1)}_0$.

A striking implication of the occurrence of $\Delta_4$ is therefore that, via~\eqref{eqn:2secG1higgs}, it encodes the charges of the Higgs field that has been used in the extremal transition from the elliptic to the genus one fibration.
At least up to physically irrelevant unimodular transformations this can be read off directly from the topological string partition function of the genus one fibration.
In fact, anticipating results from Sections~\ref{sec:n2curves} and~\ref{sec:ex2sec}, similar relations allow us to deduce the presence of a non-trivial B-field ``along'' the vanishing cycle or, in other words, that the geometry associated to a large volume point is a non-commutative resolution of a particular element of the Tate-Shafarevich group.
Before discussing the details, as well as the corresponding structures for fibrations with $N$-sections, we will briefly illustrate this result.
Our example is again the Tate-Shafarevich group of a generic genus one fibration with $2$-sections over $\mathbb{P}^2$.

The generic fiber of $X_1$ can be expressed as a quartic hypersurface in $\mathbb{P}_{112}$ and the conifold point of the fiber is a second large volume point in the stringy K\"ahler moduli space.
This is a special property of the conifold points of Calabi-Yau 1-folds, but the stringy K\"ahler moduli space of the generic fiber embeds into that of the fibration and the latter contains a corresponding large volume limit as well.
In the case of the fibration, this point is only visible after resolving a tangency between the conifold locus and the large base limit.
We calculate the transfer matrix that relates the modular parameter of the quartic hypersurface to that at the conifold and find that it acts as a Fricke involution
\begin{align}
	\tau\mapsto-\frac{1}{2\tau}\,.
	\label{eqn:frick2brief}
\end{align}
Let us point out, that a second large volume limit at $\tau=0$ already exists for generic elliptic fibrations~\cite{Candelas:1994hw}.
However, in that case it is related to the large volume limit at $\tau=i\infty$ via $\tau\rightarrow-1/\tau$, which is part of the monodromy group, and the two large volume limits are therefore physically equivalent.
On the other hand, the Fricke involution is not part of $\Gamma_1(2)$, it is not even contained in $SL(2,\mathbb{Z})$, and the two large volume points are physically distinct.

The modular properties of the partition function allow us to deduce the nature of the second large volume limit in the moduli space of the genus one fibration.
Using the behaviour of Jacobi forms under $SL(2,\mathbb{Z})$ transformations of $\tau$, it is easy to calculate the action of the Fricke involution~\eqref{eqn:frick2brief} on $\phi_{w,1}(2\tau,\tau)$,
\begin{align}
	\begin{split}
	q^{\frac12}\phi_{w,1}(2\tau,\tau)\mapsto&\exp\left(-\frac{\pi i}{2\tau}\right)\phi_{w,1}\left(-\frac{1}{\tau},-\frac{1}{2\tau}\right)=\tau^w\phi_{w,1}\left(\tau,-\frac12\right)\,.
	\label{eqn:frickejacobi}
	\end{split}
\end{align}
Note, that the factor of $q^{\frac12}$ is crucial to cancel the part of the automorphy factor of the Jacobi form that contains the elliptic parameter.
The Jacobi form on the right hand side can again be expressed in $\Gamma_1(2)$ modular forms and we find
\begin{align}
	q^{\frac12}\phi_{-2,1}\left(\tau, -\frac12\right)=&-2(\Delta_4')^{-\frac12}\,,\quad q^{\frac12}\phi_{0,1}\left(\tau, -\frac12\right)=E_{2,2}(\Delta_4')^{-\frac12}\,,
\end{align}
where $\Delta_4'\in M_\bullet(2)$ is again defined in Appendix~\ref{app:modforms} and has weight four.
The arguments of the Jacobi forms indicate that the geometry at the large volume limit $\tau=0$ arises from a transition with $m=-1/2$.
This implies a non-vanishing value of the B-field ``along'' the vanishing cycle.
We thus interpret the large volume limit $\tau=0$ in the stringy K\"ahler moduli space of the genus one fibration with $2$-sections as corresponding to the non-commutative resolution $X^{(2)}_{0,\text{nc.} 1}$ of the Jacobian fibration $J$.

In Section~\ref{sec:ex2sec} we will study the partition functions associated to $X_1$, the Jacobian fibration $J$ and the non-commutative resolution $X^{(2)}_{0,\text{nc.} 1}$ and verify the expansion in terms of Gopakumar-Vafa invariants with discrete charges that we proposed in Section~\ref{sec:gvtorsion}.
First, we discuss in detail the stringy K\"ahler moduli spaces of the fibers for $1<N\le 5$ and the analogous relations between Jacobi forms.

\section{Stringy K\"ahler moduli spaces of genus one curves}
\label{sec:strk}
At least for $N\le 5$, every genus one fibration with $N$-sections is birationally equivalent to a fibration of generic genus one curves of degree $N$~\cite{Braun:2014oya}.
The stringy K\"ahler moduli space of the fiber embeds into the moduli space of the fibration and is recovered in the large base limit.
In particular, the monodromies and transfer matrices, that one can calculate in the moduli space of the generic fiber, embed into the corresponding matrices of the fibration.
This enables us to deduce generic properties by only studying the generic genus one curves.

The monodromy group in the stringy K\"ahler moduli space of the generic degree $N$ curve for $N\le 5$ is $\Gamma_1(N)$ and this leads to the modular properties of the topological string partition function~\cite{Candelas:1994hw,Klemm:2012sx,Alim:2012ss,Schimannek:2019ijf,Cota:2019cjx,Knapp:2021vkm}.
According to homological mirror symmetry, the mirror family of tori therefore exhibits $\Gamma_1(N)$ monodromy in the complex structure moduli space.
This implies, that the family factors through the modular curve $X_1(N)$ of $\Gamma_1(N)$ and that the corresponding mirror fibers have $N$-torsion points.
Let us point out, that this can be seen as a proof of the mirror relation between genus one curves and elliptic curves with torsion points that has been observed in~\cite{Klevers:2014bqa,Braun:2014qka,Oehlmann:2016wsb,Cvetic:2016ner}.
In fact, we find that the stringy K\"ahler moduli space of generic genus one curves of degree $N$ is exactly the modular curve $X_1(N)$.
The cusps of the modular curves turn out to be in one-to-one correspondence with large volume limits of the generic fibers.
The transfer matrices relate the partition functions associated to different geometries and non-commutative resolutions that are contained in the same Tate-Shafarevich group.

For each value of $N\le 5$ we will now discuss the stringy K\"ahler moduli space of the generic degree $N$ curve, the relation to the corresponding $\Gamma_1(N)$ modular curve and the interpretation of the different large volume limits.
\subsection{$N=1$: The Weierstra{\ss} family in $\mathbb{P}_{123}$}
\begin{figure}[h!]
	\begin{tikzpicture}[remember picture,overlay,node distance=4mm]
		\draw [->,line width=1.3] (4.8,3) to [in=20,out=160] (3.5,2.7);
		\node[align=center] at (7,2.8) {Elliptic fibration $X^{(1)}_0$\\$\tau=i\infty$};
	\end{tikzpicture}
	\centering
	\includegraphics[width=.3\linewidth]{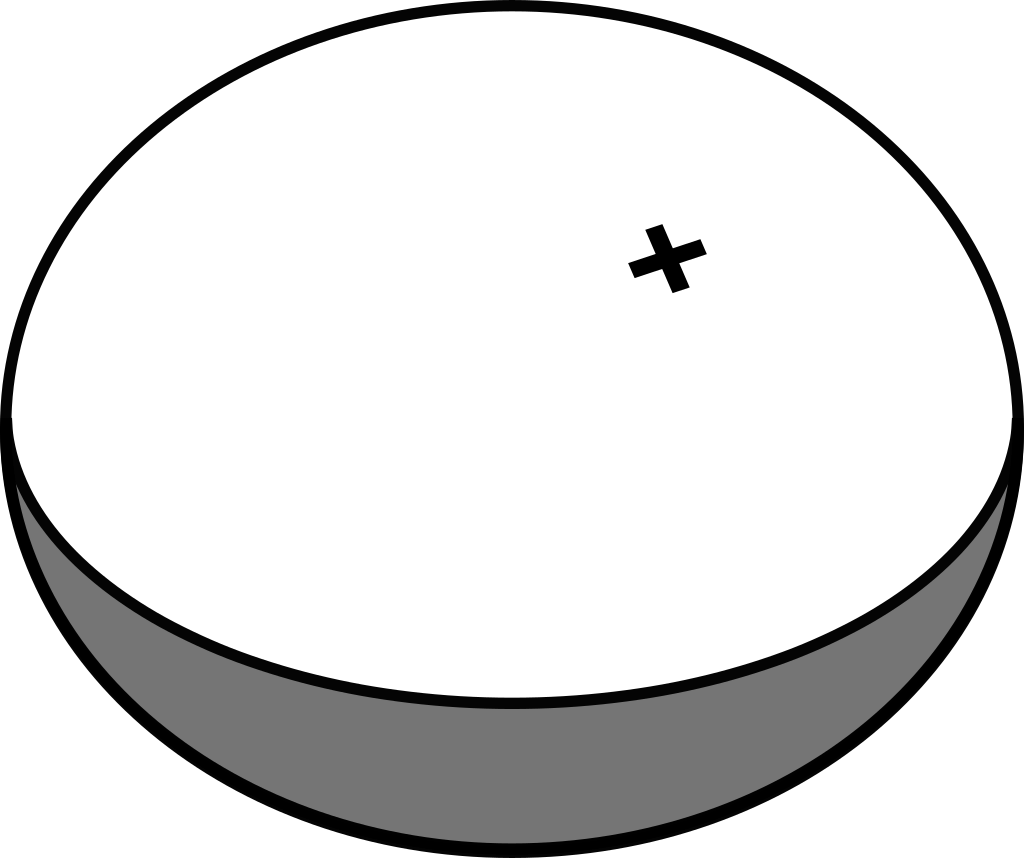}
	\caption{The stringy K\"ahler moduli space of Weierstra{\ss} curves is the modular curve $X_1(1)$. Also for a Calabi-Yau fibration of Weierstra{\ss} curves, the large volume points in the $SL(2,\mathbb{Z})$ orbit of $\tau=i\infty$ are physically equivalent.}
	\label{fig:skmoduliN1}
\end{figure}
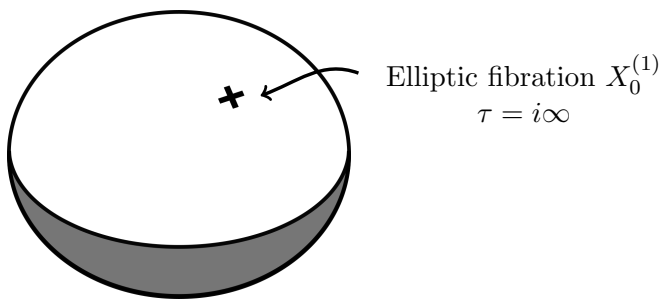
Every elliptic curve is isomorphic to an anti-canonical hypersurface in the weighted projective space $\mathbb{P}_{123}$.
Using standard techniques form mirror symmetry, see e.g.~\cite{Cox:2000vi}, the fundamental period of the mirror can be expressed as
\begin{align}
	\varpi_0=\sum\limits_{n=0}^\infty\frac{\Gamma(1+6n)}{\Gamma(1+n)\Gamma(1+2n)\Gamma(1+3n)}z^n={_2}F_1\left(\frac16,\frac56;1;432z\right)\,.
	\label{eqn:N1w0}
\end{align}
This is annihilated by the Picard-Fuchs operator
\begin{align}
	\mathcal{D}=\theta^2-12z(6\theta+1)(6\theta+5)\,,
	\label{eqn:N1pf}
\end{align}
where the logarithmic derivative is $\theta=z\partial_z$.
The logarithmic period that is annihilated by $\mathcal{D}$ takes the form
\begin{align}
	\begin{split}
		\varpi_1=&i\cdot{_2}F_1\left(\frac16,\frac56;1;1-432z\right)\\
		=&\frac{1}{2\pi i}\left(\varpi_0\log(z)+312z+77652z^2+23485136z^3+\mathcal{O}(z^4)\right)\,.
	\end{split}
	\label{eqn:N1w1}
\end{align}
The algebraic coordinate $z$ is related to the complexified volume $\tau$ of the elliptic curve via the mirror map
\begin{align}
	\tau=\frac{\varpi_1}{\varpi_0}=\frac{1}{2\pi i}\log(z)+\mathcal{O}(z)\,.
\end{align}
To calculate the monodromies, we fix the basis of periods $\vec{\Pi}=(\varpi_1,-\varpi_0)$.
A minus sign for the regular period is reflecting the convention, that the central charge of a topological 0-brane on a Calabi-Yau $d$-fold is $(-1)^d$.
From the Picard-Fuchs operator we can read off the discriminant
\begin{align}
	\Delta=1-432z\,,
\end{align}
and $z=1/432$ corresponds to a conifold point where $\tau$ goes to zero.

Geometric limits are characterized by the property that the monodromy around that point is maximally unipotent.
For a one parameter family of Calabi-Yau $d$-fold this means that the corresponding matrix $M$ satisfies that
\begin{align}
	(M-\text{id})^k=0\,,
\end{align}
if and only if $k>d$.
In the case of the family of Weierstra{\ss} curves, the large volume limit corresponds to $z=0,\,\tau= i\infty$, with the monodromy
\begin{align}
	M_\infty=\left(\begin{array}{cc}
		1&-1\\
		0&1
	\end{array}\right)\,,
\end{align}
acting on the periods $\vec{\Pi}$ and thus via M\"obius transformations on the modular parameter $\tau=\varpi_1/\varpi_0$.
This obviously satisfies $(M_\infty-\text{id})^2=0$.

However, in the case of Calabi-Yau $1$-folds, a conifold point is also of maximally unipotent monodromy (MUM).
Using the local coordinate
\begin{align}
	v=\frac{1}{432}-z\,,
\end{align}
and transforming the Picard-Fuchs operator~\eqref{eqn:N1pf}, one finds that the resulting operator is again~\eqref{eqn:N1pf} but with $z$ replaced by $v$.
This implies, that we can use a basis of periods $\vec{\Pi}_{\text{coni.}}=(\varpi_1,-\varpi_0)\big|_{z\rightarrow v}$ around $v=0$ and that this is indeed a MUM-point.
From the closed expressions for the periods~\eqref{eqn:N1w0},~\eqref{eqn:N1w0}, it is easy to read off the transfer matrix
\begin{align}
	T=i\cdot\left(\begin{array}{cc}
		0&-1\\
		1&0
	\end{array}\right)\,,
\end{align}
such that $\vec{\Pi}=T\cdot\vec{\Pi}_{\text{coni.}}$.\footnote{We thank Emanuel Scheidegger for informing us about the closed expressions for the logarithmic periods. Analogous results will also be used for the curves of degree $N$ with $N\le 4$.} 
The corresponding action on the modular parameter is $\tau\mapsto-1/\tau$.
It follows, that the monodromy around the conifold point with respect to the basis $\vec{\Pi}$ at $z=0$ is given by
\begin{align}
	M_0=TM_\infty T^{-1}=\left(\begin{array}{cc}
		1&0\\
		1&1
	\end{array}\right)\,.
\end{align}
This can also be obtained by evaluating the action of the corresponding Fourier-Mukai transform on B-brane charges, with the kernel given by the ideal sheaf of the diagonal in $X\times X$~\cite{Andreas:2001ve,Andreas:2004uf}.
Together $M_\infty$ and $M_0$ generate the modular group $SL(2,\mathbb{Z})$.

In particular, the relation $T=M_\infty M_0 M_\infty$ now implies, that the transfer matrix $T$ itself is contained in the monodromy group.
This is a consequence of T-duality along both cycles of the torus, which relates the limit of zero volume and infinite volume.
The actual stringy K\"ahler moduli space is therefore given by $X_1(1)=\overline{\mathbb{H}/SL(2,\mathbb{Z})}$, which is nothing but the $SL(2,\mathbb{Z})$ modular curve.
The modular curve contains one cusp, which can be identified with the unique geometric limit.

If one uses the Weierstrass curves to construct an elliptically fibered Calabi-Yau $d$-fold, the zero volume point also exists as a MUM-point in the stringy K\"ahler moduli space of the fibration.
However, in the algebraic coordinates on the complex structure moduli space of the mirror, this is only visible after resolving a tangency between the discriminant locus and the large base limit~\cite{Candelas:1994hw,Alim:2012ss,Cota:2017aal}.
This will be reviewed in Appendix~\ref{app:x18}.

\subsection{$N=2$: Quartic curves in $\mathbb{P}_{112}$}
\label{sec:n2curves}
\begin{figure}[h!]
	\begin{tikzpicture}[remember picture,overlay,node distance=4mm]
		\draw [->,line width=1.3] (4.8,3) to [in=20,out=160] (3.5,2.7);
		\node[align=center] at (7,2.8) {Genus one fibration $X^{(2)}_1$\\$\tau=i\infty$};
		\draw [->,line width=1.3] (-1,3) to [in=150,out=20] (1,2.8);
		\node[align=center] at (-2.7,2.2) {Non-comm. resolution $X^{(2)}_{0,\text{nc.} 1}$\\of Jacobian $X^{(2)}_0$ at $\tau=0$};
	\end{tikzpicture}
	\centering
	\includegraphics[width=.3\linewidth]{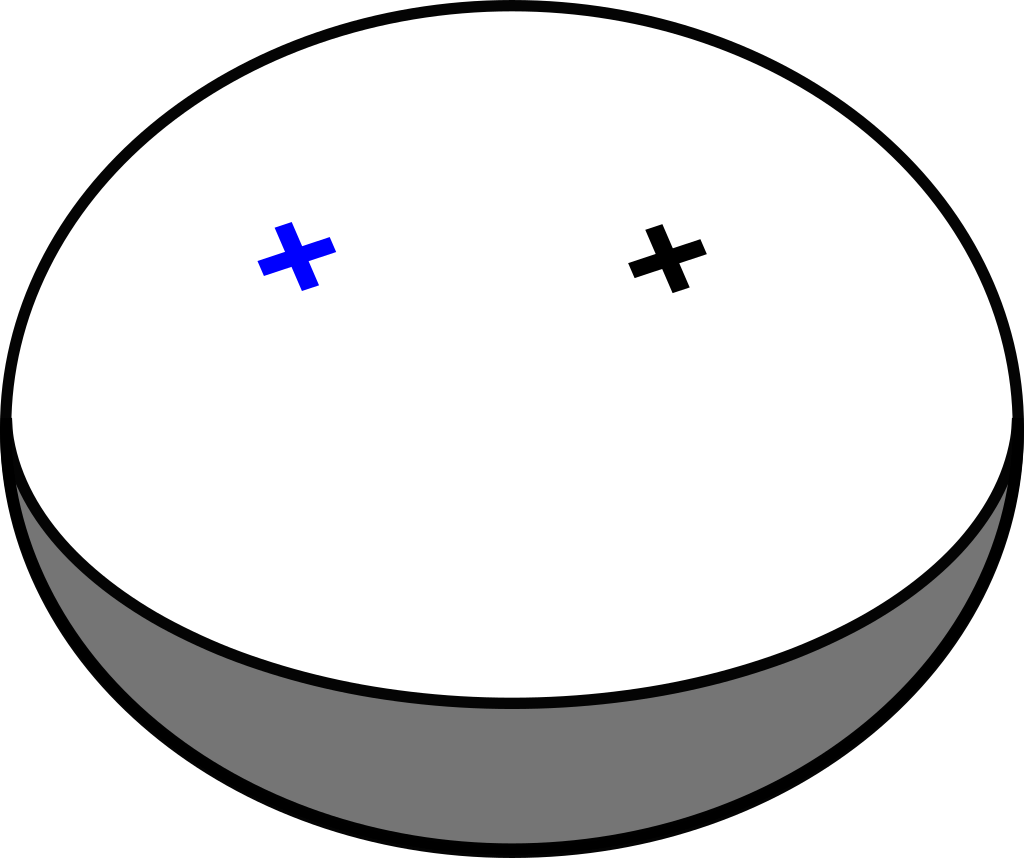}
	\caption{The stringy K\"ahler moduli space of the generic genus one curve of degree $2$ is the modular curve $X_1(2)$. For Calabi-Yau fibrations with $2$-sections, the large volume points in the $\Gamma_1(2)$ orbit of $\tau=0$ correspond to the non-commutative resolution of the Jacobian.}
	\label{fig:skmoduliN2}
\end{figure}
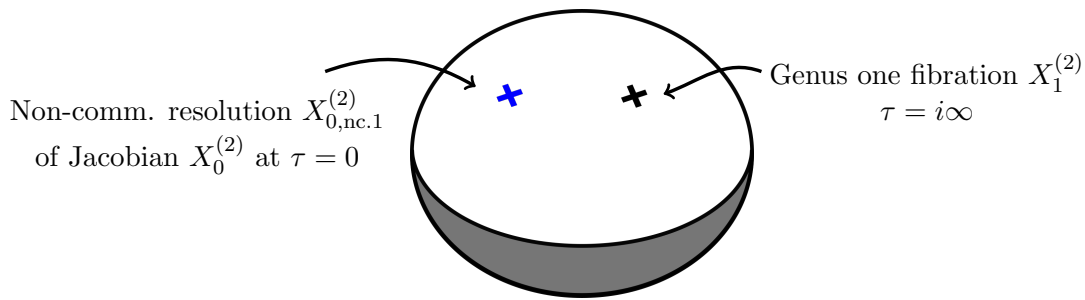
The situation becomes more interesting for genus one curves of degree $2$, which can be realized as anti-canonical hypersurfaces in the weighted projective space $\mathbb{P}_{112}$.
Here the periods of the mirror take the form
\begin{align}
	\begin{split}
		\varpi_0=&\sum\limits_{n=0}^\infty\frac{\Gamma(1+4n)}{\Gamma(1+n)^2\Gamma(1+2n)}z^n={_2}F_1\left(\frac14,\frac34;1;64z\right)\,,\\
		\varpi_1=&\frac{i}{\sqrt{2}}\cdot{_2}F_1\left(\frac14,\frac34;1;1-64z\right)\\
		=&\frac{1}{2\pi i}\left(\varpi_0\log(z)+40z+1556z^2+\frac{213232}{3}z^3+\mathcal{O}(z^4)\right)\,,
	\end{split}
	\label{eqn:N2periods}
\end{align}
and the Picard-Fuchs operator is
\begin{align}
	\mathcal{D}=\theta^2-4z(4\theta+1)(4\theta+3)\,.
	\label{eqn:ncurveN2PF}
\end{align}
Now the modular parameter $\tau=\varpi_1/\varpi_0$ is one half of the complexified volume of the curve and the discriminant polynomial is given by
\begin{align}
	\Delta=1-64z\,.
\end{align}
As a basis of periods we fix $\vec{\Pi}=(\varpi_1,-\varpi_0)$.

The discriminant locus $z=1/64$ again corresponds to the conifold point, where $\tau=0$ and the volume of the 2-brane that wraps the curve vanishes.
However, while the large volume monodromy around $z=0,\,\tau= i\infty$ is again given by
\begin{align}
	M_\infty=\left(\begin{array}{cc}
		1&-1\\
		0&1
	\end{array}\right)\,,
\end{align}
the conifold monodromy now takes the form
\begin{align}
	M_0=\left(\begin{array}{cc}
		1&0\\
		2&1
	\end{array}\right)\,.
\end{align}
Together $M_\infty$ and $M_0$ generate the congruence subgroup $\Gamma_1(2)\subset SL(2,\mathbb{Z})$.

To study the conifold point, we change coordinates
\begin{align}
	z\rightarrow v= \frac{1}{64}-z\,,
\end{align}
and find again that the Picard-Fuchs operator~\eqref{eqn:ncurveN2PF} is transformed into itself.
This implies, that the periods at the conifold locus $v=0$ of the curve are also identical to those at $z=0$ given in~\eqref{eqn:N2periods}. 
We choose the basis $\vec{\Pi}_{\text{coni.}}=(\varpi_1,-\varpi_0)\big|_{z\rightarrow v}$ and use the closed expressions from~\eqref{eqn:N2periods} to calculate the transfer matrix $T$ such that
\begin{align}
	\vec{\Pi}=T\cdot\vec{\Pi}_{\text{coni.}}\,. 
\end{align}
It turns out, that this does not act as $\tau\mapsto 1/\tau$.
Instead we find that the periods at both points are related by
\begin{align}
	T=\frac{i}{\sqrt{2}}\left(\begin{array}{cc}
		0&-1\\
		2&0
	\end{array}\right)\,,
	\label{eqn:transferN2}
\end{align}
which acts as a Fricke involution
\begin{align}
	\tau\mapsto-\frac{1}{2\tau}\,.
	\label{eqn:fricke2a}
\end{align}
This is not contained in the $\Gamma_1(2)$ monodromy group and the two large volume points are therefore physically inequivalent.
In particular, the stringy K\"ahler moduli space is isomorphic to the modular curve $X_1(2)=\overline{\mathbb{H}/\Gamma_1(2)}$ and $T$ acts as an involution on $X_1(2)$.
The two large volume limits correspond to the two $\Gamma_1(2)$-inequivalent cusps of the modular curve.

Again, the moduli space of the fiber embeds into the moduli space of a fibration and the monodromies and transfer matrices embed into the corresponding matrices of the fibration as well.
As discussed in Section~\ref{sec:higgs2}, when the large volume limit $z=0$ corresponds to a smooth genus one fibered Calabi-Yau, the conifold point corresponds to the large volume limit of the non-commutative resolution of the Jacobian fibration.
The topological string partition function on both ``spaces'' is related by the Fricke involution~\eqref{eqn:fricke2a}.

\subsection{$N=3$: Cubic curves in $\mathbb{P}^2$}
\begin{figure}[h!]
	\begin{tikzpicture}[remember picture,overlay,node distance=4mm]
		\draw [->,line width=1.3] (4.8,3) to [in=20,out=160] (3.5,2.7);
		\node[align=center] at (7,2.8) {Genus one fibration $X^{(3)}_1$\\$\tau=i\infty$};
		\draw [->,line width=1.3] (-1,3) to [in=150,out=20] (1,2.8);
		\node[align=center] at (-2.7,2.2) {Non-comm. resolution $X^{(3)}_{0,\text{nc.} 1}$\\of Jacobian $X^{(3)}_0$ at $\tau=0$};
	\end{tikzpicture}
	\centering
	\includegraphics[width=.3\linewidth]{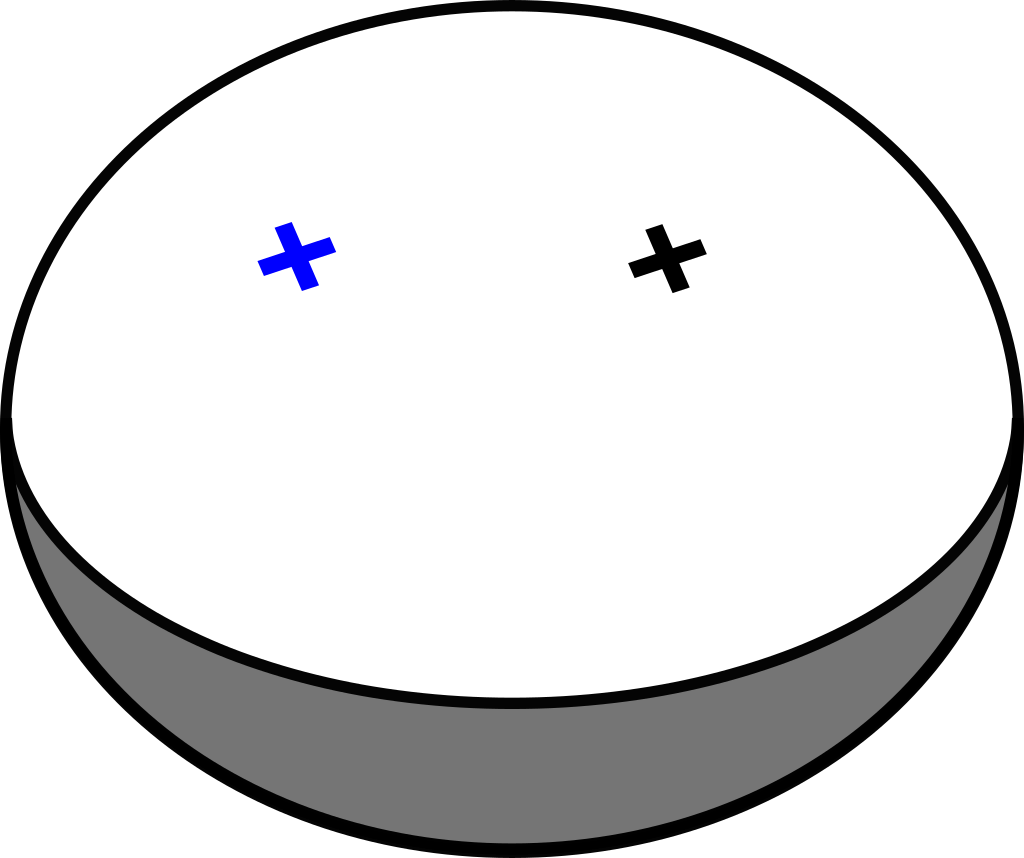}
	\caption{The stringy K\"ahler moduli space of generic genus one curves of degree $3$ is the modular curve $X_1(3)$. For Calabi-Yau fibrations with $3$-sections, the large volume points in the $\Gamma_1(3)$ orbit of $\tau=0$ correspond to the non-commutative resolution of the Jacobian.}
	\label{fig:skmoduliN3}
\end{figure}
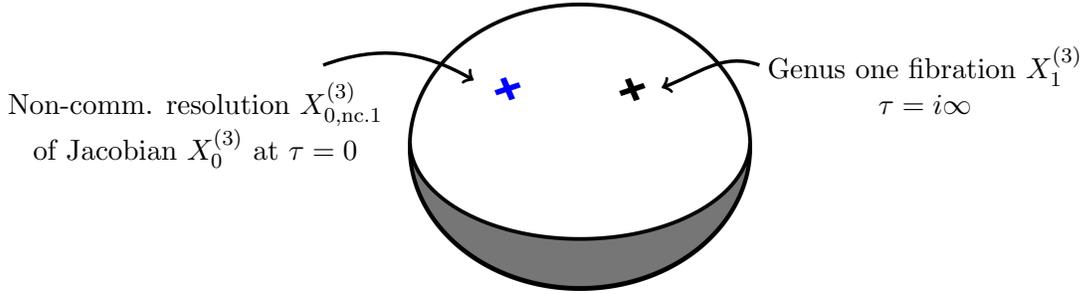
The situation for generic genus one curves of degree $3$ is largely analogous to what we found for curves of degree $2$.
They can be realized as anti-canonical hypersurfaces in $\mathbb{P}^2$ and one finds the mirror periods
\begin{align}
	\begin{split}
		\varpi_0=&\sum\limits_{n=0}^\infty\frac{\Gamma(1+3n)}{\Gamma(1+n)^3}z^n={_2}F_1\left(\frac13,\frac23;1;27z\right)\,,\\
		\varpi_1=&\frac{i}{\sqrt{3}}\cdot{_2}F_1\left(\frac13,\frac23;1;1-27z\right)\\
		=&\frac{1}{2\pi i}\left(\varpi_0\log(z)+15z+\frac{513}{2}z^2+5018z^3+\mathcal{O}(z^4)\right)\,,
	\end{split}
	\label{eqn:N3periods}
\end{align}
which are annihilated by the Picard-Fuchs operator
\begin{align}
	\mathcal{D}=\theta^2-3z(3\theta+1)(3\theta+2)\,.
\end{align}
The modular parameter $\tau=\varpi_1/\varpi_0$ is now one third of the complexified volume of the curve and the discriminant polynomial reads
\begin{align}
	\Delta=1-27z\,.
\end{align}
Our choice of basis is $\vec{\Pi}=(\varpi_1,-\varpi_0)$ and the monodromies around $z=0$ and $z=1/27$ respectively act as
\begin{align}
	M_\infty=\left(\begin{array}{cc}
		1&-1\\
		0&1
	\end{array}\right)\,,\quad
	M_0=\left(\begin{array}{cc}
		1&0\\
		3&1
	\end{array}\right)\,.
\end{align}
Again, the conifold point $z=1/27,\,\tau=0$ is another large volume point and in terms of the local coordinate
\begin{align}
	v=\frac{1}{27}-z\,,
\end{align}
a basis of periods is given by $\vec{\Pi}_{\text{coni.}}=(\varpi_1,-\varpi_0)\big|_{z\rightarrow v}$.
Using the closed expressions from~\eqref{eqn:N3periods}, we find that the transfer matrix $T$ that relates the periods via $\vec{\Pi}=T\cdot \vec{\Pi}_{\text{coni.}}$ is
\begin{align}
	T=\frac{i}{\sqrt{3}}\left(\begin{array}{cc}
		0&-1\\
		3&0
	\end{array}\right)\,,
	\label{eqn:N3fricke}
\end{align}
and acts as a Fricke involution
\begin{align}
	\tau\mapsto-\frac{1}{3\tau}\,.
\end{align}
The stringy K\"ahler moduli space is the modular curve $X_1(3)=\overline{\mathbb{H}/\Gamma_1(3)}$, which has two cusps, and the conifold point corresponds to a non-commutative resolution of the Jacobian fibration.~\footnote{The transfer matrix~\eqref{eqn:N3fricke} has also been calculated by~\cite{Alim:2013eja} in the context of local mirror symmetry on $\mathcal{O}(-3)\rightarrow\mathbb{P}^2$, and in~\cite{Scheidegger:2016ysn,Knapp:2016rec} as a toy model for the analytic continuation to the conifold point of one parameter families of Calabi-Yau $d$-folds.}

To deduce the nature of the large volume limit at $\tau=0$ in the moduli space of a genus one fibered Calabi-Yau with $3$-sections, we again study the action of the Fricke involution on the Jacobi forms and find
\begin{align}
	\begin{split}
	q^{\frac13}\phi_{w,1}(3\tau,\tau)\mapsto&\exp\left(-\frac{2\pi i}{9\tau}\right)\phi_{w,1}\left(-\frac{1}{\tau},-\frac{1}{3\tau}\right)=\tau^w\phi_{w,1}\left(\tau,-\frac13\right)\,.
	\label{eqn:frickejacobi3}
	\end{split}
\end{align}
This allows us to conclude, that the limit corresponds to a non-commutative resolution $X^{(3)}_{0,\text{nc.} 1}$ of the Jacobian fibration $J\equiv X^{(3)}_0$.
In this case, there exists a third element in the Tate-Shafarevich group $\Sh(J)\equiv \mathbb{Z}_3$ which is isomorphic to $X^{(3)}_1$ and only differs in the action of the Jacobian fibration.
This would correspond to a Higgs transition with $U(1)_{KK}\times U(1)_{6d}$ charges $(2,-3)$.
Similarly, there exist two different choices of a non-trivial B-field along a $3$-torsional $2$-cycle and we therefore expect a second non-commutative resolution $X^{(3)}_{0,\text{nc.} 2}$.
However, each choice leads to an equivalent physical theory which is reflected in the relations of the weak Jacobi forms
\begin{align}
	\begin{split}
		q^{\frac{k^2}{3}}\phi_{-2,1}(3\tau,k\tau)=&(\Delta_6)^{-\frac13}\,,\quad q^{\frac{k^2}{3}}\phi_{0,1}(3\tau,k\tau)=E_{3,1}^2(\Delta_6)^{-\frac13}\,,\\
		\phi_{-2,1}\left(\tau,\frac{k}{3}\right)=&-(\Delta_6')^{-\frac13}\,,\quad \phi_{0,1}\left(\tau,\frac{k}{3}\right)=3E_{3,1}^2(\Delta_6')^{-\frac13}\,,
	\end{split}
\end{align}
where $k\in\{1,2\}$.
This leads us to believe that the two resolutions $X^{(3)}_{0,\text{nc.} 1}$ and $X^{(3)}_{0,\text{nc.} 2}$ are equivalent.
A consequence would be, that the topological string partition functions of $X^{(3)}_0$ and $X^{(3)}_{0,\text{nc.} 1}$ are sufficient to extract the Gopakumar-Vafa invariants with $\mathbb{Z}_3$ charges.
We will verify the analogous relation in the context of genus one fibrations with $4$-sections at the hand of an explicit example in Section~\ref{sec:ex4sec}.
Again, the $\Gamma_1(3)$ modular forms $\Delta_6,\,\Delta_6'$ and $E_{3,1}$ are defined in Appendix~\ref{app:modforms}.

\subsection{$N=4$: Complete intersections of quadrics in $\mathbb{P}^3$}
\label{sec:n4fiber}
\begin{figure}[h!]
	\begin{tikzpicture}[remember picture,overlay,node distance=4mm]
		\draw [->,line width=1.3] (4.8,3) to [in=20,out=160] (3.9,2.7);
		\node[align=center] at (7.5,2.8) {Smooth genus one fibration\\$X^{(4)}_1$ at $\tau=i\infty$};
		\draw [->,line width=1.3] (-1,3) to [in=150,out=20] (.7,2.7);
		\node[align=center] at (-2.5,2.0) {Non-comm. resolution $X^{(4)}_{2,\text{nc.} 1}$\\of genus one fibration\\ $X^{(4)}_2$ at $\tau=1/2$};
		\draw [->,line width=1.3] (5,.5) to [in=-50,out=180] (2.4,2.5);
		\node[align=center] at (7.5,.5) {Non-comm. resolution $X^{(4)}_{0,\text{nc.} 1}$\\of Jacobian $X^{(4)}_0$ at $\tau=0$};
	\end{tikzpicture}
	\centering
	\includegraphics[width=.3\linewidth]{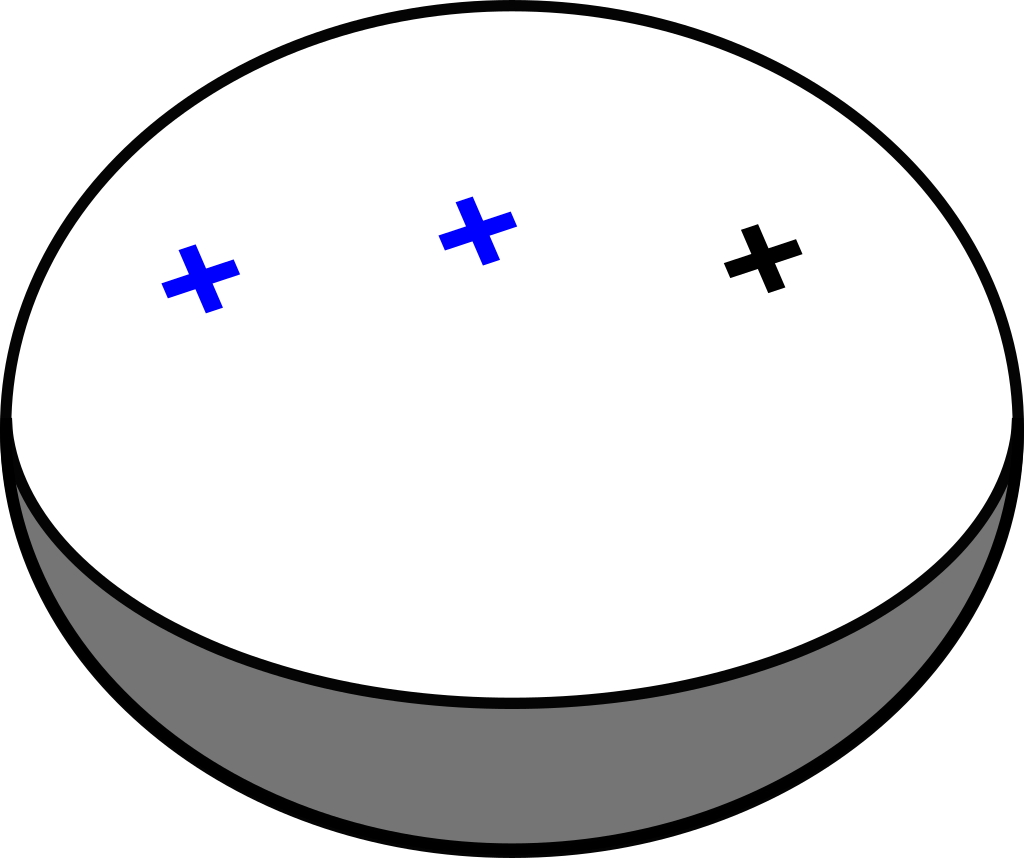}
	\caption{The stringy K\"ahler moduli space of generic genus one curves of degree $4$ is the modular curve $X_1(4)$. For a Calabi-Yau fibration $X^{(4)}_1$ with $4$-sections, the large volume points in the $\Gamma_1(4)$ orbit of $\tau=0$ correspond to the non-commutative resolution $X^{(4)}_{0,\text{nc.} 1}$ of the Jacobian $J$. On the other hand, the large volume limits at values of $\tau$ equivalent to $\tau=1/2$ are associated to a non-commutative resolution $X^{(4)}_{2,\text{nc.} 1}$ of the singular genus one fibration with $2$-sections $X^{(4)}_2\in\Sh(J)$.}
	\label{fig:skmoduliN4}
\end{figure}
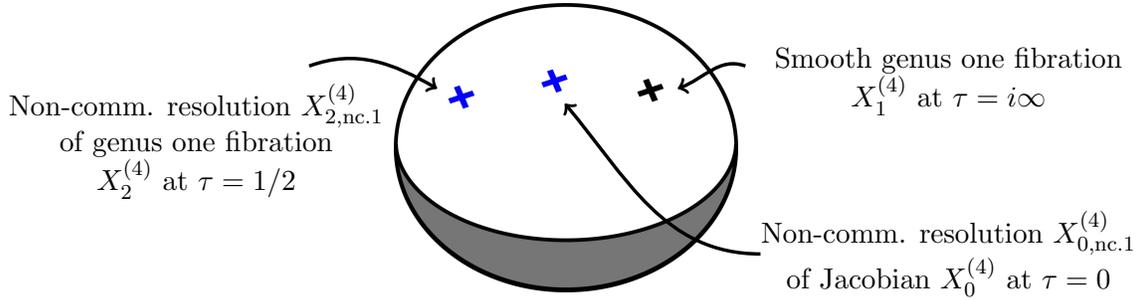
A new phenomenon occurs at $N=4$, which is related to the fact that the Tate-Shafarevich group shown in Table~\ref{tab:tselements} of a generic genus one fibered Calabi-Yau contains another singular fibration apart from the Jacobian.
This is reflected in the modular curve, which exhibits three cusps instead of two.
A generic genus one curve of degree $4$ can be realized as intersections of two quadrics in $\mathbb{P}^3$ and the mirror periods take the form
\begin{align}
	\begin{split}
		\varpi_0=&\sum\limits_{n=0}^\infty\frac{\Gamma(1+2n)^2}{\Gamma(1+n)^4}z^n={_2}F_1\left(\frac12,\frac12;1;16z\right)\,,\\
		\varpi_1=&\frac{i}{\sqrt{4}}\cdot {_2}F_1\left(\frac12,\frac12;1;1-16z\right)\\
		=&\frac{1}{2\pi i}\left(\varpi_0\log(z)+8z+84z^2+\frac{2960}{3}z^3+\mathcal{O}(z^4)\right)\,,
	\end{split}
\end{align}
while the Picard-Fuchs operator is given by
\begin{align}
	\mathcal{D}=\theta^2-4z(2\theta+1)^2\,.
\end{align}
Here the discriminant polynomial is
\begin{align}
	\Delta=1-16z\,,
\end{align}
but the moduli space contains another singular point at $z= \infty$.

Let us first discuss again the singular loci at finite values of $z$.
In terms of the usual basis of periods $\vec{\Pi}=(\varpi_1,-\varpi_0)$, the monodromies around $z=0\,,\tau= i\infty$ and $z=1/16,\,\tau=0$ act as
\begin{align}
	M_\infty=\left(\begin{array}{cc}
		1&-1\\
		0&1
	\end{array}\right)\,,\quad
	M_0=\left(\begin{array}{cc}
		1&0\\
		4&1
	\end{array}\right)\,,
\end{align}
and together they generate $\Gamma_1(4)$.
As a local coordinate at the conifold point $z=1/16$ we can choose
\begin{align}
	v=\frac{1}{16}-z\,,
\end{align}
a basis of periods is given by $\vec{\Pi}_{\text{coni.}}=(\varpi_1,-\varpi_0)\big|_{z\rightarrow v}$.
As expected, the transfer matrix $T_1$ that relates the periods at $z=0$ and $z=1/16$ via $\vec{\Pi}=T_1\cdot\vec{\Pi}_{\text{coni.}}$ is
\begin{align}
	T_1=\frac{i}{\sqrt{4}}\left(\begin{array}{cc}
		0&-1\\
		4&0
	\end{array}\right)\,,
\end{align}
and acts as a Fricke involution
\begin{align}
	\tau\mapsto-\frac{1}{4\tau}\,.
	\label{eqn:N4fricke}
\end{align}

To obtain the transfer matrix that relates the periods at $z=0$ to those at $z^{-1}=0$, we first choose the local coordinate $w=1/(256z)$ with the transformed Picard-Fuchs operator
\begin{align}
	(2\theta-1)^2-64w\theta^2\,.
\end{align}
A basis of periods is now given by $\vec{\Pi}_{\text{inf.}}=\sqrt{w}\cdot(\varpi_1,-\varpi_0)\big|_{z\rightarrow w}$ and we obtain the transfer matrix
\begin{align}
	T_2=\frac{1}{4}\left(\begin{array}{cc}1&0\\2&1\end{array}\right)\,,
\end{align}
such that $\vec{\Pi}=T_2\cdot\vec{\Pi}_{\text{inf.}}$.
This corresponds to the $\Gamma_1(2)$ transformation
\begin{align}
	\tau\mapsto\frac{\tau}{2\tau+1}\,,
	\label{eqn:n4t2}
\end{align}
and maps the cusp at $\tau= i\infty$ to the cusp $\tau=1/2$.
The monodromy around $w=0$ with respect to the basis $\vec{\Pi}$ is then given by
\begin{align}
	M_{1/2}=\left(\begin{array}{cc}-3&1\\-4&1\end{array}\right)\,.
\end{align}
This is already contained in the group $\Gamma_1(4)$ that is generated by $M_0,\,M_\infty$, and the stringy K\"ahler moduli space can be identified with the modular curve $X_1(4)=\overline{\mathbb{H}/\Gamma_1(4)}$.

However, as a consequence of the overall factor $\sqrt{w}$, the local monodromy is
\begin{align}
	\vec{\Pi}_{\text{inf.}}\rightarrow\left(\begin{array}{cc}-1&1\\0&-1\end{array}\right)\vec{\Pi}_{\text{inf.}}\,,
\end{align}
which is not quite maximally unipotent.
The appearance of $\sqrt{w}$ in $\vec{\Pi}_{\text{inf.}}$ indicates that the theory develops an enhanced quantum symmetry at this point and a better choice of coordinate is $u=\sqrt{w}$.
This is analogous to the situation at the Landau-Ginzburg point in the moduli space of the quintic~\cite{Candelas:1990rm}.
A crucial difference is that, in the appropriate coordinate, $u=0$ is a MUM-point and not regular.
Since the natural flat coordinate $\tau_u$ is then normalized such that
\begin{align}
	\tau_u=\frac{1}{2\pi i}\log(u)+\mathcal{O}(u^2)=\frac12\left(\frac{1}{2\pi i}\log(w)+\mathcal{O}(w)\right)\,,
\end{align}
we choose a basis $\vec{\Pi}_{\text{inf.}}'=u\cdot(\varpi_1/2,-\varpi_0)\big|_{z\rightarrow u^2}$ with transfer matrix
\begin{align}
	T_u=\left(\begin{array}{cc}2&0\\0&1\end{array}\right)\,,
	\label{eqn:Tu}
\end{align}
that relates the periods $\vec{\Pi}_{\text{inf.}}=T_u\cdot\vec{\Pi}_{\text{inf.}}'$.
This corresponds to the action
\begin{align}
	\tau\mapsto2\tau\,,
	\label{eqn:n4t2p}
\end{align}
on the modular parameter.

The nature of the large volume limit at $\tau=0$ in the moduli space of a genus one fibered Calabi-Yau $X_1\equiv X^{(4)}_1$ with $4$-sections can again be determined from the action of the Fricke involution on the weak Jacobi forms
\begin{align}
	\begin{split}
		q^{\frac{k^2}{4}}\phi_{w,1}(4\tau,k\tau)\mapsto\exp\left(-\frac{k^2 \pi i}{8\tau}\right)\phi_{w,1}\left(-\frac{1}{\tau},-\frac{k}{4\tau}\right)=\tau^w\phi_{w,1}\left(\tau,-\frac{k}{4}\right)\,.
	\label{eqn:frickejacobi4}
	\end{split}
\end{align}
The large volume limit $\tau=i\infty$ corresponds to the $k=1$ or, equivalently, $k=3$ element of the Tate-Shafarevich group $\Sh(J)=\mathbb{Z}_4$ with $J\equiv X^{(4)}_0$.
At $\tau=0$ we therefore expect the non-commutative resolution $X^{(4)}_{0,\text{nc.} 1}$ of $J$ that is equivalent to the resolution $X^{(4)}_{0,\text{nc.} 3}$.
Again, the physical equivalence between the theories associated to $k=1$ and $k=3$ is reflected in the relations between the Jacobi forms and $\Gamma_1(4)$ modular forms
\begin{align}
	\begin{split}
		q^{\frac{k^2}{4}}\phi_{-2,1}(4\tau,k\tau)=&(\Delta_8)^{-\frac14}\,,\quad q^{\frac{k^2}{4}}\phi_{0,1}(4\tau,k\tau)=E_{4,2}(\Delta_4)^{-\frac14}\,,\\
		\phi_{-2,1}\left(\tau,\frac{k}{4}\right)=&(\Delta_8')^{-\frac14}\,,\quad \phi_{0,1}\left(\tau,\frac{k}{4}\right)=E_{4,2}'(\Delta_8')^{-\frac14}\,,
	\end{split}
\end{align}
for $k\in\{1,3\}$.

On the other hand, the $\Gamma_1(2)$ transformation~\eqref{eqn:n4t2} acts on the Jacobi forms as
\begin{align}
	\begin{split}
	q^{\frac{k^2}{4}}\phi_{w,1}(4\tau,k\tau)\mapsto&\exp\left(2\pi i\frac{k^2}{4}\frac{\tau}{2\tau+1}\right)\phi_{w,1}\left(\frac{4\tau}{2\tau+1},\frac{k\tau}{2\tau+1}\right)\\
	=&\exp\left(2\pi i\frac{k^2}{8}\frac{\tau'-\frac12}{\tau'}\right)\phi_{w,1}\left(\frac{2\tau'-1}{\tau'},\frac{k}{2}\frac{\tau'-\frac12}{\tau'}\right)\,,\quad\tau'=\tau+\frac12\\
	=&\exp\left(2\pi i\frac{k^2}{8}\left(2\tau'-1\right)\right)\tau'^w\phi_{w,1}\left(\tau',\frac{k}{2}\left[\tau'-\frac12\right]\right)\\
		=&q^{\frac{k^2}{4}}\left(\tau+\frac12\right)^w\phi_{w,1}\left(\tau+\frac12,\frac{k}{2}\tau\right)\,.
	\end{split}
	\label{eqn:g12Jtrafo}
\end{align}
This suggests, that the large volume point at $\tau=1/2$ arises via a Higgs transition with $U(1)_{KK}\times U(1)_{6d}$ charges $(2,-4)$ and a non-vanishing B-field along the vanishing cycle.
We thus interpret it as corresponding to a non-commutative resolution $X^{(4)}_{2,\text{nc.} 1}$ of $X^{(4)}_2$.
The restrictions of the generators of weak Jacobi forms are related to $\Gamma_1(4)$ modular forms via
\begin{align}
	\begin{split}
	q^{\frac{k^2}{4}}\phi_{-2,1}\left(\tau+\frac12,\frac{k}{2}\tau\right)=&(\Delta_8'')^{-\frac14}\,,\quad q^{\frac{k^2}{4}}\phi_{0,1}\left(\tau+\frac12,\frac{k}{2}\tau\right)=E_{4,2}''(\Delta_8'')^{-\frac14}\,,
	\end{split}
	\label{eqn:n4expr2}
\end{align}
where again $k\in\{1,3\}$.
Note that the change of basis~\eqref{eqn:Tu} from $\vec{\Pi}_{\text{inf.}}$ to $\vec{\Pi}_{\text{inf.}}'$ just amounts to replacing $\tau$ with $2\tau$ on the right hand side of~\eqref{eqn:g12Jtrafo}.
The basis $\vec{\Pi}_{\text{inf.}}$ is somewhat more natural from a modular perspective, due to the expressions~\eqref{eqn:n4expr2} in terms of $\Gamma_1(4)$ modular forms and the transfer matrix acting as an element of $\Gamma_1(2)$. 
On the other hand, as we will see in Section~\ref{sec:ex4sec}, replacing $\tau\rightarrow 2\tau$ will be necessary to extract the Gopakumar-Vafa invariants with $\mathbb{Z}_2$ charges.

The large volume limit at $\tau=1/2$ has also been discussed in~\cite{Caldararu:2010ljp}, where it was conjectured to be homologically projective dual to the point at $\tau=i\infty$.
From a GLSM perspective, the theory at $\tau=1/2$ corresponds to a Landau-Ginzburg model with quadratic superpotential that is fibered over a $\mathbb{P}^1$ with non-minimal charges for the scaling relations.
An analogous hybrid point in the moduli space of the intersection of four quadrics in $\mathbb{P}^7$ was also considered in~\cite{Caldararu:2010ljp}, and argued to correspond to a non-commutative resolution of a singular space in terms of the category of matrix factorizations of the associated Landau-Ginzburg model.
Corresponding Gromov-Witten invariants have been proposed and calculated in~\cite{Sharpe:2012ji}.
However, in the moduli space of the intersection of two quadrics in $\mathbb{P}^3$ the hybrid point can be interpreted as describing an elliptic curve as a double cover over $\mathbb{P}^1$, which is regular for a generic choice of complex structure.
Singularities only occur, when this curve is fibered over a surface to construct a Calabi-Yau threefold.
As can be seen from the arguments of the Jacobi forms in~\eqref{eqn:g12Jtrafo}, the non-commutative resolution then again admits a natural interpretation in terms of a torsional flat B-field at the singularity.

\subsection{$N=5$: Pfaffian curves in $\mathbb{P}^4$ and complete intersections in $G(2,5)$}
\label{sec:n5fiber}
\begin{figure}[h!]
	\begin{tikzpicture}[remember picture,overlay,node distance=4mm]
		\draw [->,line width=1.3] (4.9,3) to [in=40,out=160] (4.0,2.75);
		\node[align=center] at (7.5,2.8) {Genus one fibration\\$X^{(5)}_1$ at $\tau=i\infty$};
		\draw [->,line width=1.3] (-1,2.5) to [in=240,out=-40] (.6,2.1);
		\node[align=center] at (-2.5,3.0) {Genus one fibration\\$X^{(5)}_2$ at $\tau=2/5$};
		\draw [->,line width=1.3] (5.4,1) to [in=-50,out=180] (3.1,2.5);
		\node[align=center] at (7.5,.5) {Non-comm. resolution $X^{(5)}_{0,\text{nc.} 1}$\\of Jacobian $X^{(5)}_0$ at $\tau=0$};
		\draw [->,line width=1.3] (-.5,1) to [in=-110,out=0] (1.6,2.6);
		\node[align=center] at (-3,1.2) {Non-comm. resolution $X^{(5)}_{0,\text{nc.} 2}$\\of Jacobian $X^{(5)}_0$ at $\tau=1/2$};
	\end{tikzpicture}
	\centering
	\includegraphics[width=.3\linewidth]{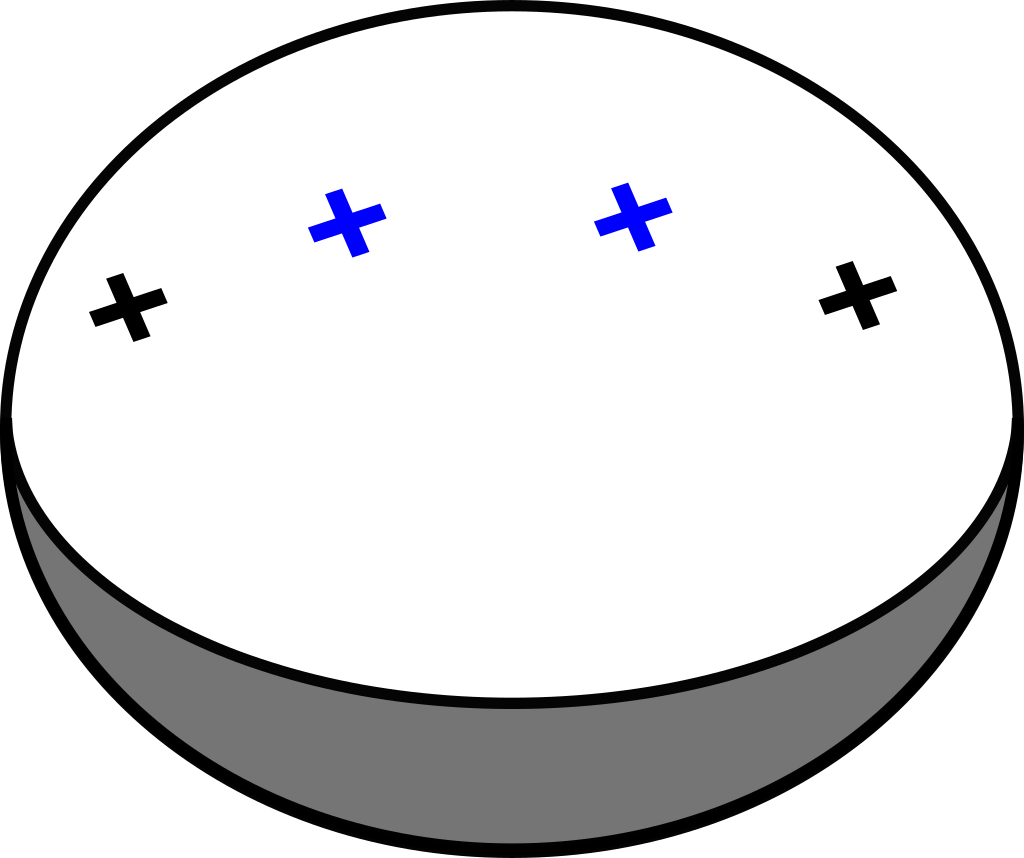}
	\caption{The stringy K\"ahler moduli space of generic genus one curves of degree $5$ is the modular curve $X_1(5)$. For a Calabi-Yau fibration $X^{(5)}_1$ with $5$-sections, the large volume points in the $\Gamma_1(5)$ orbit of $\tau=0$ correspond to the non-commutative resolution $X^{(5)}_{0,\text{nc.} 1}$ of the Jacobian $J$. The large volume limit at $\tau=2/5$ is associated to the smooth genus one fibration $X^{(5)}_2$ that is relatively homologically projective dual to $X^{(5)}_1$.  On the other hand, the large volume limits at values in the orbit of $\tau=1/2$ correspond to the non-commutative resolution $X^{(5)}_{0,\text{nc.} 2}$ of the Jacobian $J$.}
	\label{fig:skmoduliN5}
\end{figure}
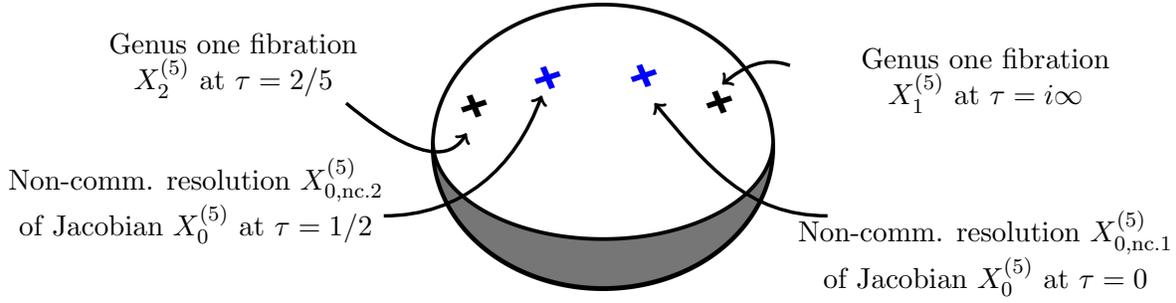
Genus one curves of degree $5$ are special in that they cannot be realized as complete intersections in a toric ambient space.
In fact, there are two isomorphic descriptions in terms of a Pfaffian curve in $\mathbb{P}^4$ and a complete intersection curve in the Grassmannian $G(2,5)$.
This is closely related to the fact that the Tate-Shafarevich group of the corresponding fibrations contains in general two smooth genus one fibrations.
The stringy K\"ahler moduli space has been discussed in~\cite{Knapp:2021vkm}.

The fundamental period of the mirror can not be expressed directly as a hypergeometric function but takes the form~\cite{Hori:2013gga}
\begin{align}
	\varpi_0=\sum\limits_{n=0}^\infty {_3}F_2(-n,-n,1+n;1,1;1)z^n\,,
\end{align}
while the logarithmic period reads
\begin{align}
	\varpi_1=\frac{1}{2\pi i}\left(\varpi_0\log(z)+5z+\frac{75}{2}z^2+\frac{1855}{6}z^3+\mathcal{O}(z^4)\right)\,.
\end{align}
In this case we are not aware of any closed expression in terms of common special functions.
The modular parameter $\tau=\varpi_1/\varpi_0$ is now one fifth of the complexified volume of the curve and the Picard-Fuchs operator is given by
\begin{align}
	\mathcal{D}=\theta^2-z(11\theta^2+11\theta+3)-z^2(\theta+1)^2\,,
\end{align}
with the discriminant polynomial
\begin{align}
	\Delta=1-11z-z^2\,.
\end{align}
This vanishes at the two conifold points
\begin{align}
	z_{\pm} = -\frac12\left(11\pm5\sqrt{5}\right)\,,
	\label{eqn:n5coniroots}
\end{align}
with $z_-\approx 0.09$ and $z_+\approx -11.1$.
Again, there is another large volume limit at $z= \infty$.

If the large volume point $z=0$ is constructed as a complete intersection curve in $G(2,5)$, the homologically projective dual Pfaffian curve is associated to the large volume limit $z=\infty$.
Starting with a basis of periods $\vec{\Pi}=(\varpi_1,-\varpi_0)$ at $z=0$, the monodromy matrices around $z=0$, $z_-$ and $z_+$ are respectively found to be
\begin{align}
	M_\infty=\left(\begin{array}{cc}
		1&-1\\
		0&1
	\end{array}\right)\,,\quad
	M_+=\left(\begin{array}{cc}
		1&0\\
		5&1
	\end{array}\right)\,,\quad
	M_-=\left(\begin{array}{cc}
		-9&5\\
		-20&11
	\end{array}\right)\,,
\end{align}
and together they generate $\Gamma_1(5)$~\cite{Knapp:2021vkm}.
The stringy K\"ahler moduli space is therefore equivalent to the modular curve $X_1(5)=\overline{\mathbb{H}/\Gamma_1(5)}$.
On the other hand, the transfer matrix that transports the periods from $z=0$ to $z=\infty$ is given by
\begin{align}
	T=\left(\begin{array}{cc}
		2&-1\\
		5&-2
	\end{array}\right)\,.
	\label{eqn:n5lrtrafo}
\end{align}
This satisfies $T^2=1$, therefore acts on $\tau$ as an involution, and maps the cusp $\tau=i\infty$ to that at $\tau=2/5$.

To calculate the transfer matrix from $z=0,\,\tau=i\infty$ to $z=z_-,\,\tau=0$, we choose a local coordinate $v$ that is related to $z$ via
\begin{align}
	z=\frac52\left(25-11\sqrt{5}\right)\left(\frac{1}{5\sqrt{5}}-v\right)\,,
	\label{eqn:5sevvcoord}
\end{align}
and a basis of periods $\vec{\Pi}_{\text{coni.}}=(\omega_{1,c},-\omega_{0,c})$, with leading terms
\begin{align}
	\begin{split}
		\omega_{0,c}=&1+v \left(z_-+3\right)+v^2 \left(20-5 z_-\right)+\mathcal{O}(v^3)\,,\\
		\omega_{1,c}=&\frac{1}{2\pi i}\left(\omega_{0,c}\log(v)+\frac12 v (10 + 2 z_-)+\mathcal{O}(v^2)\right)\,.
	\end{split}
\end{align}
Using numerical analytic continuation, we can then obtain the transfer matrix
\begin{align}
	T_1=i\sqrt{\frac{25-11\sqrt{5}}{2}}\left(\begin{array}{cc}0&-1\\5&0\end{array}\right)\,,
\end{align}
such that $\vec{\Pi}_{\text{coni.}}=T_1\cdot \vec{\Pi}$, and this acts on the modular parameter again as a Fricke involution
\begin{align}
	\tau\mapsto-\frac{1}{5\tau}\,.
\end{align}
Note that the normalization of the coordinate $v$ in~\eqref{eqn:5sevvcoord} is fixed by requiring that the transfer matrix, up to an overall scaling, is integral.
The same Fricke involution also connects the cusps at $z=\infty,\,\tau=2/5$ and $z=z_+,\,\tau=1/2$.

Let us now consider again the interpretation of the images of the large volume limits in the moduli spaces of genus one fibered Calabi-Yau manifolds with $5$-sections.
As discussed in~\cite{Knapp:2021vkm}, the large volume limits at $\tau=i\infty$ and $\tau=2/5$ correspond to smooth genus one fibrations that we respectively denote as $X_1\equiv X^{(5)}_1$ and $X_2\equiv X^{(5)}_2$.
We will refer to the singular Jacobian fibration as $J\equiv X^{(5)}_0$.
The action of the Fricke involution on the generators of weak Jacobi forms
\begin{align}
	\begin{split}
		q^{\frac{k^2}{5}}\phi_{w,1}(5\tau,k\tau)\mapsto\exp\left(-\frac{2k^2 \pi i}{25\tau}\right)\phi_{w,1}\left(-\frac{1}{\tau},-\frac{k}{5\tau}\right)=\tau^w\phi_{w,1}\left(\tau,-\frac{k}{5}\right)\,,
	\label{eqn:frickejacobi5}
	\end{split}
\end{align}
is analogous to the previous cases and implies that the large volume limit $\tau=0$ corresponds to the non-commutative resolution $X^{(5)}_{0,\text{nc.} 1}=X^{(5)}_{0,\text{nc.} 4}$ of $J$.
On the other hand, the transfer matrix~\eqref{eqn:n5lrtrafo} that connects the large volume limits associated to $X^{(5)}_1$ and $X^{(5)}_2$ acts as
\begin{align}
	\begin{split}
		q^{\frac{1}{5}}\phi_{w,1}(5\tau,\tau)\mapsto&\exp\left(\frac{2\pi i}{5}\frac{2\tau-1}{5\tau-2}\right)\phi_{w,1}\left(\frac{10\tau-5}{5\tau-2},\frac{2\tau-1}{5\tau-2}\right)\\
		=&\exp\left(2\pi i\frac{1}{25} \frac{2\tau'-5}{\tau'-2}\right)\phi_{w,1}\left(\frac{2\tau'-5}{\tau'-2},\frac{1}{5}\frac{2\tau'-5}{\tau'-2}\right)\,,\quad \tau'=5\tau\\
		=&e^{-\frac{4\pi i}{5}}q^{\frac45}\phi_{w,1}\left(5\tau,2\tau\right)\,.
	\end{split}
\end{align}
The arguments of the Jacobi forms reflect that the theory which arises from a Higgs transition with $U(1)_{KK}\times U(1)_{6d}$ charges $(1,-5)$ is transformed into one that results from a transition using a field with charges $(2,-5)$.
Combining this with~\eqref{eqn:frickejacobi5}, we conclude that the large volume limit at $z_+\,,\tau=1/2$ corresponds to the non-commutative resolution $X^{(5)}_{0,\text{nc.} 2}=X^{(5)}_{0,\text{nc.} 3}$.

\section{Example 1: Genus one fibrations with $2$-sections}
\label{sec:ex2sec}
We will now illustrate our results at the example of a genus one fibered Calabi-Yau threefold $X^{(2)}_1$ over $\mathbb{P}^2$ with $2$-sections.
There are a total of four geometries involved in this example.
Following the discussion in Section~\ref{sec:strk}, the stringy K\"ahler moduli space of $X^{(2)}_1$ contains a second large volume limit that is associated to the non-commutative resolution $X^{(2)}_{0,\text{nc.} 1}$ of the Jacobian fibration $X^{(2)}_0$.
On the other hand, the singular Jacobian fibration itself is related via a complex structure deformation to the generic elliptic fibration $X^{(1)}_0$ over $\mathbb{P}^2$.
The relation among these fibrations is illustrated in Figure~\ref{fig:ex1geos}.
All of these geometries are connected via Higgs/extremal transitions to an elliptically fibered Calabi-Yau $X^{(2)}_{\text{el.}}$.
The corresponding effective theories in M- and F-theory as well as their relations via Higgs transitions are summarized in Figure~\ref{fig:z2fieldtheories}.
\begin{figure}[h!]
	\begin{tikzpicture}[remember picture,node distance=4mm]
		\node[align=center] at (5,3) {$X^{(2)}_{0,\text{n.c.}1}$};
		\node[align=center] at (9,3) {$X^{(2)}_{1}$};
		\node[align=center] at (7,3.3) {\tiny str. K\"ahler def.};
		\draw [<->] (6,3) to (8,3);
		\node[align=center] at (5,5) {$X^{(2)}_{0,\text{n.c.}0}$};
		\node[align=center] at (9,5) {$X^{(1)}_{0}$};
		\node[align=center] at (7,5.3) {\tiny cplx. def.};
		\draw [<->,dashed] (6,5) to (8,5);
		\draw [<->,dotted] (5,3.5) to (5,4.5);
		\node[align=center,rotate=90] at (4.7,4) {\tiny B-field};
	\end{tikzpicture}
	\centering
	\caption{The relations between the smooth genus one fibration $X^{(2)}_1$ with two sections, the non-commutative resolutions of the Jacobian fibration $X^{(2)}_{0}$ and the generic elliptic fibration $X^{(1)}_0$ over $\mathbb{P}^2$.}
	\label{fig:ex1geos}
\end{figure}
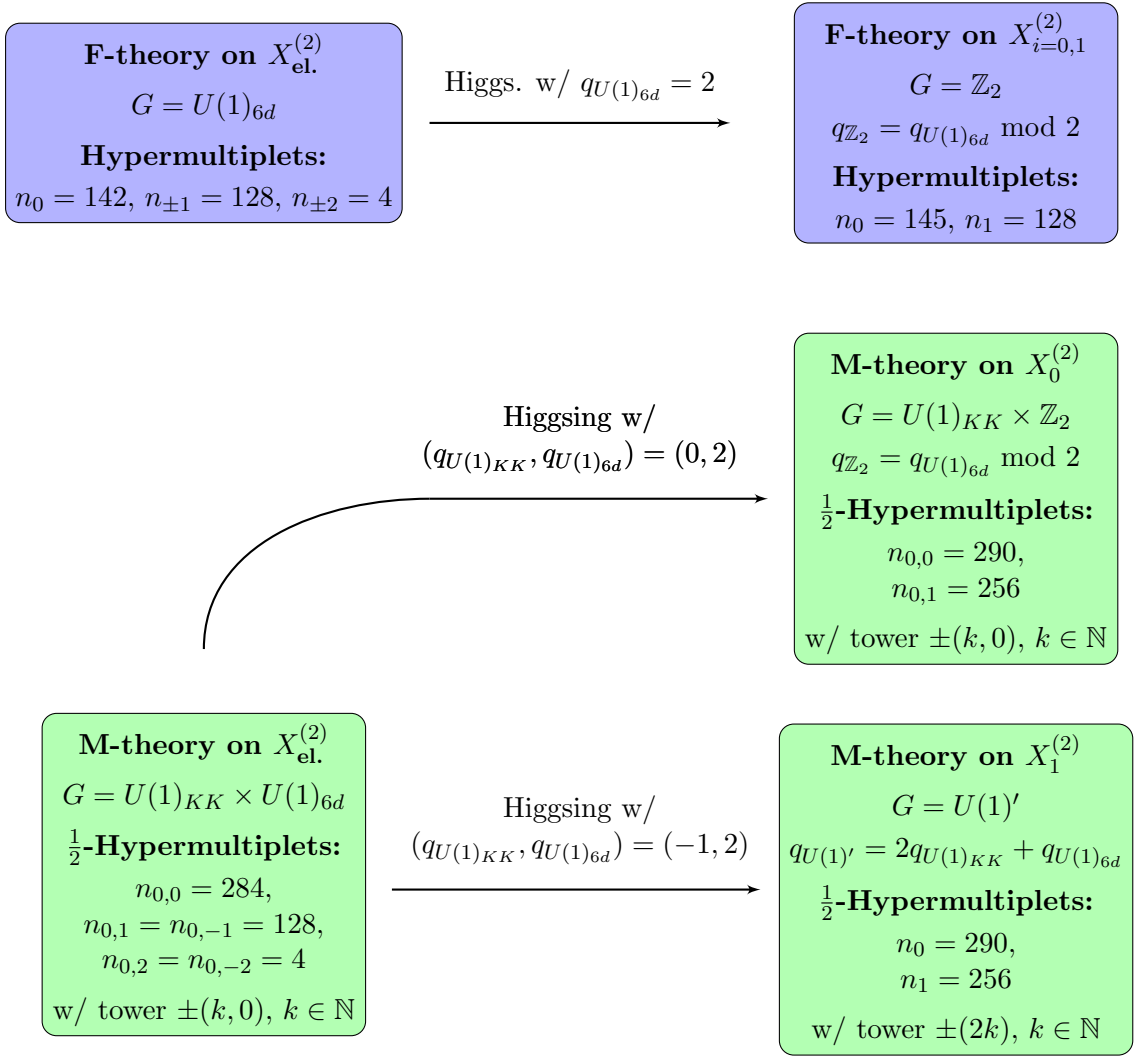
\begin{figure}[p!]
	\centering
\begin{tikzpicture}[remember picture,node distance=4mm, >=latex',
bblock/.style = {draw, rectangle, rounded corners=0.5em,align=center,minimum height=2cm,minimum width=4.3cm,fill=blue!30},
gblock/.style = {draw, rectangle, rounded corners=0.5em,align=center,minimum height=2cm,minimum width=4.3cm,fill=green!30},
                        ]            
	\node [bblock,align=center] at ( 0, 12) {
		\textbf{F-theory on $X^{(2)}_{\text{el.}}$}\\[.5em]
		$G=U(1)_{6d}$\\[.5em]
		\textbf{Hypermultiplets:}\\[.2em]
		$n_0=142,\,n_{\pm1}=128,\,n_{\pm2}=4$
		};
	\node [bblock,align=center] at ( 10, 12) {
		\textbf{F-theory on $X^{(2)}_{i=0,1}$}\\[.5em]
		$G=\mathbb{Z}_2$\\[.2em]
		$q_{\mathbb{Z}_2}=q_{U(1)_{6d}}\text{ mod }2$\\[.5em]
		\textbf{Hypermultiplets:}\\[.2em]
		$n_0=145,\,n_{1}=128$
		};
	\node [gblock,align=center] at ( 0, 2) {
		\textbf{M-theory on $X^{(2)}_{\text{el.}}$}\\[.5em]
		$G=U(1)_{KK}\times U(1)_{6d}$\\[.5em]
		\textbf{$\frac12$-Hypermultiplets:}\\[.2em]
		$n_{0,0}=284,$\\
		$n_{0,1}=n_{0,-1}=128,$\\
		$n_{0,2}=n_{0,-2}=4$\\[.5em]
		w/ tower $\pm(k,0),\,k\in\mathbb{N}$
		};
	\node [gblock,align=center] at ( 10, 7) {
		\textbf{M-theory on $X^{(2)}_{0}$}\\[.5em]
		$G=U(1)_{KK}\times \mathbb{Z}_2$\\[.2em]
		$q_{\mathbb{Z}_2}=q_{U(1)_{6d}}\text{ mod }2$\\[.5em]
		\textbf{$\frac12$-Hypermultiplets:}\\[.2em]
		$n_{0,0}=290,$\\$n_{0,1}=256$\\[.5em]
		w/ tower $\pm(k,0),\,k\in\mathbb{N}$
		};
	\node [gblock,align=center] at ( 10, 1.8) {
		\textbf{M-theory on $X^{(2)}_{1}$}\\[.5em]
		$G=U(1)'$\\[.2em]
		$q_{U(1)'}=2q_{U(1)_{KK}}+q_{U(1)_{6d}}$\\[.5em]
		\textbf{$\frac12$-Hypermultiplets:}\\[.2em]
		$n_{0}=290,$\\$n_{1}=256$\\[.5em]
		w/ tower $\pm(2k),\,k\in\mathbb{N}$
		};
		\draw [->,thick] (3,12) to (7,12);
		\node[align=center] at (5,12.5) {Higgs. w/ $q_{U(1)_{6d}}=2$};
		\draw [-,thick] (0,5) to [in=180,out=90](3,7);
		\draw [->,thick] (3,7) to (7.5,7);
		\node[align=center] at (5,7.8) {Higgsing w/\\$(q_{U(1)_{KK}},q_{U(1)_{6d}})=(0,2)$};
		\node[align=center] at (5,7.8) {Higgsing w/\\$(q_{U(1)_{KK}},q_{U(1)_{6d}})=(0,2)$};
		\draw [->,thick] (2.5,1.8) to (7.4,1.8);
		\node[align=center] at (5,2.6) {Higgsing w/\\$(q_{U(1)_{KK}},q_{U(1)_{6d}})=(-1,2)$};
\end{tikzpicture}
	\caption{The six- and five-dimensional gauge groups and matter spectra of F- and M-theory compactifications on $X^{(2)}_{\text{el.}}$ and $X^{(2)}_{i=0,1}$, as well as the relations via Higgs transitions.}
	\label{fig:z2fieldtheories}
\end{figure}

First, we are going to discuss the elliptic fibration and calculate the modular expressions for the base degree one partition function.
Specializing the geometric elliptic parameter as required by the charges of the Higgs field and the B-field then allows us to obtain corresponding expressions for $X^{(2)}_0,\,X^{(2)}_1$ and the non-commutative resolution of $X^{(2)}_0$.
The results for $X^{(2)}_0$ are equivalent to those for $X^{(1)}_0$, which have been calculated in~\cite{Huang:2015sta}, and base degree one partition function on $X^{(2)}_1$ can also be calculated directly using standard techniques from mirror symmetry and~\cite{Cota:2019cjx}.
However, the Higgs transition will fix the normalization of the free energies associated to the non-commutative resolution $X^{(2)}_{0,\text{n.c.}1}$ of $X^{(2)}_0$ and also provides a strong consistency check.

We then study the stringy K\"ahler moduli space of $X^{(2)}_1$ and find that the cusp at $\tau=0$ lies in a tangency between the large base limit and the conifold locus.
A coordinate transformation that resolves the tangency into normal crossing divisors relates the large volume limits associated to $X^{(2)}_1$ and $X^{(2)}_{0,\text{nc.} 1}$.
This enables us to obtain the Picard-Fuchs system associated to $X^{(2)}_{0,\text{nc.} 1}$ from that of $X^{(2)}_1$ and to verify that the free energies match those that we obtained via Higgs transition.
We also verify that the base degree one topological string partition functions of $X^{(2)}_1$ and $X^{(2)}_{0,\text{nc.} 1}$ transform into each other under Fricke involutions.
Finally, using the expansion proposed in Section~\ref{sec:gvtorsion}, we will combine the free energies and base degree one partition function of $X^{(2)}_{0,\text{nc.} 1}$ and $X^{(2)}_{0}$ to extract Gopakumar-Vafa invariants with $\mathbb{Z}_2$ charges and compare them to the Gopakumar-Vafa invariants of $X^{(2)}_1$.

In the following discussion we will assume a basic familiarity with mirror symmetry for hypersurfaces and complete intersections in toric ambient spaces.
Introductions can be found for example in~\cite{Hosono:1994av,Cox:2000vi,Paul-KonstantinOehlmann:2019jgr}.
Readers that are not interested in details of the calculation can safely skip ahead to the next example in Section~\ref{sec:ex4sec}.

\subsection{The elliptic fibration $X^{(2)}_{\text{el.}}$ with $2$ sections}
\label{sec:ell2sec}
The elliptically fibered Calabi-Yau threefold $X^{(2)}_{\text{el.}}$ is a generic anti-canonical hypersurface in the toric variety that is defined by the data
\begin{align}
\begin{blockarray}{crrrrrrrl}
	&&&&&C_1&C_2&C_3\\
\begin{block}{c(rrrr|rrr)l}
	  X&-1& 1& 0& 0& 1&-1&-3&\leftarrow\text{holomorphic section }S_0\\
	  Y&-1&-1& 0& 0& 0& 1& 0&\leftarrow\text{two-section}\\
	  Z& 1& 0& 0& 0& 1& 0& 0&\leftarrow\text{three-section}\\
	  e& 0& 1& 0& 0&-1& 2& 1&\leftarrow\text{rational section }S_1\\
	x_1& 0& 0& 1& 0& 0& 0& 1&\leftarrow\text{vertical divisor }D_b\\
	x_2& 0& 0& 0& 1& 0& 0& 1&\phantom{x}\hspace{1.5cm}\text{\ditto}\\
	x_3&-3& 2&-1&-1& 0& 0& 1&\phantom{x}\hspace{1.5cm}\text{\ditto}\\
	   & 0& 0& 0& 0&-1&-2&-1&\\
\end{block}
\end{blockarray}\,.
\label{eqn:f6overf1toricData}
\end{align}
The fibration has two sections and contains $I_2$ fibers that lead to charge $1$ and $2$ matter in F-theory~\cite{Klevers:2014bqa}.
In the notation of~\cite{Klevers:2014bqa}, the fiber corresponds to the $F_6$ polytope and the fibration parameters are $\mathcal{S}_7=D_b,\,\mathcal{S}_9=-2D_b$.

Using TOPCOM~\cite{Rambau:TOPCOM-ICMS:2002} we find that the points in the first four columns of~\eqref{eqn:f6overf1toricData} admit two different fine regular star triangulations but the flop that connects the corresponding toric varieties does not intersect the Calabi-Yau.
The last three rows thus contain the linear relations that generate the intersection of the Mori cones that one obtains from the two toric phases.
To the left of each row we define the homogeneous coordinate associated to a point and the corresponding divisor is given on the right.

Before we discuss the structure of the fibration and perform the modular bootstrap let us provide the intersection numbers and basic topological invariants.
We take
\begin{align}
	J_1=[Z]\,,\quad J_2=[Y]\,,\quad J_3=[x_1]\,,
\end{align}
as generators of the K\"ahler cone and use SageMath~\cite{sagemath} to calculate the triple intersection numbers
\begin{align}
	\begin{split}
		c_{111}=107\,,&\quad c_{112}=50\,,\quad c_{113}=19\,,\quad c_{122}=20\,,\quad c_{123}=10\,,\\
		c_{133}=3\,,&\quad c_{222}=8\,,\quad c_{223}=4\,,\quad c_{233}=2\,,\quad c_{333}=0\,.
	\end{split}
	\label{eqn:intersectionsF6fibration}
\end{align}
The independent Hodge numbers are $h^{1,1}=3$ and $h^{2,1}=141$ which fixes the Euler characteristic $\chi=-276$\,.
Moreover, the intersections of the K\"ahler cone generators with the second Chern class are
\begin{align}
	c_2\cdot \vec{J}=(170,\,68,\,36)^\intercal\,.
\end{align}
We introduce complexified K\"ahler parameters $t^i,\,i\in\{1,2,3\}$ such that the complexified K\"ahler form $\omega$ is parametrized as
\begin{align}
	\omega=\sum_{i=1}^3t^iJ_i\,.
	\label{eqn:kaehlerEllQ2}
\end{align}

We choose $s_0=\{e=0\}\cap X^{(2)}_{\text{el.}}$ as the zero-section and $s_1=\{X=0\}\cap X^{(2)}_{\text{el.}}$ as the generator of the Mordell-Weil group $\text{MW}(X^{(2)}_{\text{el.}})=\mathbb{Z}$.
The image of $s_1$ under the Shioda map is then
\begin{align}
	\sigma(s_1)=S_1-S_0-4D_b\,,
	\label{eqn:shiodaEllQ2}
\end{align}
with the classes $S_0,\,S_1,\,D_b$ of the two sections and the vertical divisor being
\begin{align}
	S_0=J_1-J_2-3\cdot J_3\,,\quad S_1=-J_1+2\cdot J_2+J_3\,,\quad D_b=J_3\,.
	\label{eqn:sectJellQ2}
\end{align}
Using the intersection numbers~\eqref{eqn:intersectionsF6fibration} one can calculate the height pairing
\begin{align}
	b_{11}=\pi^{-1}\pi_*\left(\sigma(s_1)\cdot\sigma(s_1)\right)=-8\cdot D_b\,.
	\label{eqn:heightPairingEllQ2}
\end{align}
To perform the modular bootstrap we need a modular parametrization of the complexified K\"ahler form $\omega$ that takes the form
\begin{align}
	\omega=\tau\cdot\left(S_0+\frac{1}{2}D\right)+m\cdot\sigma(s_1)+t\cdot D_b\,,
	\label{eqn:modularKaehlerEllQ2}
\end{align}
where $D$ is the height pairing of the zero section with itself
\begin{align}
	D=-\pi^{-1}\pi_*\left(S_0\cdot S_0\right)=\pi^{-1}c_1(B)=3D_b\,.
\end{align}
The second equality is generally true for sections of elliptic Calabi-Yau manifolds.

The genus zero Gopakumar-Vafa invariants can be calculated using mirror symmetry~\cite{Hosono:1993qy} and some invariants with base degree zero and one are provided in the Tables~\ref{tab:gvg0bd0ellQ2} and~\ref{tab:gvg0bd1ellQ2}.
Here the degree $(d_1,d_2,d_3)$ of a curve $C$ is determined by the intersections with the K\"ahler cone generators, i.e. $d_i=C\cdot J_i,\,i=1,\ldots,3$.
\begin{table}[h!]
\begin{align*}
\begin{array}{c|ccccc}
	n^0_{d_1,d_2,0}&d_2=0&1&2&3&4\\\hline
	d_1=0& 0 & 0 & 0 & 0 & 0 \\
	1& 4 & 128 & 0 & 0 & 0 \\
	2& 0 & 128 & 4 & 0 & 0 \\
	3& 0 & 0 & 276 & 0 & 0 \\
	4& 0 & 0 & 4 & 128 & 0 \\
	5& 0 & 0 & 0 & 128 & 4 \\
	6& 0 & 0 & 0 & 0 & 276 \\
\end{array}
\end{align*}
	\caption{Low degree genus zero Gopakumar-Vafa invariants at base degree zero for the elliptically fibered Calabi-Yau threefold with 2 sections that arises from the toric data~\eqref{eqn:f6overf1toricData}.}
	\label{tab:gvg0bd0ellQ2}
\end{table}
\begin{table}[h!]
\begin{align*}
\begin{array}{c|ccccc}
	n^0_{d_1,d_2,1}&d_2=0&1&2&3&4\\\hline
	d_1=0& 3 & 0 & 0 & 0 & 0 \\
	1& -8 & 0 & 0 & 0 & 0 \\
	2& 6 & -256 & 0 & 0 & 0 \\
	3& -8 & 512 & -552 & 0 & 0 \\
	4& 3 & 512 & 9216 & -256 & 0 \\
	5& 0 & -256 & 182368 & 35328 & -8 \\
	6& 0 & 0 & 9216 & 3120896 & 53246 \\
\end{array}
\end{align*}
	\caption{Low degree genus zero Gopakumar-Vafa invariants at base degree one for the elliptically fibered Calabi-Yau threefold with 2 sections that arises from the toric data~\eqref{eqn:f6overf1toricData}.}
	\label{tab:gvg0bd1ellQ2}
\end{table}
From the fiber GV-invariants in Table~\ref{tab:gvg0bd0ellQ2} we can deduce that the fibration exhibits two different types of isolated $I_2$ fibers that respectively lead to hypermultiplets of charge $1$ and $2$ in F-theory~\cite{Paul-KonstantinOehlmann:2019jgr}.
We denote the respective multiplicities by $n_1,\,n_2$ with
\begin{align}
	n_1=128\,,\quad n_2=4\,,
	\label{eqn:I2multiesEllQ2}
\end{align}
and thus we can indeed break $U(1)\rightarrow\mathbb{Z}_2$ in the F-theory vacuum associated to $M$.
Note that for a Higgs transition one needs at least two equally charged hypermultiplets to form a D-flat direction in the superpotential~\cite{Honecker:2006qz}.
The number of neutral hypermultiplets is
\begin{align}
	n_0=h^{2,1}+1=142\,.
\end{align}
The negative of the Euler characteristic is also encoded in the fiber GV-invariants.

With~\eqref{eqn:I2multiesEllQ2} and Table~\ref{tab:gvg0bd0ellQ2} we can deduce that the class of the curves that lead to charge $2$ matter are $C^{(2)}_1$ and $C^{(2)}_2$ with
\begin{align}
	C^{(2)}_1\cdot\vec{J}=(1,\,0,\,0)^\intercal\,,\quad C^{(2)}_2\cdot\vec{J}=(2,\,2,\,0)^\intercal\,.
\end{align}
From these degrees and using~\eqref{eqn:shiodaEllQ2} and~\eqref{eqn:sectJellQ2} we calculate the charges of the corresponding five-dimensional half-hypermultiplets
\begin{align}
	(q_{KK},q_{6d})_1=(1,\,-2)\,,\quad (q_{KK},q_{6d})_2=(0,\,2)\,.
\end{align}
The two limits in which either of these multiplets become massless both lie at the boundary of the K\"ahler cone and, at the level of the modular parameters in~\eqref{eqn:modularKaehlerEllQ2}, are
\begin{align}
	\begin{split}
		1:&\,\quad \tau\rightarrow 2\tau\,,\quad m\rightarrow \tau\,,\\
		2:&\,\quad \tau\rightarrow \tau\,,\quad m\rightarrow 0\,.
		\label{eqn:q2higgsLimits1}
	\end{split}
\end{align}
The first transition connects the elliptic fibration to the smooth genus one fibration $X^{(2)}_1$ with $2$-sections while the second leads to the Jacobian fibration $J\equiv X^{(2)}_0$.
As was discussed in~\cite{Mayrhofer:2014laa}, the second limit, when combined with a subsequent complex structure deformation that removes the singularity from the $I_2$ singular curves that lead to charge $2$ matter, leaves a terminal singularity in the $I_2$ singular curves that realized the charge $1$ matter.
This renders the Jacobian singular.
In order to perform a transition to the non-commutative resolution $X^{(2)}_{0,\text{nc.} 1}$, we turn on a non-trivial B-field along the vanishing cycle setting
\begin{align}
	3:&\,\quad \tau\rightarrow\tau\,,\quad m\rightarrow \frac12\,.
	\label{eqn:q2higgsLimits2}
\end{align}
The value of the B-field is motivated from the analytic small resolution $\bar{J}$ of $J$, that contains 2-torsional 2-cycles and is conjecturally related to $J$ by an irrelevant deformation in the worldsheet theory.

To study the modular implications of each of these transitions we perform the modular bootstrap.
The general ansatz for the base degree one partition functions is
\begin{align}
	Z_B(\tau,m,\lambda)=\frac{\phi_{4,16}(\tau,m)}{\eta(\tau)^{36}\phi_{-2,1}(\tau,\lambda)}\,,
\end{align}
where the numerator $\phi_{4,16}(\tau,m)$ is an $\text{SL}(2,\mathbb{Z})$ weak Jacobi form of index $4$ and weight $16$.
The index is determined from the height pairing~\eqref{eqn:heightPairingEllQ2} as $-b_{11}/2$.
We use genus zero Gopakumar-Vafa invariants to fix the numerator and find
\begin{align}
	\begin{split}
		\phi_{4,16}(\tau,m)=&-\frac{1}{12^6}\left(3E_4(31E_4^3+113E_6^2)A^4+12E_6(115E_4^3+29E_6^2)A^3B\right.\\
		&+6E_4^2(125E_4^3+307E_6^2)A^2B^2+108E_4E_6(11E_4^3+5E_6^2)AB^3\\
		&\left.+(37E_4^6+355E_4^3E_6^2+40E_6^4)B^4\right)\,,
	\end{split}
\end{align}
with $A=\phi_{0,1}(\tau,m)$ and $B=\phi_{-2,1}(\tau,m)$.

Having found an expression for the base degree one topological string partition function in terms of modular and Jacobi forms we can now look at the effect of the
three limits in~\eqref{eqn:q2higgsLimits1} and~\eqref{eqn:q2higgsLimits2}.
The resulting partition functions are
\begin{align}
	\begin{split}
		Z^{(2)}_{1,d_B=1}(\tau,\lambda)=&Z_B(2\tau,\tau,\lambda)=\frac{1}{9}\frac{\Delta_4^2\left(44e_2^4+29e_2^2e_4-37e_4^2\right)}{\eta(2\tau)^{36}\phi_{-2,1}(2\tau,\lambda)}\,,\\
		Z^{(2)}_{0,d_B=1}(\tau,\lambda)=&Z_B(\tau,0,\lambda)=-\frac{1}{48}\frac{e_4\left(31e_4^3+113e_6^2\right)}{\eta(\tau)^{36}\phi_{-2,1}(\tau,\lambda)}\,,\\
		Z^{(2)}_{0,\text{nc.},d_B=1}(\tau,\lambda)=&Z_B(\tau,1/2,\lambda)=\frac{256}{9}\frac{{\Delta_4'}^2\left(736e_2^4-341e_2^2e_4+37e_4^2\right)}{\eta(\tau)^{36}\phi_{-2,1}(\tau,\lambda)}\,,
	\end{split}
	\label{eqn:ellQ2topRestr}
\end{align}
where the shorthands $e_2(\tau)=E_{2,2}(\tau)$, $e_4(\tau)=E_4(\tau)$ and $e_6(\tau)=E_6(\tau)$ are used for readability.
As expected, the first limit results in the partition function associated to the genus one fibrations with 2-sections that we study in the next section.
The second limit reproduces the result for the degree $18$ hypersurface $X^{(1)}_0$ in $\mathbb{P}_{11169}$, which is the generic elliptic fibration over $\mathbb{P}^2$ and was studied in~\cite{Huang:2015sta}.
Using the transformation of the $\Gamma_1(2)$ modular forms that are derived in Appendix~\ref{app:modforms}, it is easy to check that the Fricke involution
\begin{align}
	\tau\mapsto-\frac{1}{2\tau}\,,
\end{align}
transforms the partition functions $Z^{(2)}_{1,d_B=1}$ and $Z^{(2)}_{0,\text{nc.},d_B=1}$ into each other.

The Gopakumar-Vafa invariants at low degree and genus $g=0,1$ that correspond to the limits 1 and 2 in~\eqref{eqn:q2higgsLimits1},~\eqref{eqn:q2higgsLimits2} are listed in the Tables~\ref{tab:sec2lim1gv1} and~\ref{tab:sec2lim1gv2}.
Let us stress, that the ordinary Gopakumar-Vafa invariants that one would obtain from transition 3 do not contain any enumerative information.
The corresponding partition function of the non-commutative resolution $X^{(2)}_{0,\text{nc.} 1}$ has to be combined with that of $X^{(2)}_0$ in order to extract the Gopakumar-Vafa invariants with $\mathbb{Z}_2$ charges.
This will be done in Section~\ref{sec:q2jacsec}.
However, the leading terms of the instanton contributions to the genus zero free energy in this limit are
\begin{align}
	F_{0,\text{inst.}}\big|_{X^{(2)}_{0,\text{nc.} 1}}=28 q_1+3 q_2+\frac{191}{2} q_1^2-56 q_1 q_2-\frac{45 }{8}q_2^2+\mathcal{O}(q^3)\,.
	\label{eqn:f02secncresHiggs}
\end{align}
We later use this to normalize the higher degree expansion that we obtain from the Picard-Fuchs operators associated to the large volume limit of $X^{(2)}_{0,\text{nc.} 1}$.
\begin{table}[h!]
\begin{align}
\begin{array}{c|ccccc}
	n^{(0)}_{d1,d2}&d_2=0&1&2&3&4\\\hline
	d_1=0& 0 & -4 & -4 & -12 & -48 \\
	1&256 & 512 & 768 & 2560 & 12544 \\
	2&284 & 199696 & -62440 & -252032 & -1526508 \\
	3&256 & 6311936 & 15098880 & 33786368 & 177694720 \\
	4&284 & 104656856 & 11018163216 & -2740916912 & -15447666984 \\
\end{array}
\end{align}
	\caption{Genus zero Gopakumar-Vafa invariants that arise from the elliptic fibration with toric data~\eqref{eqn:f6overf1toricData} in the limit $\tau\rightarrow 2\tau$ and $m\rightarrow \tau$.
	This corresponds to $q_1\rightarrow 1$ with $q_1=\exp(2\pi i t^1)$ in the parametrization~\eqref{eqn:kaehlerEllQ2}.}
	\label{tab:sec2lim1gv1}
\end{table}
\begin{table}[h!]
\begin{align}
\begin{array}{c|ccccc}
	n^{(0)}_{d1,d2}&d_2=0&1&2&3&4\\\hline
	d_1=0& 0 & 3 & -6 & 27 & -192 \\
	1&540 & -1080 & 2700 & -17280 & 154440 \\
	2&540 & 143370 & -574560 & 5051970 & -57879900 \\
	3&540 & 204071184 & 74810520 & -913383000 & 13593850920 \\
	4&540 & 21772947555 & -49933059660 & 224108858700 & -2953943334360 \\
\end{array}
\end{align}
	\caption{Genus zero Gopakumar-Vafa invariants that arise from the elliptic fibration with toric data~\eqref{eqn:f6overf1toricData} in the limit $m\rightarrow 0$.
	This corresponds to $q_2\rightarrow q_1^{-1}$ with $q^1=\exp(2\pi i t^1)$ in the parametrization~\eqref{eqn:kaehlerEllQ2}.}
	\label{tab:sec2lim1gv2}
\end{table}

\subsection{The genus one fibration $X^{(2)}_1$ with $2$-section}
\label{sec:g12sec}
Let us now study the stringy K\"ahler moduli space of the genus one fibration $X^{(2)}_1$, which arises from the elliptic fibration that we studied above via transition one in~\eqref{eqn:q2higgsLimits1}.
As we already discussed, the latter physically amounts to a Higgs transition with the five-dimensional scalar fields in half-hypermultiplets of charge $(q_{KK},q_{6d})=(1,-2)$ becoming massless and getting a non-zero vacuum expectation value.

The toric data associated to the ambient space is
\begin{align}
\begin{blockarray}{crrrrrrl}
	&&&&&C_1&C_2\\
\begin{block}{c(rrrr|rr)l}
	  X&-1& 1& 0& 0& 1&-2&\leftarrow\text{two-section }E_0\\
	  Y&-1&-1& 0& 0& 1& 0&\leftarrow\text{two-section}\\
	  Z& 1& 0& 0& 0& 2& 1&\leftarrow\text{four-section}\\
	x_1& 0& 0& 1& 0& 0& 1&\leftarrow\text{vertical divisor }D_b\\
	x_2& 0& 0& 0& 1& 0& 1&\phantom{x}\hspace{1.0cm}\text{\ditto}\\
	x_3&-3& 2&-1&-1& 0& 1&\phantom{x}\hspace{1.0cm}\text{\ditto}\\
	   & 0& 0& 0& 0&-4&-2&\\
\end{block}
\end{blockarray}\,.
\label{eqn:f4overf1toricData}
\end{align}
Again, the points of the polytope admit two different fine regular star triangulations and the corresponding toric phases are connected by a flop that does not intersect the Calabi-Yau.
The relations in the last two columns in~\eqref{eqn:f4overf1toricData} thus correspond to the intersection of the two toric Mori cones.

We denote the generators of the K\"ahler cone by
\begin{align}
	J_1=[Y]\,,\quad J_2=[x_1]\,,
\end{align}
and use PALP~\cite{Kreuzer:2002uu} and SageMath~\cite{sagemath} to calculate the topological invariants
\begin{align}
	\begin{split}
		h^{1,1}=&2\,,\quad h^{2,1}=144\,,\quad \chi=-284\,,\\
		c_{111}=8\,,\quad &c_{112}=4\,,\quad c_{122}=2\,,\quad c_{222}=0\,,
	\end{split}
	\label{eqn:g1sec2top1}
\end{align}
as well as the intersections of the K\"ahler cone generators with the second Chern class
\begin{align}
	\vec{b}=c_2\cdot\vec{J}=(68,\,36)^\intercal\,.
	\label{eqn:g1sec2top2}
\end{align}

It is easy to perform the modular bootstrap and confirm that the base degree one topological string partition function indeed coincides with $Z_1^{(2)}(\tau,\lambda)$ in~\eqref{eqn:ellQ2topRestr}.
The details for an analogous example can for example be found in~\cite{Cota:2019cjx}.
At this point our interest concerns the stringy K\"ahler moduli space of this geometry.
In particular, we need the discriminant polynomial and the Picard-Fuchs operators that annihilate the periods on the mirror.

Using the classical results from~\cite{Hosono:1993qy} and the Mori cone generators in~\eqref{eqn:f4overf1toricData}, we can write down the fundamental period on the mirror of $X^{(2)}_1$ as
\begin{align}
	\varpi_0=\sum\limits_{k_1,k_2=0}^\infty\frac{\Gamma(1+4k_1+2k_2)}{\Gamma(1+k_1-2k_2)\Gamma(1+k_1)\Gamma(1+2k_1+k_2)\Gamma(1+k_2)^3}z_1^{k_1}z_2^{k_2}\,.
	\label{eqn:genus1sec2w0}
\end{align}
The large volume limit of $X^{(2)}_1$ corresponds to the origin $z_1=z_2=0$.
To determine the Picard-Fuchs system we make an ansatz and fix the coefficients by demanding that the operators annihilate~\eqref{eqn:genus1sec2w0}.
In this way we obtain the operators
\begin{align}
	\begin{split}
	\mathcal{D}_1=&\theta _1 \left(\theta _1-2 \theta _2\right)-4 z_1 \left(1+4 \theta _1+2 \theta _2\right) \left(3+4 \theta _1+2 \theta _2\right)\,,\\
	\mathcal{D}_2=&\theta _2^3-2 z_2 \left(-1+\theta _1-2 \theta _2\right) \left(\theta _1-2 \theta _2\right) \left(1+4 \theta _1+2 \theta _2\right)\,.
	\end{split}
	\label{eqn:2secpf}
\end{align}
The inverse mirror map can be calculated from the logarithmic periods and is given by
\begin{align}
	\begin{split}
	z_1(q_1,q_2)=&q_1-40 q_1^2+2 q_1 q_2 + 1324 q_1^3+3 q_2^2 q_1-416 q_2 q_1^2+\mathcal{O}(q^4)\,,\\
	z_2(q_2,q_2)=&q_2-56 q_1 q_2-4 q_2^2+1812 q_1^2 q_2+624 q_1 q_2^2+6 q_2^3+\mathcal{O}(q^4)\,,
	\end{split}
\end{align}
where $q_i=\exp(2\pi i t^i)$.
As discussed in~\cite{Cota:2019cjx}, a possible choice of basis for the normalized periods takes the form
\begin{align}
	\begin{split}
		\Pi^{(6)}=&\frac16c_{ijk}t^it^jt^k+\frac{1}{24}b_it^i+\chi\frac{\zeta(3)}{(2\pi i)^3}\,,\\
		\Pi^{(4)}_i=&-\frac12c_{ijk}t^jt^k-\frac12c_{iij}t^j-\frac16c_{iii}-\frac{1}{24}b_i\,,\\
		\Pi^{(2)}_i=&t^i\,,\quad \Pi^{(0)}=-1\,.
	\end{split}
	\label{eqn:periodbasis}
\end{align}
in terms of the topological invariants~\eqref{eqn:g1sec2top1},~\eqref{eqn:g1sec2top2}.

With the operators~\eqref{eqn:2secpf} we can also determine the Yukawa couplings
\begin{align}
	\begin{split}
		C_{111}=&\frac{8(1+68 z_1+768 z_1^2-16 z_2-960 z_1 z_2-6400 z_1^2 z_2)}{z_1^3\Delta}\,,\\
		C_{112}=&\frac{4(1-32 z_1-2048 z_1^2-16 z_2+640 z_1 z_2+25600 z_1^2 z_2)}{z_1^2z_2\Delta}\,,\\
		C_{122}=&\frac{2(1-128 z_1+4096 z_1^2-16 z_2+2240 z_1 z_2-102400 z_1^2 z_2)}{z_1z_2^2\Delta}\,,\\
		C_{222}=&\frac{-16z_2(1+80z_1)(1-320z_1)}{z_2^3\Delta}\,,
	\end{split}
\end{align}
which fixes the discriminant polynomial $\Delta$ to be
\begin{align}
	\begin{split}
	\Delta=&1-192 z_1+12288 z_1^2-262144 z_1^3-16 z_2\\
	&+3200 z_1 z_2-1024000 z_1^2 z_2+12800000 z_1^2 z_2^2\\
		=&(1-64z_1)^3+\mathcal{O}(z_2)\,.
	\end{split}
	\label{eqn:2secdisc}
\end{align}

Note that the associated singular locus $\{\Delta=0\}$ in the complex structure moduli space of the mirror intersects the large base complex structure limit $z_2\rightarrow 0$ in a triple tangency
As discussed in~\cite{Cota:2019cjx}, such a tangency is a generic feature of the mirrors of elliptic and genus one fibered Calabi-Yau manifolds and leads to a relation between the fiberwise and generic conifold monodromy.
The fiberwise conifold monodromy together with the large volume monodromy associated to $z_1=0$ generates a $\Gamma_1(2)$ action on the complexified volumes of 2-branes in the genus one fibrations which, together with the holomorphic anomaly equations, leads to the $\Gamma_1(2)$ modular structure of the topological string partition function.

By modifying the involution~\eqref{eqn:x18inv} in the moduli space of the degree $18$ hypersurface $X^{(1)}_0$ in $\mathbb{P}_{11169}$, we find that changing coordinates to $v_1,v_2$ with
\begin{align}
	z_1=\frac{1-64v_1}{64}\,,\quad z_2= -\frac{1}{8}v_2\left(\frac{64v_1}{1-64v_1}\right)^3\,,
	\label{eqn:2sectrafo}
\end{align}
blows up the tangency and maps the large volume limit of the genus one fibration to a second point of maximally unipotent monodromy.
As argued in Section~\ref{sec:higgs2}, the latter corresponds to the large volume limit of the non-commutative resolution of the Jacobian fibration $X^{(2)}_{0,\text{nc.} 1}$.
Note that in this case the involution is not induced by a toric automorphism because the large volume limits that are transformed into each other are not equivalent.

\subsection{The Jacobian fibration with discrete torsion $X^{(2)}_{0,\text{n.c.}1}$}
\label{sec:q2jacsec}
We are now going to study the large volume point that is associated to the non-commutative resolution of the Jacobian fibration $X^{(2)}_{0,\text{nc.} 1}$.
To our knowledge, mirror symmetry has not been applied in this setting before.
We will thus provide more details on the individual steps and confirm that our usual geometric intuitions about the periods still apply.

Applying the transformation~\eqref{eqn:2sectrafo} to the Picard-Fuchs operators~\eqref{eqn:2secpf} produces a Picard-Fuchs system that is generated by the new operators
\begin{align}
	\begin{split}
	\mathcal{D}_1=&\theta _1 \left(\theta _1-3 \theta _2\right)+256 v_1^2 \left(1+4 \theta _1+2 \theta _2\right) \left(3+4 \theta _1+2 \theta _2\right)\\
		&-4 v_1 \left(3+16 \theta _1+32 \theta _1^2+8 \theta _2-32 \theta _1 \theta _2-44 \theta _2^2\right)\,,\\
	\mathcal{D}_2=&3 \theta _2^3-576 v_1 \theta _2^3+36864 v_1^2 \theta _2^3-786432 v_1^3 \theta _2^3\\
		&-3 v_2 \left(\theta _1-3 \theta _2\right) \left(2+3 \theta _1+\theta _1^2+9 \theta _2+3 \theta _1 \theta _2+9 \theta _2^2\right)\\
		&+8192 v_1^3 v_2 \left(1+4 \theta _1+2 \theta _2\right) \left(270+375 \theta _1+76 \theta _1^2+408 \theta _2\right.\\
		&\left.+170 \theta _1 \theta _2+216 \theta _2^2\right)-128 v_1^2 v_2 \left(339+1991 \theta _1+2816 \theta _1^2+704 \theta _1^3\right.\\
		&\left.+814 \theta _2+1720 \theta _1 \theta _2+832 \theta _1^2 \theta _2-3484 \theta _2^2+580 \theta _1 \theta _2^2-3000 \theta _2^3\right)\\
		&+4 v_1 v_2 \left(54+468 \theta _1+764 \theta _1^2+248 \theta _1^3-315 \theta _2-2700 \theta _2^2-2700 \theta
   _2^3\right)\,.
	\end{split}
\end{align}
Solutions that are annihilated by the Picard-Fuchs system can easily be found by making an ansatz and we choose the basis
\begin{align}
	\begin{split}
		\varpi_0=&1+12 v_1+420 v_1^2+18480 v_1^3+900900 v_1^4-10080 v_1^3 v_2+\mathcal{O}(v^5)\,,\\
		\varpi_1=&\varpi_0\cdot\log(v_1)+40v_1+2v_2+1556v_1^2+132v_1v_2-15v_2^2+\mathcal{O}(v^3)\,,\\
		\varpi_2=&\varpi_0\cdot\log(v_2)+68v_1-6v_2+3298v_1^2-396v_1v_2+45v_2^2+\mathcal{O}(v^3)\,,\\
		\varpi_3=&\varpi_0\cdot\log(v_2)^2+\mathcal{O}(v)\,,\\
		\varpi_4=&\varpi_0\cdot\left[\log(v_1)^2+\frac23\log(v_1)\log(v_2)\right]+\mathcal{O}(v)\,,\\
		\varpi_5=&\varpi_0\cdot\left[\log(v_1)^3+\log(v_1)^2\log(v_2)+\frac13\log(v_1)\log(v_2)^2\right]+\mathcal{O}(v)\,.
	\end{split}
	\label{eqn:triplog}
\end{align}
From this one can easily check that $v_1=v_2=0$ is indeed a point of maximally unipotent monodromy.
Note that not all of the coefficients in the fundamental period $\varpi_0$ are positive.
What might appear as an error in the choice of coordinates is indeed necessary to reproduce the prepotential~\eqref{eqn:f02secncresHiggs} that we obtained via Higgs transition.

Using the flat coordinates $t^1=\varpi_1/\varpi_0$ and $t^2=\varpi_2/\varpi_0$ as well as their exponentials $q_k=\exp(2\pi i t^k)$ we can calculate and invert the mirror map to get
\begin{align}
	\begin{split}
		z_1(q_1,q_2)=&q_1-40 q_1^2-2 q_1 q_2+1324 q_1^3+188 q_1^2 q_2+5 q_1 q_2^2+\mathcal{O}(q^4)\,,\\
		z_2(q_1,q_2)=&q_2-68 q_1 q_2+6 q_2^2+2550 q_1^2 q_2-356 q_1 q_2^2+9 q_2^3+\mathcal{O}(q^4)\,,
	\end{split}
	\label{eqn:q2mirror}
\end{align}
and the coefficients turn out to be integral.
By inserting this into the triple logarithmic period $\varpi_5$ in~\eqref{eqn:triplog} and matching the normalization with~\eqref{eqn:f02secncresHiggs} we find the prepotential
\begin{align}
	\begin{split}
	F_0=&\frac{1}{3!}c_{ijk}t^it^jt^k+p_2(t^1,t^2)+28 q_1+3 q_2+\frac{191}{2} q_1^2-56 q_1 q_2-\frac{45}{8} q_2^2\\
		&+\frac{784}{27} q_1^3+10 q_1^2 q_2+140 q_1 q_2^2+\frac{244}{9} q_2^3+\mathcal{O}(q^4)\,,
\end{split}
	\label{eqn:f0d2jac}
\end{align}
with the triple intersection numbers being
\begin{align}
	c_{111}=9\,,\quad c_{112}=3\,,\quad c_{122}=1\,,\quad c_{222}=0\,,
	\label{eqn:jac2secTriple}
\end{align}
and $p_2(t^1,t^2)$ some quadratic polynomial in $t^1,t^2$ that we do not need to fix.
The intersections $c_{122}=1$ and $c_{222}=0$ confirm that we are indeed studying an elliptic fibration.
In particular, the intersection numbers match those of the unresolved Jacobian fibration $X^{(2)}_0$ which is complex structure deformation equivalent to $X^{(1)}_0$.
In the following we denote the corresponding divisors by $J_i,\,i\in\{1,2,3\}$ such that
\begin{align}
	c_{ijk}= J_i\cdot J_j\cdot J_k\,.
\end{align}

We can now calculate the GV invariants with $U(1)\times\mathbb{Z}_2$ charges using the expansion~\eqref{eqn:torsiongvexs}.
To this end we are using the interpretation in terms of the analytic small resolution $\bar{J}$ of the Jacobian $J=X^{(2)}_0$ with a B-field along 2-torsional 2-cycles.
Recall that to extract the invariants we need the free energies of the topological A-models on the Jacobian fibration with both choices for the torsion part of the B-field homomorphism
\begin{align}
	b:\,\mathbb{Z}_2\rightarrow\mathbb{C}^\times\,.
\end{align}
The prepotential~\eqref{eqn:f0d2jac} corresponds to the choice $b(-1)=-1$ while the topological string with $b(-1)=1$ is equivalent to that on $X^{(1)}_0$ with the results provided in Appendix~\ref{app:x18}.
Together with~\eqref{eqn:torsiongvexs} this leads to two expansions and we can extract the genus zero invariants listed in Table~\ref{tab:d2jacz2q0} and~\ref{tab:d2jacz2q1}.
\begin{table}[h!]
	\centering
	\begin{align*}
\begin{array}{c|ccccc}
	n^{(0)}_{(d_1,d_2),0}&d_2=0&1&2&3&4\\\hline
	d_1=0&0 & 3 & -6 & 27 & -192 \\
	1&284 & -568 & 1420 & -9088 & 81224 \\
	2&284 & 71690 & -287328 & 2528834 & -28983900 \\
	3&284 & 102020368 & 37395096 & -456631384 & 6796521000 \\
	4&284 & 10886398051 & -24966323788 & 112053235020 & -1476958051288 \\
	5&284 & 538258103416 & 3886249778448 & -21356075563104 & 301888990988900 \\
\end{array}
	\end{align*}
	\caption{Genus $0$ torsion refined Gopakumar-Vafa invariants with $\mathbb{Z}_2$ charge $0$.}
	\label{tab:d2jacz2q0}
\end{table}
\begin{table}[h!]
	\centering
	\begin{align*}
\begin{array}{c|ccccc}
	n^{(0)}_{(d_1,d_2),1}&d_2=0&1&2&3&4\\\hline
	d_1=0&0 & 0 & 0 & 0 & 0 \\
	1&256 & -512 & 1280 & -8192 & 73216 \\
	2&256 & 71680 & -287232 & 2523136 & -28896000 \\
	3&256 & 102050816 & 37415424 & -456751616 & 6797329920 \\
	4&256 & 10886549504 & -24966735872 & 112055623680 & -1476985283072 \\
	5&256 & 538260148736 & 3886245092352 & -21356060043264 & 301889011932928 \\
\end{array}
	\end{align*}
	\caption{Genus $0$ torsion refined Gopakumar-Vafa invariants with $\mathbb{Z}_2$ charge $1$.}
	\label{tab:d2jacz2q1}
\end{table}
Due to the structure of the expansion~\eqref{eqn:torsiongvexs} the sum of the invariants with different discrete charges $n^{(0)}_{(d_1,d_2),0}+n^{(0)}_{(d_1,d_2),1}$ reproduces the Gopakumar-Vafa invariants on the generic elliptic fibration $X^{(1)}_0$ over $\mathbb{P}^2$.
Note that the invariants $n^{(0)}_{(d_1,0),0}$ and $n^{(0)}_{(d_1,0),1}$ respectively correspond to the multiplicity of hypermultiplets with $\mathbb{Z}_2$ charge $0$ and $1$ that arise in the associated F-theory compactification.
These multiplicities match those that we obtained from the genus one fibration with 2-section in Section~\ref{sec:g12sec}.
This provides a highly non-trivial check of our definition of the torsion refined Gopakumar-Vafa invariants as traces of multiplicities of five-dimensional BPS states with discrete charges.

We have already obtained the modular expression for the base degree one topological string partition function $Z^{(2)}_{0,\text{nc.},d_B=1}(\tau,\lambda)$ in~\eqref{eqn:ellQ2topRestr} by using the Higgs transition in~\ref{sec:ell2sec}.
Together with the corresponding expression $Z^{(2)}_{0,d_B=1}(\tau,\lambda)$ for the general elliptic fibration we can again extract the corresponding torsion refined GV invariants in Table~\ref{tab:bd1jacz2q0} and~\ref{tab:bd1jacz2q1}.
\begin{table}[h!]
	\begin{align*}
	\begin{array}{c|cccccc}
		n^{(g)}_{(d_1,1),0}&g=0&1&2&3&4&5\\\hline
		d_1=0&	 3 & 0 & 0 & 0 & 0 & 0 \\
		1& -568 & -6 & 0 & 0 & 0 & 0 \\
		2& 71690 & 1118 & 9 & 0 & 0 & 0 \\
		3& 102020368 & -139996 & -1656 & -12 & 0 & 0 \\
		4& 10886398051 & -204466374 & 206117 & 2182 & 15 & 0 \\
		5& 538258103416 & -22385483914 & 307182872 & -270080 & -2696 & -18 \\
	\end{array}
	\end{align*}
	\caption{Torsion refined Gopakumar-Vafa invariants at base degree one with $\mathbb{Z}_2$ charge $0$.}
	\label{tab:bd1jacz2q0}
\end{table}
\begin{table}[h!]
	\begin{align*}
	\begin{array}{c|cccccc}
		n^{(g)}_{(d_1,1),1}&g=0&1&2&3&4&5\\\hline
		d_1=0&	 0 & 0 & 0 & 0 & 0 & 0 \\
		1& -512 & 0 & 0 & 0 & 0 & 0 \\
		2& 71680 & 1024 & 0 & 0 & 0 & 0 \\
		3& 102050816 & -140288 & -1536 & 0 & 0 & 0 \\
		4& 10886549504 & -204527616 & 206848 & 2048 & 0 & 0 \\
		5& 538260148736 & -22385970176 & 307276288 & -271360 & -2560 & 0 \\
	\end{array}
	\end{align*}
	\caption{Torsion refined Gopakumar-Vafa invariants at base degree one with $\mathbb{Z}_2$ charge $1$.}
	\label{tab:bd1jacz2q1}
\end{table}

To calculate the genus one free energy we use the general ansatz
\begin{align}
	\begin{split}
		F_1=&-\frac12\left(3+h^{1,1}-\frac{\chi}{12}\right)K-\frac12\log\,\text{det}\,G\\
		&-\frac{1}{24}\sum\limits_{i=1}^{h^{1,1}}(b_i+12)\log\,z_i-\frac{1}{12}\sum\limits_ic_i\log\,\Delta_i\,,
	\end{split}
	\label{eqn:f1ansatz}
\end{align}
where $K$ is the K\"ahler potential of the moduli space metric which in the holomorphic limit takes the form $K=-\log(\varpi_1)$ and the determinant of the Weil-Petersson metric becomes $\det\,G=\det(\partial_{v_i}t^j)$.
The coefficients $b_i,\,i=1,2$ are geometrically defined as
\begin{align}
	b_i= c_2\cdot J_i\,,
\end{align}
and the coefficients $c_i$ encode the difference of the number of vector- and hypermultiplets that become massless at the corresponding conifold locus~\cite{Vafa:1995ta}.
However, since we are dealing with a non-commutative resolution of a singular Calabi-Yau, the precise definition of $b_i$ as well as the Euler characteristic $\chi$ might be more subtle.
For the purpose of this work we remain agnostic about this question and use the results from the modular bootstrap to fix the ambiguities in~\eqref{eqn:f1ansatz}.

Using the prepotential~\eqref{eqn:f0d2jac} we can calculate the Yukawa couplings
\begin{align}
	\begin{split}
	C_{111}=&\frac{1}{v_1^3\Delta_1\Delta_2^2}\left(9-1520 v_1+243 v_2+73728 v_1^2-26784 v_1 v_2-393216 v_1^3\right.\\
		&+486144 v_1^2 v_2-16777216 v_1^4+13115392 v_1^3 v_2-805306368 v_1^5\\
		&\left.+83886080 v_1^4 v_2+838860800 v_1^5 v_2\right)\,,\\
		C_{112}=&\frac{1}{v_1^2v_2\Delta_1\Delta_2}\left(3-512 v_1+81 v_2+24576 v_1^2-9072 v_1 v_2+184576 v_1^2 v_2\right.\\
		&\left.-16777216 v_1^4+3440640 v_1^3 v_2+26214400 v_1^4 v_2\right)\,,\\
		C_{122}=&\frac{1}{v_1v_2^2\Delta_1}\left(1-192 v_1+27 v_2+12288 v_1^2-3072 v_1 v_2\right.\\
		&\left.-262144 v_1^3+69120 v_1^2 v_2+819200 v_1^3 v_2\right)\,,\\
		C_{222}=&\frac{1}{v_2^2\Delta_1}(1-64v_1)(1-80v_1)(9-320v_1)\,,
	\end{split}
\end{align}
which provides is with two components of the discriminant
\begin{align}
	\begin{split}
	\Delta_1=&1-256 v_1+27 v_2+24576 v_1^2-5328 v_1 v_2-1048576 v_1^3+358400 v_1^2 v_2\\
		&+16777216 v_1^4-8192000 v_1^3 v_2+12800000 v_1^3 v_2^2\,,\\
	\Delta_2=&1-64v_1\,.
	\end{split}
\end{align}
The torsion refined GV invariants in Table~\ref{tab:bd1jacz2q0} and~\ref{tab:bd1jacz2q1} that we extracted from the base degree one topological string partition functions now fix the coefficients
\begin{align}
	h^{1,1}=2\,,\quad \chi=-284\,,\quad c_1=1\,,\quad c_2=-18\,,\quad b_1=102\,,\quad b_2=36\,.
	\label{eqn:f1q2amb}
\end{align}
It is reassuring that the ``Euler characteristic'' matches that of the genus one fibration in Section~\ref{sec:g12sec} which is twisted derived equivalent to the non-commutative resolution $X^{(2)}_{0,\text{nc.} 1}$ of the Jacobian.
Moreover, the intersection with the second Chern class $b_2=36$ is generic for genus one fibrations over $\mathbb{P}^2$.

Combining the ansatz~\eqref{eqn:f1ansatz} with the values~\eqref{eqn:f1q2amb}, the mirror map~\eqref{eqn:q2mirror} and the corresponding expressions for the generic elliptic fibration $X^{(1)}_0$ over $\mathbb{P}^2$ from Appendix~\ref{app:x18}, we can extract the torsion refined Gopakumar-Vafa invariants~\eqref{eqn:torsiongvexs}.
The results are listed in Table~\ref{eqn:g1jacz2q0} and~\ref{eqn:g1jacz2q1}.
\begin{table}[h!]
	\begin{align*}
\begin{array}{c|ccccc}
	n^{(1)}_{(d_1,d_2),0}&d_2=0&1&2&3&4\\\hline
	d_1=0&0 & 0 & 0 & -10 & 231 \\
	1&3 & -6 & 15 & 2460 & -80934 \\
	2&3 & 1118 & -4478 & -538114 & 24463911 \\
	3&3 & -139996 & 1063446 & 76106186 & -4878920934 \\
	4&3 & -204466374 & 260890596 & -8351959904 & 795522313810 \\
	5&3 & -22385483914 & 561105104884 & -1664224356396 & -93207724436675 \\
\end{array}
	\end{align*}
	\caption{Genus $1$ torsion refined Gopakumar-Vafa invariants with $\mathbb{Z}_2$ charge $0$.}
	\label{eqn:g1jacz2q0}
\end{table}
\begin{table}[h!]
	\begin{align*}
\begin{array}{c|ccccc}
	n^{(1)}_{(d_1,d_2),1}&d_2=0&1&2&3&4\\\hline
	d_1=0&0 & 0 & 0 & 0 & 0 \\
	1&0 & 0 & 0 & 2304 & -73728 \\
	2&0 & 1024 & -4096 & -541184 & 24443904 \\
	3&0 & -140288 & 1062912 & 76172800 & -4880498688 \\
	4&0 & -204527616 & 260966400 & -8352126976 & 795540107264 \\
	5&0 & -22385970176 & 561107998208 & -1664243043072 & -93207516623872 \\
\end{array}
	\end{align*}
	\caption{Genus $1$ torsion refined Gopakumar-Vafa invariants with $\mathbb{Z}_2$ charge $1$.}
	\label{eqn:g1jacz2q1}
\end{table}
We observe that all of the genus $1$ invariants with $\mathbb{Z}_2$ charge $0$, base degree $0$ and non-vanishing fiber degree are equal to minus one times the Euler characteristic of the base of the fibration.
As discussed in~\cite{Paul-KonstantinOehlmann:2019jgr}, this is a generic feature of the regular GV invariants on elliptically fibered Calabi-Yau threefolds.
Together with the vanishing of the corresponding invariants for non-zero discrete charges this also provides a strong constraint to fix the ambiguities in the ansatz for the genus one free energy and thus obtain the topological invariants associated to a non-commutative resolution.

Finally, let us check that the involution~\eqref{eqn:2sectrafo} acts on the modular parameter $\tau$ by a Fricke involution
\begin{align}
	\tau\rightarrow -\frac{1}{2\tau}\,.
\end{align}
To calculate the transfer matrix from the large volume limit associated with $X^{(2)}_1$ to the large volume limit of $X^{(2)}_{0,\text{n.c.} 1}$, we need to fix a basis of the periods.
On a smooth Calabi-Yau threefold, the generic basis~\eqref{eqn:periodbasis} can be determined by using the Gamma class formula~\cite{Iritani_2009} to calculate the asymptotic behaviour of the central charges of a basis of topological B-branes~\cite{Cota:2019cjx}.
The divisors, curves and points that support the 4-, 2-, and 0-branes can be chosen such that they do not intersect the singularities.
We therefore use the corresponding expressions for the normalized periods from~\eqref{eqn:periodbasis} with the topological invariants~\eqref{eqn:jac2secTriple} and~\eqref{eqn:f1q2amb}. 
This implies, that the classical terms of these quantum periods are identical to those of the generic elliptic fibration $X^{(1)}_0$.
However, the 6-brane necessarily intersects the $\mathbb{Q}$-factorial terminal singularities and potentially receives corrections.
Special geometry fixes all contributions except for the constant term, that multiplies $\zeta(3)$ and on a smooth Calabi-Yau is proportional to the Euler characteristic~\cite{Hosono:1994av}.
It turns out, that there is a unique choice for this term that leads to a rational transfer matrix.
The corresponding basis is $\vec{\Pi}_{\text{n.c.}}=\left(\Pi^{(6)}_{\text{n.c.}},\Pi^{(4)}_{\text{n.c.},i=1,2},\Pi^{(2)}_{\text{n.c.},i=1,2},\Pi^{(0)}_{\text{n.c.}}\right)$, where
\begin{align}
	\begin{split}
		\Pi^{(6)}_{\text{n.c.}}=&\frac16c_{ijk}t^it^jt^k+\frac{1}{24}b_it^i-92\frac{\zeta(3)}{(2\pi i)^3}\,,\\
		\Pi^{(4)}_{\text{n.c.},i}=&-\frac12c_{ijk}t^jt^k-\frac12c_{iij}t^j-\frac16c_{iii}-\frac{1}{24}b_i\,,\\
		\Pi^{(2)}_{\text{n.c.},i}=&t^i\,,\quad \Pi^{(0)}_{\text{n.c.}}=-1\,.
	\end{split}
	\label{eqn:sec2jacbasis}
\end{align}
and with $c_{ijk},\,b_i$ given in~\eqref{eqn:jac2secTriple} and~\eqref{eqn:f1q2amb}.

By performing a numerical analytic continuation of the periods~\eqref{eqn:triplog}, we obtain the transfer matrix
\begin{align}
	T=\frac{i}{\sqrt{2}}\left(\begin{array}{cccccc}
 0 & -\frac{1}{2} & 2 & -1 & -\frac{1}{2} & -\frac{5}{2} \\
 1 & 0 & -2 & 8 & -3 & 6 \\
 0 & 0 & 0 & 2 & -1 & 1 \\
 0 & 0 & 0 & 0 & 0 & -1 \\
 0 & 0 & 1 & 0 & 0 & 0 \\
 0 & 0 & 0 & 2 & 0 & 0 \\
	\end{array}\right)\,,
	\label{eqn:fullTransferN2}
\end{align}
that relates the period basis~\eqref{eqn:periodbasis} of the smooth genus one fibration $X^{(2)}_1$ to the periods~\eqref{eqn:sec2jacbasis} of the non-commutative resolution $X^{(2)}_{0,\text{n.c.} 1}$ via
\begin{align}
	\vec{\Pi}_{\text{n.c.}}=T\cdot \vec{\Pi}\,.
\end{align}
This implies the relation
\begin{align}
	\tau_{\text{n.c.}}=\frac{\Pi^{(2)}_{\text{n.c.},1}}{\Pi^{(0)}_{\text{n.c.}}}=-\frac{\Pi^{(0)}}{2\Pi^{(2)}_1}=-\frac{1}{2\tau}\,.
\end{align}
Note that as expected, the transfer matrix~\eqref{eqn:transferN2} that we calculated in the moduli space of the generic fiber embeds into the corresponding transfer matrix~\eqref{eqn:fullTransferN2} in the stringy K\"ahler moduli space of the fibration.

\section{Example 2: Genus one fibrations with $4$-sections}
\label{sec:ex4sec}
As our second example, we will study the large volume limits in the stringy K\"ahler moduli space of a particular genus one fibered Calabi-Yau threefold $X^{(4)}_1$ with $4$-sections over $\mathbb{P}^2$.
This turns out to be closely connected to our previous example.

Our starting point is again an elliptic fibration $X^{(4)}_{\text{el.}}$ with two independent sections.
The F-theory effective theory has a $U(1)$ gauge symmetry and charge $4$ hypermultiplets that allow for a Higgs transition $U(1)\rightarrow \mathbb{Z}_4$.
In the five-dimensional M-theory vacuum, the inequivalent values $q_{KK}\in\{0,1,2\}$ for the Kaluza-Klein charge of the Higgs field and the corresponding Wilson lines respectively lead to a gauge group $U(1)_{KK}\times \mathbb{Z}_4$, $U(1)'$ and $U(1)''\times \mathbb{Z}_2$.
They arise from compactifications on the singular Jacobian fibration $X^{(4)}_0$, the smooth genus one fibration with 4-sections $X^{(4)}_1$ itself and the singular genus one fibration with 2-sections $X^{(4)}_2$.
An overview of the six- and five-dimensional gauge groups and matter spectra, as well as the relations via Higgs transitions, is provided in Figure~\ref{fig:z4fieldtheories}.
\begin{figure}[p!]
	\centering
\begin{tikzpicture}[remember picture,node distance=4mm, >=latex',
bblock/.style = {draw, rectangle, rounded corners=0.5em,align=center,minimum height=2cm,minimum width=4.3cm,fill=blue!30},
gblock/.style = {draw, rectangle, rounded corners=0.5em,align=center,minimum height=2cm,minimum width=4.3cm,fill=green!30},
                        ]            
	\node [bblock,align=center] at ( 0, 12) {
		\textbf{F-theory on $X^{(4)}_{\text{el.}}$}\\[.5em]
		$G=U(1)_{6d}$\\[.5em]
		\textbf{Hypermultiplets:}\\[.2em]
		$n_0=64,\,n_{\pm1}=104,\,n_{\pm2}=76,$\\
		$n_{\pm3}=24,\,n_{\pm 4}=6$
		};
	\node [bblock,align=center] at ( 10, 12) {
		\textbf{F-theory on $X^{(4)}_{i=0,\ldots,2}$}\\[.5em]
		$G=\mathbb{Z}_4$\\[.2em]
		$q_{\mathbb{Z}_4}=q_{U(1)_{6d}}\text{ mod }4$\\[.5em]
		\textbf{Hypermultiplets:}\\[.2em]
		$n_0=69,\,n_{\pm1}=128,\,n_{\pm2}=76$
		};
	\node [gblock,align=center] at ( 0, 2) {
		\textbf{M-theory on $X^{(4)}_{\text{el.}}$}\\[.5em]
		$G=U(1)_{KK}\times U(1)_{6d}$\\[.5em]
		\textbf{$\frac12$-Hypermultiplets:}\\[.2em]
		$n_{0,0}=128,$\\
		$n_{0,1}=n_{0,-1}=104,$\\
		$n_{0,2}=n_{0,-2}=76,$\\
		$n_{0,3}=n_{0,-3}=24,$\\
		$n_{0,4}=n_{0,-4}=6$\\[.5em]
		w/ tower $\pm(k,0),\,k\in\mathbb{N}$
		};
	\node [gblock,align=center] at ( 10, 7) {
		\textbf{M-theory on $X^{(4)}_{0}$}\\[.5em]
		$G=U(1)_{KK}\times \mathbb{Z}_4$\\[.2em]
		$q_{\mathbb{Z}_4}=q_{U(1)_{6d}}\text{ mod }4$\\[.5em]
		\textbf{$\frac12$-Hypermultiplets:}\\[.2em]
		$n_{0,0}=138,$\\$n_{0,1}=n_{0,-1}=128,$\\$n_{0,2}=152$\\[.5em]
		w/ tower $\pm(k,0),\,k\in\mathbb{N}$
		};
	\node [gblock,align=center] at ( 10, 1.8) {
		\textbf{M-theory on $X^{(4)}_{1}$}\\[.5em]
		$G=U(1)'$\\[.2em]
		$q_{U(1)'}=4q_{U(1)_{KK}}+q_{U(1)_{6d}}$\\[.5em]
		\textbf{$\frac12$-Hypermultiplets:}\\[.2em]
		$n_{0}=138,$\\$n_{1}=n_{-1}=128,$\\$n_{2}=152$\\[.5em]
		w/ tower $\pm(4k),\,k\in\mathbb{N}$
		};
	\node [gblock,align=center] at ( 10, -3.5) {\textbf{M-theory on $X^{(4)}_{2}$}\\[.5em]
		$G=U(1)''\times\mathbb{Z}_2$\\[.2em]
		$q_{U(1)''}=2q_{U(1)_{KK}}+q_{U(1)_{6d}}$\\[.2em]
		$q_{\mathbb{Z}_2}=q_{U(1)_{KK}}\text{ mod }2$\\[.5em]
		\textbf{$\frac12$-Hypermultiplets:}\\[.2em]
		$n_{0,0}=138,\,n_{1,0}=128,$\\
		$n_{0,1}=152,\,n_{1,1}=128$\\[.5em]
		w/ tower $\pm(2,1),\,k\in\mathbb{N}$
		};
		\draw [->,thick] (3,12) to (7,12);
		\node[align=center] at (5,12.5) {Higgs. w/ $q_{U(1)_{6d}}=4$};
		\draw [-,thick] (0,5) to [in=180,out=90](3,7);
		\draw [->,thick] (3,7) to (7.5,7);
		\node[align=center] at (5,7.8) {Higgsing w/\\$(q_{U(1)_{KK}},q_{U(1)_{6d}})=(0,4)$};
		\node[align=center] at (5,7.8) {Higgsing w/\\$(q_{U(1)_{KK}},q_{U(1)_{6d}})=(0,4)$};
		\draw [->,thick] (2.5,1.8) to (7.4,1.8);
		\node[align=center] at (5,2.6) {Higgsing w/\\$(q_{U(1)_{KK}},q_{U(1)_{6d}})=(-1,4)$};
		\draw [-,thick] (0,-1) to [in=180,out=-90](3,-3);
		\draw [->,thick] (3,-3) to (7.4,-3);
		\node[align=center] at (5,-2.2) {Higgsing w/\\$(q_{U(1)_{KK}},q_{U(1)_{6d}})=(-2,4)$};
\end{tikzpicture}
	\caption{The six- and five-dimensional gauge groups and matter spectra of F- and M-theory compactifications on $X^{(4)}_{\text{el.}}$ and $X^{(4)}_{i=0,\ldots,2}$, as well as the relations via Higgs transitions.}
	\label{fig:z4fieldtheories}
\end{figure}

In Type IIA compactifications, the Jacobian fibration allows for the choice of a flat torsional B-field $b:\,\mathbb{Z}_4\rightarrow\mathbb{C}^\times$ at the $\mathbb{Q}$-factorial terminal singularity.
The B-field is determined by the image of the generator $b(1)=\exp(k\cdot 2\pi i/4),\,k=0,\ldots,3$.
This leads to three non-commutative resolutions $X^{(4)}_{0,\text{n.c.}k},\,k=1,\ldots,3$ of which $X^{(4)}_{0,\text{n.c.}1}$ and $X^{(4)}_{0,\text{n.c.}3}$ are equivalent.
The singular fibration itself corresponds to a vanishing B-field, i.e. $X^{(4)}_0=X^{(4)}_{0,\text{n.c.}0}$.
The genus one fibration $X^{(4)}_2$ with $2$-sections also contains $\mathbb{Q}$-factorial terminal singularities and admits a flat torsional B-field $b:\,\mathbb{Z}_2\rightarrow\mathbb{C}^\times$.
We denote the non-commutative resolution with a non-zero B-field by $X^{(4)}_{0,\text{n.c.}1}$ while $X^{(4)}_2=X^{(4)}_{2,\text{n.c.}0}$.

\begin{figure}[h!]
	\begin{tikzpicture}[remember picture,node distance=4mm]
		\node[align=center] at (1,3) {$X^{(4)}_{2,\text{n.c.}1}$};
		\node[align=center] at (5,3) {$X^{(4)}_{0,\text{n.c.}1}$};
		\node[align=center] at (9,3) {$X^{(4)}_{1}$};
		\node[align=center] at (3,3.3) {\tiny str. K\"ahler def.};
		\node[align=center] at (7,3.3) {\tiny str. K\"ahler def.};
		\draw [<->] (2,3) to (4,3);
		\draw [<->] (6,3) to (8,3);
		\node[align=center] at (5,5) {$X^{(4)}_{0,\text{n.c.}0}$};
		\node[align=center] at (9,5) {$X^{(1)}_{0}$};
		\node[align=center] at (7,5.3) {\tiny cplx. def.};
		\draw [<->,dashed] (6,5) to (8,5);
		\node[align=center] at (1,1) {$X^{(4)}_{2,\text{n.c.}0}$};
		\node[align=center] at (5,1) {$X^{(4)}_{0,\text{n.c.}2}$};
		\node[align=center] at (1,-1) {$X^{(2)}_{1}$};
		\node[align=center] at (5,-1) {$X^{(2)}_{0,\text{n.c.}1}$};
		\draw [<->,dashed] (1,.5) to (1,-.5);
		\draw [<->,dashed] (5,.5) to (5,-.5);
		\node[align=center,rotate=90] at (.7,0) {\tiny cplx. def.};
		\node[align=center,rotate=90] at (4.7,0) {\tiny cplx. def.};
		\draw [<->] (2,-1) to (4,-1);
		\node[align=center] at (3,-.8) {\tiny str. K\"ahler def.};
		\draw [<->,dotted] (1,1.5) to (1,2.5);
		\node[align=center,rotate=90] at (.7,2) {\tiny B-field};
		\draw [<->,dotted] (5,1.5) to (5,2.5);
		\node[align=center,rotate=90] at (4.7,2) {\tiny B-field};
		\draw [<->,dotted] (5,3.5) to (5,4.5);
		\node[align=center,rotate=90] at (4.7,4) {\tiny B-field};
	\end{tikzpicture}
	\centering
	\caption{The inequivalent geometries $X^{(4)}_i,\,i=0,\ldots,2$ from the $\mathbb{Z}_4$ Tate-Shafarevich group of the smooth genus one fibered Calabi-Yau $X^{(4)}_1$, as well as their non-commutative resolutions and smooth deformations.
	Deformations of the stringy K\"ahler structure, the complex structure or a change of a flat torsional B-field are respectively indicated by solid, dashed and dotted arrows.}
	\label{fig:ex2geos}
\end{figure}
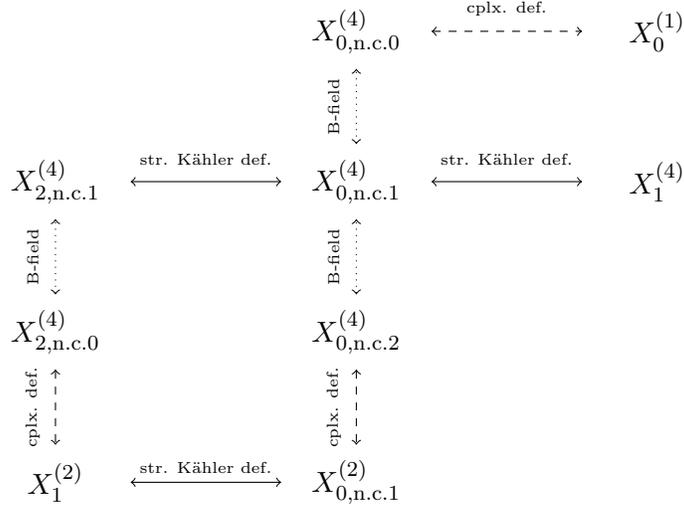

As we show below, the resolutions $X^{(4)}_{0,\text{n.c.}1}=X^{(4)}_{0,\text{n.c.}3}$ and $X^{(4)}_{2,\text{n.c.}1}$ correspond to large volume limits in the stringy K\"ahler moduli space of the genus one fibration $X^{(4)}_1$.
However, in order to extract the Gopakumar-Vafa invariants associated to M-theory compactifications on $X^{(4)}_0$ or $X^{(4)}_2$, we also need the topological string partition functions for other values of the B-field, that is on $X^{(4)}_{0,\text{n.c.}0}$ and $X^{(4)}_{0,\text{n.c.}1}$ or $X^{(4)}_{2,\text{n.c.}0}$.
It turns out, that $X^{(4)}_0=X^{(4)}_{0,\text{n.c.}0}$ can again be deformed into the generic elliptic fibration $X^{(1)}_0$ over $\mathbb{P}^2$.
On the other hand, we will find that $X^{(4)}_{0,\text{n.c.}2}$ can be deformed into $X^{(2)}_{0,\text{n.c.}1}$, while $X^{(4)}_{2,\text{n.c.}0}$ admits a deformation to $X^{(2)}_{1}$.
Both of the geometries have been discussed in the previous example.
This will allow us to determine the Gopakumar-Vafa invariants associated to M-theory compactifications on all of the elements of the Tate-Shafarevich group.
The relations among the various geometries are illustrated in Figure~\ref{fig:ex2geos}.

\subsection{The elliptic fibration $X^{(4)}_{\text{el.}}$ with $2$ sections}
Starting with the elliptic fibration $X^{(4)}_{\text{el.}}$ will help us again to choose appropriate normalizations for the free energies and the algebraic coordinates at the various large volume limits.
It will also allow us to identify the smooth deformations of the singular genus one fibration $X^{(4)}_{2}$ and the non-commutative partial resolution $X^{(4)}_{0,\text{n.c.}2}$.
Much of the standard geometric discussion is analogous to what has already been done in~\ref{sec:ex2sec} and will be kept brief.
The toric data of the ambient space is given by
\begin{align}
\begin{blockarray}{rrrrrrrrl}
	&&&&&C_1&C_2&C_3\\
\begin{block}{(rrrrr|rrr)l}
	-1&-1&-1& 0& 0& 0& 1& 0&\\
	 0& 0& 1& 0& 0& 0& 1& 0&\leftarrow\text{three-section }J_2\\
	 0& 1& 0& 0& 0& 0& 1&-1&\\
	 1& 0& 0& 0& 0& 1& 0& 0&\leftarrow\text{four-section }J_1\\
	-1& 0& 0& 0& 0& 1&-1& 0&\\
	 0& 0& 0& 1& 0& 0& 0& 1&\leftarrow\text{vertical divisor }J_3\\
	 0& 0& 0& 0& 1& 0& 0& 1&\phantom{x}\hspace{1.5cm}\text{\ditto}\\
	 0& 1& 0&-1&-1& 0& 0& 1&\phantom{x}\hspace{1.5cm}\text{\ditto}\\
	 0& 0& 0& 0& 0&-1&-1&-2&\leftarrow\text{nef divisor }-D_{\nabla_1}\\
	 0& 0& 0& 0& 0&-1&-1& 0&\leftarrow\text{nef divisor }-D_{\nabla_2}\\
\end{block}
\end{blockarray}\,.
\label{eqn:nef51overf1toricData}
\end{align}
The fiber corresponds to the codimension two nef partition $(5,1)$ that has been studied in~\cite{Paul-KonstantinOehlmann:2019jgr} and, in the notation of that paper, the fibration is determined by the choice of bundles
\begin{align}
	\mathcal{S}_2=H\,,\quad \mathcal{S}_6=3H\,,\quad\mathcal{S}_7=2H\,,\quad\mathcal{S}_9=H\,,
	\label{eqn:z4fpars}
\end{align}
where $H$ is the hyperplane class on $\mathbb{P}^2$.
Using the base independent expressions for the multiplicities of hypermultiplets from~\cite{Paul-KonstantinOehlmann:2019jgr}, we obtain the spectra that are summarized in Figure~\ref{fig:z4fieldtheories}.
The classes of the divisors that determine the nef partition are given in the last two rows of~\eqref{eqn:nef51overf1toricData}.

We use the basis of the K\"ahler cone $J_i,\,i=1,\ldots,3$, with the triple intersection numbers $c_{ijk}=J_i\cdot J_j\cdot J_k$ given by 
\begin{align}
	\begin{split}
		c_{111}=8\,,&\quad c_{112}=8\,,\quad c_{113}=8\,,\quad c_{122}=8\,,\quad c_{123}=8\,,\\
		c_{133}=4\,,&\quad c_{222}=5\,,\quad c_{223}=5\,,\quad c_{233}=3\,,\quad c_{333}=0\,,
	\end{split}
	\label{eqn:intersectionsnef51fibration}
\end{align}
while the usual topological invariants are
\begin{align}
	h^{1,1}=3\,,\quad h^{2,1}=63\,,\quad \chi=-120\,,\quad b_i=c_2\cdot\vec{J}=(50,\,56,\,36)^\intercal\,.
\end{align}
The classes of two independent sections $s_0,\,s_1$ are
\begin{align}
	S_0=J_1-J_2\,,\quad S_1=3J_2-2J_1+8J_3\,,
\end{align}
while $D_b=J_3$ is the vertical divisor that is induced by the hyperplane class of the base.
We choose $s_0$ as the zero-section and this determines the image of $s_1$ under the Shioda map~\eqref{eqn:shiodagen} to be
\begin{align}
	\sigma(s_1)=-3J_1+4J_2-12J_3\,,
\end{align}
with height pairing $b=-40$.
The modular parametrization of the K\"ahler form is given by
\begin{align}
	\omega=\tau\cdot\left(S_0+\frac32D_b\right)+m\cdot\sigma(s_1)+t\cdot D_b\,.
\end{align}
However, the height pairing implies that the index of the base degree zero partition function with respect to the elliptic parameter $m$ is $r_m=20$.
This makes the general ansatz rather unwieldy.

To determine the relevant normalizations, it will be sufficient to study the base degree zero contributions to the genus zero free energy.
Using $q=e^{2\pi i\tau},\,\zeta=e^{2\pi im}$ and $Q=e^{2\pi it}$ we can expand the instanton contributions to the free energy as
\begin{align}
	F_{0,\text{inst.}}=\sum\limits_{n=-1}^\infty\sum\limits_{\substack{k=-4\\4n+k\ge0}}^\infty \tilde{n}_k\text{Li}_3(q^n\zeta^k)+\mathcal{O}(Q)\,,
	\label{eqn:4secEllf00}
\end{align}
with the coefficients
\begin{align}
	\tilde{n}_0=120\,,\quad \tilde{n}_{\pm1}=104\,,\quad \tilde{n}_{\pm2}=76\,,\quad \tilde{n}_{\pm3}=24\,,\quad \tilde{n}_{\pm4}=6\,.
\end{align}
Except for $\tilde{n}_0$, which is the minus the Euler characteristic, these are just the multiplicities of six-dimensional hypermultiplets.

The limits of the sums in~\eqref{eqn:4secEllf00} reflect, that the elliptic fibration only allows for an extremal transition to the genus one fibration $X^{(4)}_1$, which is possible on the boundary of the K\"ahler cone
\begin{align}
	\tau\rightarrow 4\tau\,,\quad m\rightarrow \tau\,.
\end{align}
It is not possible to shrink the curves that correspond to the other Higgs fields in Figure~\ref{fig:z4fieldtheories} without leaving the K\"ahler cone or, equivalently, choosing a different chamber of the Coulomb branch in M-theory.
Nevertheless, after restricting the K\"ahler parameters, the difference between the phases only affects the lowest order terms of the free energy in an expansion in $\text{Li}_3(q^n)$.
The coefficients in this expansion are periodic and the large order behaviour is therefore sufficient to reconstruct the entire base degree zero contribution to the free energy of the Higgsed theory.

The specializations of the complexified K\"ahler parameters that correspond to the various Higgs transitions in Figure~\ref{fig:z4fieldtheories} are listed in Table~\ref{tab:z4khlhiggs}.
\begin{table}[h!]
	\begin{align*}
	\def\arraystretch{1.4}
	\begin{array}{c|cccc}
		&q_{KK}&q_{6d}&\tau& m\\\hline
		X^{(4)}_{0,\text{n.c.}k}&0&4&\tau&k/4\\
		X^{(4)}_{1}&-1&4&4\tau&\tau\\
		X^{(4)}_{2,\text{n.c.}k}&-2&4&2\tau+k/2&\tau\\
	\end{array}
	\end{align*}
	\label{tab:z4khlhiggs}
	\caption{The restrictions of K\"ahler parameters that correspond to the Higgs transitions in Figure~\ref{fig:z4fieldtheories}. The choices of B-fields follow from the analysis in Section~\ref{sec:n4fiber}.}
\end{table}
Here we have already chosen the appropriate B-fields for the complexified K\"ahler parameters in the Type IIA compactification, to reproduce the results that we derived from the transformation of the Jacobi forms in Section~\ref{sec:n4fiber}.
Applying the various restrictions to~\eqref{eqn:4secEllf00}, we obtain corresponding expressions for the genus zero free energy of the Higgsed theory
\begin{align}
	F'_{0,\text{inst.}}=\sum\limits_{k=0}^\infty n'_k\text{Li}_3(q^k)+\mathcal{O}(Q)\,,
	\label{eqn:z4f0higgs}
\end{align}
We use the large order behaviour to reconstruct the coefficients $n'_k$ for all $k\in\mathbb{N}$ and obtain the results in Table~\ref{tab:z4f0leading}.
All of the coefficients are at least 8-periodic, i.e. $n'_k=n'_{k+8}$.
\begin{table}[h!]
	\begin{align*}
	\def\arraystretch{1.4}
	\begin{array}{c|cccccccc}
		&n'_1&n'_2&n'_3&n'_4&n'_5&n'_6&n'_7&n'_8\\\hline
		X^{(4)}_{0,\text{n.c.}0}&540&540&540&540&540&540&540&540\\
		X^{(4)}_{0,\text{n.c.}1}&-20&-14&-20&-6&-20&-14&-20&-6\\
		X^{(4)}_{0,\text{n.c.}2}& 28& 92& 28& 92& 28& 92& 28& 92\\
		X^{(4)}_{1}&128&152&128&132&128&152&128&132\\
		X^{(4)}_{2,\text{n.c.}0}&256&284&256&284&256&284&256&284\\
		X^{(4)}_{2,\text{n.c.}1}& 0& 52& 0&13& 0& 52& 0& 18\\
	\end{array}
	\end{align*}
	\label{tab:z4f0leading}
	\caption{The coefficients of the base degree zero contributions to the genus zero free energies~\eqref{eqn:z4f0higgs} for the fibrations in $\Sh(X^{(4)}_0)$ and their non-commutative resolutions.}
\end{table}

As expected, the result for the singular Jacobian fibration $X^{(4)}_0=X^{(4)}_{0,\text{n.c.}0}$ corresponds to the free energy of the smooth deformation $X^{(1)}_0$.
On the other hand, we recognize the coefficients for $X^{(4)}_{2}=X^{(4)}_{2,\text{n.c.}0}$ as the base degree zero contributions to the free energy of the genus one fibration with $2$-sections $X^{(2)}_1$ that we studied in Section~\ref{sec:g12sec}.
Correspondingly, the coefficients for the non-commutative resolution $X^{(4)}_{0,\text{n.c.}2}$ of the Jacobian fibration lead us to identify the smooth deformation as $X^{(2)}_{0,\text{n.c.}1}$.
This suggests, that $X^{(4)}_{0,\text{n.c.}2}$ is only a non-commutative \textit{partial} resolution.
Recall that, while the generic elliptic fibration $X^{(1)}_0$ over $\mathbb{P}^2$ is unique, there are various different smooth genus one fibrations with $N$-sections over $\mathbb{P}^2$ for $N\ge 2$.
Of course we have constructed this example specifically such that the smooth deformation of $X^{(4)}_2$ is $X^{(2)}_1$.
For other genus one fibrations with $4$-sections there will be a different fibration with $2$-sections that is the smooth deformation of the corresponding element of the Tate-Shafarevich group.
As we will see now, the results for $X^{(4)}_{0,\text{n.c.}1}$ and $X^{(4)}_{2,\text{n.c.}1}$ correspond to the free energies at large volume limits in the stringy K\"ahler moduli space of $X^{(4)}_1$ and will allow us to fix the correct normalization of the local coordinates.

\subsection{The smooth genus one fibration $X^{(4)}_1$ with $2$-sections}
The smooth genus one fibered Calabi-Yau threefold $X^{(4)}_1$ with $4$-sections is constructed as a codimension two complete intersection in a fibration of $\mathbb{P}^3$ over $\mathbb{P}^2$.
The toric data of the ambient space is given by
\begin{align}
\begin{blockarray}{rrrrrrrl}
	&&&&&C_1&C_2\\
\begin{block}{(rrrrr|rr)l}
	-1&-1&-1& 0& 0& 1& 0&\leftarrow\text{four-section }J_1\\
	 0& 0& 1& 0& 0& 1& 0&\\
	 0& 1& 0& 0& 0& 1&-1&\\
	 1& 0& 0& 0& 0& 1& 0&\\
	 0& 0& 0& 1& 0& 0& 1&\leftarrow\text{vertical divisor }J_2\\
	 0& 0& 0& 0& 1& 0& 1&\phantom{x}\hspace{1.5cm}\text{\ditto}\\
	 0& 1& 0&-1&-1& 0& 1&\phantom{x}\hspace{1.5cm}\text{\ditto}\\
	 0& 0& 0& 0& 0&-2&-2&\leftarrow\text{nef divisor }-D_{\nabla_1}\\
	 0& 0& 0& 0& 0&-2& 0&\leftarrow\text{nef divisor }-D_{\nabla_2}\\
\end{block}
\end{blockarray}\,.
\label{eqn:nef00overf1toricData}
\end{align}
This corresponds to a fibration of the nef partition $(0,0)$ that was studied in~\cite{Paul-KonstantinOehlmann:2019jgr} and the fibration parameters are again~\eqref{eqn:z4fpars}.
Using the base independent expressions for the multiplicities of the reducible fibers from~\cite{Paul-KonstantinOehlmann:2019jgr}, we obtain the six- and five-dimensional matter spectra that were summarized in Figure~\ref{fig:z4fieldtheories}.

In terms of the K\"ahler cone basis $J_1,\,J_2$ we obtain the triple intersection numbers $c_{ijk}=J_i\cdot J_j\cdot J_k$ with
\begin{align}
	c_{111}=8\,,\quad c_{112}=8\,,\quad c_{122}=4\,,\quad c_{222}=0\,,
\end{align}
and the topological invariants
\begin{align}
	h^{1,1}=2\,,\quad h^{2,1}=68\,,\quad \chi=-132\,,\quad b_i=c_2\cdot\vec{J}=(56,\,36)^\intercal\,.
\end{align}
The Picard-Fuchs system is generated by the operators
\begin{align}
	\begin{split}
	\mathcal{D}_1=&\theta _1^2-2 \theta _1 \theta _2+2 \theta _2^2-4 z_1 \left(1+2 \theta _1\right) \left(1+2 \theta _1+2 \theta _2\right)\\
		&-4 z_2 \left(\theta _1-\theta _2\right) \left(1+2 \theta _1+2 \theta _2\right)\,,\\
	\mathcal{D}_2=&\theta _2^3-2 z_2 \left(\theta _1-\theta _2\right) \left(1+\theta _1+\theta _2\right) \left(1+2 \theta _1+2 \theta _2\right)\,,
	\end{split}
	\label{eqn:z4g1pfs}
\end{align}
and the discriminant takes the form $\Delta=\Delta_1\cdot \Delta_2$ with
\begin{align}
	\Delta_1=&\left(1-16 z_1\right)^3+4 z_2 \left(1-320 z_1-2048 z_1^2\right)-4096 z_1 z_2^2\,,\quad\Delta_2=1+4z_2\,.
\end{align}
The genus one free energy is given by the ansatz~\eqref{eqn:f1ansatz} with $c_1=c_2=1$.
Some Gopakumar-Vafa invariants at genus zero and one are listed in Table~\ref{tab:X41gvg0}.
\begin{table}[h!]
	\begin{align*}
	\begin{array}{c|cccccc}
		n^{(0)}_{d_1,d_2} & d_2=0 & 1 & 2 & 3 & 4 & 5 \\\hline
	 d_1=0 & 0 & 24 & -2 & 0 & 0 & 0 \\
	 1 & 128 & 1024 & 128 & 0 & 0 & 0 \\
	 2 & 152 & 12960 & 65552 & 15008 & -3944 & 6144 \\
	 3 & 128 & 106496 & 2745856 & 9884672 & 3008000 & -417792 \\
	 4 & 132 & 642288 & 58013096 & 771268912 & 2213502440 & 813801776 \\
	 5 & 128 & 3186688 & 817603840 & 29000105984 & 254905497600 & 619268484096 \\
	\end{array}
	\end{align*}
	\label{tab:X41gvg0}
	\caption{Genus zero Gopakumar-Vafa invariants of the smooth genus one fibration $X^{(4)}_1$.}
\end{table}
Note that the base degree zero invariants match with the result from Table~\ref{tab:z4f0leading} and reproduce the correct spectrum of hypermultiplets in the five-dimensional M-theory compactification.

In terms of the $4$-section $S_0=J_1$ with height pairing
\begin{align}
	D=-\pi^*\pi(S_0\cdot S_0)=-8\,,
\end{align}
and the vertical divisor $D_b=J_2$, the modular parametrization of the K\"ahler is given by
\begin{align}
	\omega=\tau\cdot\left(S_0+\frac{1}{8}D\right)+t\cdot D_b\,.
\end{align}
We denote the corresponding exponentiated parameters by $q=e^{2\pi i\tau}$ and $Q=e^{2\pi i t}$.
The genus zero Gopakumar-Vafa invariants allow us to determine the base degree one partition function
\begin{align}
	Z^{(4)}_{1,d_B=1}(\tau,\lambda)=\frac{1}{64}\frac{\Delta_8(e_1^2-e_2)^2(2e_1^4-9e_1^2e_2+e_2^2)}{\eta(4\tau)^{36}\phi_{-2,1}(4\tau,\lambda)}\,,
	\label{eqn:z4g1bd1}
\end{align}
where we use the shorthand notation $e_1(\tau)=E_{4,1}(\tau)$ and $e_2=E_{2,2}(\tau)$.

\subsection{The Jacobian fibration with discrete torsion $X^{(4)}_{0,\text{n.c.}1}$}
\label{sec:x40nc1}
The large volume limit that corresponds to the non-commutative resolution $X^{(4)}_{0,\text{n.c.}1}$ can be obtained by choosing coordinates $w_1, w_2$ with
\begin{align}
	z_1=\frac{1-16v_1}{16}\,,\quad z_2= -\frac{1}{2}v_2\left(\frac{16v_1}{1-16v_1}\right)^3\,.
	\label{eqn:4sectrafo1}
\end{align}
Applying this transformation to the operators~\eqref{eqn:z4g1pfs} leads to a new Picard-Fuchs system that is generated by
\begin{align}
	\begin{split}
	\mathcal{D}_1=&\theta _1 \left(\theta _1-3 \theta _2\right)-4 v_1 \left(1+4 \theta _1+20 \theta _1^2+2 \theta _2-44 \theta _1 \theta _2-20 \theta _2^2\right)\\
		&+256 v_1^2 \left(1+4 \theta _1+10 \theta _1^2+2 \theta _2-14 \theta _1 \theta _2-15 \theta _2^2\right)-2048 v_1^3 \left(3+12 \theta _1+20 \theta _1^2\right.\\
		&\left.+6 \theta _2-12 \theta _1 \theta _2-30 \theta _2^2\right)+1024 v_1^2 v_2 \left(3 \theta _2-\theta_1+1\right) \left(3 \theta _2-\theta_1\right)\\
		&+65536 v_1^4 \left(1+4 \theta _1+5 \theta _1^2+2 \theta _2+\theta _1 \theta _2-5 \theta _2^2\right)-8192 v_1^3 v_2 \left(4 \theta _1-1\right) \left(\theta _1-3 \theta _2\right)\\
		&-262144 v_1^5 \left(1+2 \theta _1\right) \left(1+2 \theta _1+2 \theta _2\right)+131072 v_1^4 v_2 \left(\theta _1-\theta _2\right) \left(1+2 \theta _1+2 \theta _2\right)\,,\\
	\mathcal{D}_2=&\theta _1 \left(\theta _1-3 \theta _2-1\right) \left(\theta _1-3 \theta _2\right)-4 v_1 \left(\theta _1-3 \theta _2\right) \left(1+12 \theta _1^2+2 \theta _2-4 \theta _1 \theta _2-20 \theta _2^2\right)\\
		&+64 v_1^2 \left(1+6 \theta _1+12 \theta _1^2+12 \theta _1^3-8 \theta _1^2 \theta _2-4 \theta _2^2-40 \theta _1 \theta _2^2-12 \theta _2^3\right)\\
		&-1024 v_1^3 \left(1+2 \theta _1\right) \left(1+\theta _1+\theta _2\right) \left(1+2 \theta _1+2 \theta _2\right)\,.
	\end{split}
\end{align}
The corresponding regular period now takes the form
\begin{align}
	\varpi_0=1+4 v_1+36 v_1^2+400 v_1^3+4900 v_1^4+288 v_1^3 v_2+\mathcal{O}(v^5)\,.
\end{align}
and the remaining five periods can be easily obtained as well.
After calculating, inverting and inserting the mirror map, we obtain the genus zero free energy
\begin{align}
	\begin{split}
	F_0=&\frac{1}{3!}c_{ijk}t^it^jt^k+p_2(t^1,t^2)-20 q_1+3 q_2-\frac{33}{2} q_1^2+40 q_1 q_2-\frac{45}{8} q_2^2\\
	&-\frac{560}{27} q_1^3+266 q_1^2 q_2-100 q_1 q_2^2+\frac{244}{9} q_2^3+\mathcal{O}(q^4)\,,
	\end{split}
	\label{eqn:z4jacr1f0}
\end{align}
with $p_2(t^1,t^2)$ being some quadratic polynomial and the triple intersection numbers
\begin{align}
	c_{111}=9\,,\quad c_{112}=3\,,\quad c_{122}=1\,,\quad c_{222}=0\,,
\end{align}
again being equal to those of the generic elliptic fibration $X^{(1)}_0$.
The normalization is determined from the base degree zero expansion~\eqref{eqn:z4f0higgs} with the coefficients from Table~\ref{tab:z4f0leading} that have been obtained using the Higgs transition.
This also fixes the factor $1/2$ in the choice of coordinates~\eqref{eqn:4sectrafo1}.

Applying the Fricke involution~\eqref{eqn:N4fricke}
\begin{align}
	\tau\mapsto-\frac{1}{4\tau}\,,
\end{align}
to~\eqref{eqn:z4g1bd1} should provide us with the base degree one partition function on $X^{(4)}_{0,\text{n.c.}1}$.
However, this would require us to understand the transformation of the topological string coupling $\lambda$ and perform a careful analysis of the automorphic properties of the partition function.
The latter are a consequence of the holomorphic anomaly equations~\cite{Bershadsky:1993cx,Witten:1993ed,Aganagic:2006wq,Gunaydin:2006bz} and were used in~\cite{Huang:2015sta,Cota:2019cjx} to derive the expansion of the topological string partition function in weak Jacobi forms.
We will leave such an analysis for future work and, for now, take the more pragmatic approach and fix the argument such that it is consistent with the Higgs transition.
There remains an undetermined scale factor $a$ such that $\phi_{w,i}(N\tau,\lambda)$ is replaced by $a^w\phi_{w,i}(\bullet,\lambda)$, which can be fixed using the properly normalized genus zero free energy.
Using the action of the Fricke involution on the $\Gamma_1(4)$ modular forms that we derive in Appendix~\ref{sec:mfG14}, we obtain the base degree one partition function
\begin{align}
	\begin{split}
	Z^{(4)}_{0,\text{n.c.}1,d_B=1}(\tau,\lambda)=&-4\frac{\Delta_8'(2e_1^2-e_2)^2(16e_1^4-3e_1^2e_2-e_2^2)}{\eta(\tau)^{36}\phi_{-2,1}(\tau,\lambda)}\,.
	\end{split}
	\label{eqn:z4jacr1bd1}
\end{align}

We find that the discriminant takes the form $\Delta=\Delta_1\cdot\Delta_2\cdot\Delta_3\cdot\Delta_4$ with
\begin{align}
	\begin{split}
	\Delta_1=&1-80 v_1+27 v_2+2560 v_1^2-1440 v_1 v_2-40960 v_1^3+27392 v_1^2 v_2\\
		&+327680 v_1^4-212992 v_1^3 v_2-1048576 v_1^5+524288 v_1^4 v_2-65536 v_1^3 v_2^2\,,\\
	\Delta_2=&1-48 v_1+768 v_1^2-4096 v_1^3-4096 v_1^3 v_2\,,\\
	\Delta_3=&1+v_2\,,\quad \Delta_4=1-16v_1\,.
	\end{split}
\end{align}
The ambiguities in the ansatz~\eqref{eqn:f1ansatz} for the genus one free energy can be fixed by using the base degree one partition function~\eqref{eqn:z4jacr1bd1} and we find
\begin{align}
	\chi=-132\,,\quad b_1=102\,,\quad b_2=36\,,\quad c_1=1\,,\quad c_2=1\,,\quad c_3=0\,,\quad c_4=-28\,.
\end{align}
Again, the Euler characteristic $\chi=-132$ matches that of the genus one fibration $X^{(4)}_1$ while the intersections with the second Chern class
\begin{align}
	b_i=c_2\cdot J_i\,,
\end{align}
are those of the smooth deformation, which is the generic elliptic fibration $X^{(1)}_0$.

We now want to extract the Gopakumar-Vafa invariants associated to the M-theory compactification on $X^{(4)}_0$ with gauge group $G=U(1)_{KK}\times\mathbb{Z}_4$ from the expansion~\eqref{eqn:torsiongvexs}.
To this end, we have to combine the information from all of the non-commutative resolutions $X^{(4)}_{0,\text{n.c.}i},\,i=0,\ldots,2$.
Since the A-model is invariant under complex structure deformations, we can use the respective smooth deformations $X^{(1)}_0$ and $X^{(2)}_{0,\text{n.c.}1}$ of $X^{(4)}_0$ and $X^{(4)}_{0,\text{n.c.}2}$, that have been discussed in Appendix~\ref{app:x18} and Section~\ref{sec:q2jacsec}.
Tables that contain some of the invariants are provided in Appendix~\ref{app:gvX40}.
Note that the base degree zero invariants at base degree zero correctly reproduce the spectrum of hypermultiplets from Figure~\ref{fig:z4fieldtheories}.

\subsection{The genus one fibration with discrete torsion $X^{(4)}_{2,\text{n.c.}1}$}
\label{sec:x42nc1}
We have now discussed all of the relevant phenomena and nothing unexpected happens for $X^{(4)}_{2,\text{n.c.}1}$.
The large volume limit can be found after choosing coordinates $w_1,w_2$ with
\begin{align}
	z_1=\frac{1}{(16w_1)^2}\,,\quad z_2= 4w_2\,,
	\label{eqn:4sectrafo2}
\end{align}
where $z_1,z_2$ are the algebraic complex structure coordinates for the mirror of $X^{(4)}_1$.
The corresponding Picard-Fuchs system is generated by the operators
\begin{align}
	\begin{split}
	\mathcal{D}_1=&\theta _1 \left(\theta _1-2 \theta _2\right)-16 w_1^2 \left(1+2 \theta _1+\theta _1^2+4 \theta _2+4 \theta _1 \theta _2+8 \theta _2^2\right)\\
		&+512 w_1^2 w_2 \left(\theta _1-2 \theta _2\right) \left(1+\theta _1+2 \theta _2\right)\,,\\
	\mathcal{D}_2=&\theta _2^3+2 w_2 \left(\theta _1-2 \theta _2-1\right) \left(\theta _1-2 \theta _2\right) \left(\theta _1+2 \theta _2+1\right)\,,
	\end{split}
\end{align}
and the leading terms of the fundamental period are
\begin{align}
	\varpi_0=1+4 w_1^2-48 w_1^2 w_2+36 w_1^4-4320 w_1^4 w_2+400 w_1^6+15120 w_1^4 w_2^2+\mathcal{O}(w^7)\,.
\end{align}
Again using the information from Table~\ref{tab:z4f0leading}, we find the correctly normalized genus zero free energy is
\begin{align}
	\begin{split}
	F_0=&\frac{1}{3!}c_{ijk}t^it^jt^k+p_2(t^1,t^2)+4 q_2+52 q_1^2-\frac{9}{2} q_2^2+\frac{328}{27} q_2^3-368 q_1^2 q_2\\
	&+\frac{39}{2} q_1^4+520 q_1^2 q_2^2-\frac{777}{16} q_2^4+\mathcal{O}(q^5)\,,
	\end{split}
\end{align}
with the triple intersection numbers
\begin{align}
	c_{111}=8\,,\quad c_{112}=4\,,\quad c_{122}=2\,,\quad c_{222}=0\,.
\end{align}
These match the intersection numbers~\eqref{eqn:g1sec2top1} of the genus one fibration $X^{(2)}_1$, that we identify as the smooth deformation of $X^{(4)}_2$.
Applying the $\Gamma_1(2)$ transformation~\eqref{eqn:n4t2} and the change of basis~\eqref{eqn:n4t2p} to~\eqref{eqn:z4g1bd1} leads to the base degree one partition function $Z^{(4)}_{2,\text{n.c.}1,d_B=1}(2\tau,\lambda)$ on the non-commutative resolution $X^{(4)}_{2,\text{nc.}1}$, with
\begin{align}
	Z^{(4)}_{2,\text{n.c.}1,d_B=1}(\tau,\lambda)=&-4\frac{\Delta_8''e_1^4(2e_1^4+5e_1^2e_2-6e_2^2)}{\eta\left(\tau+\frac12\right)\phi_{-2,1}\left(\tau+\frac12,\lambda\right)}\,,
	\label{eqn:z42bd1}
\end{align}
where the ambiguity in the overall scale is fixed by the genus zero free energy.
The discriminant takes the form $\Delta=\Delta_1\cdot \Delta_2$ with the factors
\begin{align}
	\begin{split}
		\Delta_1=&1-48 w_1^2+2048 w_1^2 w_2+768 w_1^4+81920 w_1^4 w_2\\
		&-4096 w_1^6+1048576 w_1^4 w_2^2-65536 w_1^6 w_2\,,\\
		\Delta_2=&1+16w_2\,,
	\end{split}
\end{align}
and the genus one free energy is given by~\eqref{eqn:f1ansatz}, while the base degree one partition function~\eqref{eqn:z42bd1} fixes the ambiguities
\begin{align}
	\chi=-132\,,\quad b_1=68\,,\quad b_2=36\,,\quad c_1=1\,,\quad c_2=1\,.
\end{align}
We again observe, that the Euler characteristic matches that of the genus one fibration $X^{(4)}_1$, which lies in the same stringy K\"ahler moduli space, while the intersections with the second Chern class $b_i=c_2\cdot J_i$ are identical to those of the smooth deformation $X^{(2)}_1$ of $X^{(4)}_2$.

Combining the information from $X^{(4)}_{2,\text{n.c.}1}$ and the smooth deformation $X^{(2)}_1$ of $X^{(4)}_2=X^{(4)}_{2,\text{n.c.}0}$ allows us to extract the Gopakumar-Vafa invariants, using the expansion~\eqref{eqn:torsiongvexs}.
Tables with invariants are provided in Appendix~\ref{app:gvX42}.
The base degree zero invariants at genus zero exactly reproduce the matter spectrum of the M-theory compactification on $X^{(4)}_2$ from Figure~\ref{fig:z4fieldtheories}.
This also serves as a highly non-trivial cross check for our choice of coordinates.

\section{Example 3: Irrational MUM-points and 5-sections}
\label{sec:ex5sec}
Having studied fibrations with $2$-sections and $4$-sections in significant detail, there seems to be little need for a third example.
However, one novel phenomenon motivates us to also discuss a genus one fibered Calabi-Yau threefold with $5$-section.

There it turns out, that the coefficients of the instanton contributions to the free energies at the large volume limits that correspond to the non-commutative resolutions of the Jacobian fibrations are not rational but elements of $\mathbb{Q}[\sqrt{5}]$.
The appearance of such irrational MUM-points has previously been observed for open topological strings~\cite{Laporte:2012hv,Jefferson:2013vfa} and led to the question of an enumerative interpretation.~\footnote{We thank Albrecht Klemm for pointing this out to us.}
Here we provide an explicit example of irrational MUM-points in the closed string case and show that the irrationality is a consequence of the roots of unity that appear in the expansion~\eqref{eqn:torsiongvexs}. 
As a result, the Gopakumar-Vafa invariants with discrete charges that are encoded in the irrational free energies turn out to be integral and match the field theoretic expectation.

To avoid redundancy, we will only summarize the results.
We also omit the discussion of an elliptic fibration, which can be easily constructed from the data provided in~\cite{Knapp:2021vkm}.
To again have a concrete example, we choose $1_a$ and $1_b$ as our smooth genus one fibered Calabi-Yau threefolds $X^{(5)}_1,\,X^{(5)}_2$ with $5$-sections over $\mathbb{P}^2$.
They are related by relative homological projective duality and respectively correspond to the cusps at $\tau=0$ and $\tau=2/5$ of the $\Gamma_1(5)$ modular curve.
All of the topological invariants, as well as the Picard-Fuchs systems, can be found in~\cite{Knapp:2021vkm}.
On the other hand, we argued in Section~\ref{sec:n5fiber} that the remaining two cusps of the modular curve at $\tau=0$ and $\tau=1/2$ respectively correspond to the non-commutative resolutions of the Jacobian fibration $X^{(5)}_{0,\text{n.c.}1}$ and $X^{(5)}_{0,\text{n.c.}2}$.

The M-theory compactification on the Jacobian fibration $X^{(5)}_0$ leads to a five-dimensional theory with gauge group
\begin{align}
	G=U(1)_{KK}\times\mathbb{Z}_5\,,
\end{align}
and the numbers $n_q$ of $(0,q)$ charged half-hypermultiplets are given by
\begin{align}
	n_{\pm 1}=100\,,\quad n_{\pm 2}=125\,,
	\label{eqn:z5multipls}
\end{align}
while the Euler characteristic $\chi=-90$ of $X^{(5)}_1$ determines the number of massless uncharged Dirac fields to be $n_0=90$.
The corresponding Kaluza-Klein towers are generated by adding multiples of $(\pm1,0)$ to the charges.

The Picard-Fuchs system at the large volume point associated to $X^{(5)}_1$ is, in the mirror dual algebraic coordinates $z_1,z_2$, generated by the operators
\begin{align}
	\begin{split}
		\mathcal{D}_1=&\Theta _1^2-3 \Theta _1 \Theta _2+7 \Theta _2^2+z_1 \left(-3-11 \Theta _1-11 \Theta _1^2\right)\\
		&-z_1^2 \left(1+\Theta _1+\Theta _2\right) \left(1+\Theta _1+2 \Theta _2\right)-z_2 \left(1+\Theta _1+2 \Theta _2\right) \left(14+15 \Theta _1+14 \Theta _2\right)\,,\\
   		\mathcal{D}_2=&\Theta _2^3-z_2 \left(1+\Theta _1+\Theta _2\right) \left(1+\Theta _1+2 \Theta _2\right) \left(2+\Theta _1+2 \Theta _2\right)\,,
	\end{split}
\end{align}
and the discriminant takes the form
\begin{align}
	\Delta=(1-11z_1-z_1^2)^3+\mathcal{O}(z_2)\,,
\end{align}
with the roots in the large base limit given by~\eqref{eqn:n5coniroots},
\begin{align}
	z_+=-\frac12\left(11+5\sqrt{5}\right)\,,\quad z_-=-\frac12\left(11-5\sqrt{5}\right)\,.
\end{align}
To study the large volume limit corresponding to $X^{(5)}_{0,\text{n.c.}1}$, we resolve the triple tangency at $z_-=z_2=0$ by choosing coordinates $v_1,v_2$ such that
\begin{align}
	z_1=\frac52\left(25-11\sqrt{5}\right)\left(\frac{1}{5\sqrt{5}}-v_1\right)\,,\quad z_2=-\frac{v_1^3v_2}{\left(\frac{1}{5\sqrt{5}}-v_1\right)^3}\,.
\end{align}
The particular normalization can again be found by constructing an appropriate elliptic fibration, using the results from~\cite{Knapp:2021vkm}, and studying the Higgs transitions with B-fields.
On the other hand, the correct choice of coordinates to find the large volume limit $X^{(5)}_{0,\text{n.c.}2}$ inside the triple tangency at $z_+=z_2=0$ is given by $w_1,w_2$ with
\begin{align}
	z_1=-\frac52\left(25+11\sqrt{5}\right)\left(\frac{1}{5\sqrt{5}}+w_1\right)\,,\quad z_2=\frac{w_1^3w_2}{\left(\frac{1}{5\sqrt{5}}+w_1\right)^3}\,.
\end{align}
Due to their size, we do not provide the generators of the respective Picard-Fuchs systems.
However, the coefficients of the operators are not rational but contained in $\mathbb{Q}[\sqrt{5}]$ and the leading terms of the genus zero free energies are
\begin{align}
	\begin{split}
		F_{0,X^{(5)}_{0,\text{n.c.}1}}=&\frac{1}{3!}c_{ijk}t^it^jt^k+p_2(t^1,t^2)-5 q_1 \left(10+z_-\right)+3 q_2-\frac{5}{8} q_1^2 \left(79+7 z_-\right)\\
		&+10 q_1 q_2 \left(10+z_-\right)-\frac{45}{8} q_2^2+\mathcal{O}(q^3)\,,\\
		F_{0,X^{(5)}_{0,\text{n.c.}2}}=&\frac{1}{3!}c_{ijk}t^it^jt^k+p_2(t^1,t^2)-5 q_1 \left(10+z_+\right)+3 q_2-\frac{5}{8} q_1^2 \left(79+7 z_+\right)\\
		&+10 q_1 q_2 \left(10+z_+\right)-\frac{45}{8} q_2^2+\mathcal{O}(q^3)\,,
	\end{split}
	\label{eqn:x50ncf0}
\end{align}
with the triple intersection numbers in both cases being
\begin{align}
	c_{111}=9\,,\quad c_{112}=3\,,\quad c_{122}=1\,,\quad c_{222}=0\,.
\end{align}
As we by now expect, these are the same as those of the smooth deformation $X^{(1)}_0$ of $X^{(5)}_0$.
Comparing the expansions~\eqref{eqn:x50ncf0}, it turns out that the free energies of $X^{(5)}_{0,\text{n.c.}1}$ and $X^{(5)}_{0,\text{n.c.}2}$ are transformed into each other by the action of the Galois group of the discriminant polynomial of the fiber, that exchanges the roots $z_+$ and $z_-$.

The base degree one partition function of $X^{(5)}_1$ and $X^{(5)}_2$ are given by
\begin{align}
	\begin{split}
	Z^{(5)}_{1,d_B=1}(\tau,\lambda)=&\frac{-5\Delta_{10}e_{1,1}^2e_{1,2}\left(10e_{1,1}^3-108e_{1,1}^2e_{1,2}+343e_{1,1}e_{1,2}^2-2e_{1,2}^3\right)}{\eta(5\tau)^{36}\phi_{-2,1}(5\tau,\lambda)}\,,\\
	Z^{(5)}_{2,d_B=1}(\tau,\lambda)=&\frac{-5\Delta_{10}'e_{1,2}^2e_{1,1}\left(10e_{1,2}^3-108e_{1,2}^2e_{1,1}+343e_{1,2}e_{1,1}^2-2e_{1,1}^3\right)}{\eta(5\tau)^{36}\phi_{-2,1}(5\tau,\lambda)}\,,
	\end{split}
\end{align}
where we use the shorthands $e_{1,1}=E_{5,1,a}$ and $e_{1,2}=E_{5,1,b}$.
After applying the Fricke involution $\tau\mapsto-1/5\tau$ we obtain the corresponding results for $X^{(5)}_{0,\text{n.c.}1}$ and $X^{(5)}_{0,\text{n.c.}2}$,
\begin{align}
	Z^{(5)}_{0,\text{n.c.}k,d_B=1}=\frac{\phi_k(\tau)}{\eta(\tau)^{36}\phi_{-2,1}(\tau,\lambda)}\,,\quad\text{for }k\in\{1,2\}\,,
\end{align}
with
\begin{align}
	\begin{split}
	\phi_1(\tau)=&5 \left(e_{1,1}^2-11 e_{1,1} e_{1,2}-e_{1,2}^2\right) \left(-3 e_{1,1}^4 \left(2-11 z_-\right)+e_{1,1}^3 e_{1,2} \left(18-349 z_-\right)\right.\\
		&\left.+15 e_{1,1}^2 e_{1,2}^2 \left(11+2 z_-\right)+e_{1,1} e_{1,2}^3 \left(3857+349 z_-\right)+3 e_{1,2}^4 \left(123+11 z_-\right)\right)\,,\\
	\phi_2(\tau)=&5 \left(e_{1,1}^2-11 e_{1,1} e_{1,2}-e_{1,2}^2\right) \left(-3 e_{1,1}^4 \left(2-11 z_+\right)+e_{1,1}^3 e_{1,2} \left(18-349 z_+\right)\right.\\
		&\left.+15 e_{1,1}^2 e_{1,2}^2 \left(11+2 z_+\right)+e_{1,1} e_{1,2}^3 \left(3857+349 z_+\right)+3 e_{1,2}^4 \left(123+11 z_+\right)\right)\,.
	\end{split}
\end{align}
Again, the results are exchanged under the action of the Galois group that permutes the roots of the discriminant of the fiber.

Trying to apply the ordinary Gopakumar-Vafa expansion to the partition functions of $X^{(5)}_{0,\text{n.c.}1}$ or $X^{(5)}_{0,\text{n.c.}2}$ will obviously lead to irrational numbers that take values in $\mathbb{Q}[\sqrt{5}]$.
This makes it clear, that those numbers do not contain any enumerative information.
However, the expansion~\eqref{eqn:torsiongvexs} in terms of Gopakumar-Vafa invariants with $\mathbb{Z}_5$ charges takes the form
\begin{align}
	\begin{split}
	&\log\left[Z_{\text{top.},\text{n.c.}k}(t_1,t_2,\lambda)\right]\\
	=&\sum\limits_{g=0}^\infty\sum\limits_{d_1,d_2=0}^\infty\sum\limits_{q=0}^4\sum\limits_{m=1}^\infty n_{(d_1,d_2),q}^{(g)}\cdot\frac{1}{m}\left(2\sin\frac{m\lambda}{2}\right)^{2g-2} e^{mq\frac{2\pi i k}{5}}q_1^{md_1}q_2^{md_2}\,.
	\end{split}
\end{align}
The fifth roots of unity themselves introduce an irrationality that exactly cancels with the square roots in the partition function and we obtain integral invariants.
Tables with some of the invariants are provided in Appendix~\ref{sec:gvX50}.
As a highly non-trivial check of the enumerative interpretation, one can see that the base degree zero invariants at genus zero again match with the field theoretic expectation from~\eqref{eqn:z5multipls}.

\section{Outlook}
\label{sec:outlook}
We have found a rich network of relationships among smooth torus fibered Calabi-Yau manifolds and non-commutative resolutions of fibrations with $\mathbb{Q}$-factorial terminal singularities.
At the heart of this network is a correspondence between the cusps of the $\Gamma_1(N)$ modular curve and the various elements of the Tate-Shafarevich group of a genus one fibered Calabi-Yau manifold with $N$-sections.
This is hidden from classical geometry and only manifests itself in the stringy K\"ahler moduli space.
It requires the B-field also to obtain a non-commutative resolution of the singular elements.
The correspondence is highly surprising and hints at an exciting new connection between string theory and number theory that warrants further investigation.

A beautiful instance of this connection that we studied in this paper, is the interplay between Higgs transitions, modular- and Jacobi forms and the cusps of the modular curve.
Generically, we have found that the Fricke involution $\tau\mapsto-1/N\tau$, which permutes the cusps of the modular curve at $\tau=i\infty$ and $\tau=0$, relates the restrictions of Jacobi forms
\begin{align}
	q^{\frac{mk^2}{N}}\phi_{w,m}(N\tau,k\tau)\xrightarrow{\tau\mapsto-\frac{1}{N\tau}}\tau^w\phi_{w,m}\left(\tau,\frac{k}{N}\right)\,.
\end{align}
Applied to the stringy K\"ahler moduli space of Calabi-Yau manifolds with $N$-sections, this transforms the topological string partition functions on the genus one fibration and a non-commutative resolution of the Jacobian fibration into each other.

For torus fibered Calabi-Yau threefolds that also exhibit a K3-fibration, one can often construct a dual Heterotic compactification on $(K3\times T^2)/\mathbb{Z}_N$~\cite{Banlaki:2019bxr}.
Using the standard dictionary of Type IIA-Heterotic duality~\cite{Kachru:1995wm}, the Fricke involution is expected to manifest itself on the Heterotic side as a Fricke T-duality.
A similar situation arises in compactifications of Type IIA strings on $(K3\times T^2)/\mathbb{Z}_N$~\cite{Persson:2015jka}.
In that case, the Fricke T-duality of the IIA strings translates into a Fricke S-duality of the dual Heterotic CHL compactifications.

Fricke involutions have also appeared in the context of non-compact Calabi-Yau manifolds~\cite{Alim:2013eja}.
There, the origin of the modular properties is quite different and stems from the mirror essentially being an elliptic curve with a choice of meromorphic differential.
Nevertheless, they also relate theories at different points in the K\"ahler moduli space and facilitate the integration of the holomorphic anomaly equation to obtain all genus results.
The modular transformations between partition functions on compact Calabi-Yau manifolds should equally impose strong conditions.
We leave a careful investigation of the utility of these constraints in solving the topological string to higher genus or base degree for future work.

Even the construction of different elements of the Tate-Shafarevich group for a given smooth genus one fibered Calabi-Yau manifold is non-trivial and few examples are known~\cite{dolgachev1992elliptic,Mayrhofer:2014laa,Cvetic:2015moa}.
Perhaps the richest set of examples has recently been obtained in~\cite{Knapp:2021vkm}, and consists of pairs of smooth genus one fibrations with $5$-sections that are related by relative homological projective duality.
This motivated our current investigation and is a special case of the structure that we uncovered.
Our general results now make it possible to not only find a large volume limit for any element of the Tate-Shafarevich group of a generic torus fibered Calabi-Yau manifold with $N$-sections and $N\le 5$, but also for their various non-commutative resolutions.
This opens up a vast new class of string compactifications for exploration.

By assuming that the fibration is generic, we have excluded fibrations with fibral divisors, multiple fibers or non-flat fibers.
The exclusion of fibral divisors has only been made to simplify an already lengthy exposition.
Our arguments involving Higgs transitions and the modular properties of the topological string partition function can easily be generalized and therefore we expect our results to directly carry over.
Potential problems that would arise from multiple fibers or non-flat fibers are more difficult to estimate, given that not even the modular bootstrap on smooth fibrations has been worked out in this generality.
One immediate consequence of multiple fibers would be, that we have to consider the full Weil-Ch\^atelet group and can not restrict ourselves to the Tate-Shafarevich subgroup.
It would be desirable to get a better understanding of the modular implications and, ideally, to get rid of these technical assumptions in the future.

Another obvious restriction is that we only consider genus one fibrations with $N$-sections for $N\le5$.
This is partly a technical issue and we plan to address the $N=6$ case in~\cite{sec6wip}.
More fundamentally, only the $\Gamma_1(N)$ modular curves for $N\le 10$ and $N=12$ have genus zero.
It is conjectured that the moduli space of a consistent theory of quantum gravity is simply connected~\cite{Ooguri:2006in}.
This suggests, that there are no Calabi-Yau varieties with $N$-sections for $N=11$ or $N>12$ that are sufficiently smooth to be used in string compactifications.
In the mirror dual picture this excludes fibrations with corresponding $N$-torsional sections.
However, a careful analysis shows that the bound on $N$-torsional sections is actually lower and the upper limit on Calabi-Yau $d$-folds with $d\ge 3$ is $N=6$~\cite{Hajouji:2019vxs,Dierigl:2020lai}.
The relation between $N$-sections and $N$-torsional sections under mirror symmetry, at least at this point, rests on studying the stringy K\"ahler moduli space of generic degree $N$ curves.
Nevertheless, assuming that the correspondence holds in general, we expect that genus one fibered Calabi-Yau varieties with $N$-sections for $N\ge 7$ are also too singular to lead to regular physics in string theory.
The relationship between the cusps of the modular curve and the elements of the Tate-Shafarevich group imposes strong constraints on the structure of the stringy K\"ahler moduli space of the fibration and can potentially be used to prove such a bound.

On the other hand, a natural avenue to try to construct an independent proof of the mirror relation between genus one curves of degree $N$ and elliptic curves with $N$-torsional sections is the Greene-Plesser mirror construction~\cite{GREENE199015}.
This is based on a group action $G$ on the worldsheet theory $K$, such that the quotient theory $K/G$ is isomorphic to $K$ up to a change of sign for the left moving $U(1)$ symmetry.
If the group action is compatible with the target space geometry $M$ of $K$, this leads to a mirror Calabi-Yau $M/K$.
Particular examples are given by toric hypersurfaces that exhibit a Fermat point in the complex structure moduli space, which is the case for the mirrors of the generic genus one curves of degree $2$ and $3$.
Another way to interpret the $\mathbb{Z}_N$ symmetry is in terms of the shift symmetry that the $N$-torsional section induces on the mirror elliptic curve.
The quotient of the elliptic curve by this action is a genus one curve of degree $N$ and this does not depend on a particular toric realization.
More difficult seems to be the opposite, and for our purpose more important conclusion, namely that the mirror of a genus one curve of degree $N$ is an elliptic curve with an $N$-torsion point.
Perhaps a careful study of the associated worldsheet theories leads to a general argument.

We also introduced Gopakumar-Vafa invariants with discrete charges. 
These correspond to traces over the BPS spectrum of five-dimensional M-theory compactifications on Calabi-Yau threefolds with $\mathbb{Q}$-factorial terminal singularities.
Perhaps somewhat surprisingly, they are encoded not only in the topological string partition functions of the singular fibration itself, but extracting them requires also the partition function on the various non-commutative resolutions.
We have performed highly non-trivial checks in various examples and also demonstrated that the invariants can associate an enumerative theory to irrational MUM-points.
A method to calculate genus zero Gopakumar-Vafa invariants on certain local Calabi-Yau threefolds in terms of open string zero modes has recently been proposed in~\cite{Collinucci:2021wty}.
This was also applied to calculate invariants associated to certain singularities that do not admit a small crepant resolution, and additional T-brane data was required to obtain a well-defined counting problem.
It would be interesting to see, if a similar approach can also be used to calculate the invariants associated to the local singularities that are stabilized by a flat torsional B-field.

Another immediate application of the Gopakumar-Vafa invariants of fibrations with $\mathbb{Q}$-factorial terminal singularities, is the calculation of BPS spectra of five-dimensional superconformal field theories.
This involves a decoupling of gravity, which for compactifications on smooth Calabi-Yau manifolds leads to the possibility of refining the invariants~\cite{Huang:2013yta,Haghighat:2014vxa,Gu:2017ccq,DelZotto:2017mee,Cota:2019cjx}.
The refined expansion involves a second topological string coupling and an ansatz in terms of Jacobi forms of half-integral index.
An analogous ansatz can be made also for the corresponding non-compact limits of the partition functions on non-commutative resolutions.
Studying the corresponding refined invariants would be another interesting subject for future work.
Let us point out, that currently the most efficient technique of calculating the refined invariants on non-compact elliptic fibrations is the use of so-called blowup equations~\cite{Grassi:2016nnt,Gu:2018gmy,Gu:2019dan,Gu:2019pqj,Gu:2020fem}.
This has recently been extended to genus one fibrations in~\cite{Duan:2021ges}.
Developing this technique also for non-commutative resolutions of torus fibrations is an interesting related problem.

For the generic fibrations that we studied in this paper, the structure of the Jacobian fibration is rather similar to that of the genus one fibration.
This makes it straightforward to reconstruct the massless spectrum of the M-theory compactification on the Jacobian directly from the genus one fibration.
In general, it is known that the structure of the Jacobian fibration can be quite different from the genus one fibration~\cite{Braun:2014oya,Oehlmann:2016wsb,Baume:2017hxm,Anderson:2018heq,Anderson:2019kmx}.
Particularly drastic examples arise when an $N$-section is folding the fibers of fibral divisors such that the gauge algebra is twisted in the compactification on the genus one fibration~\cite{larastringmath2021}.
It then becomes a non-trivial task to determine the exact effective theory of the M-theory compactification on the Jacobian.
On smooth fibrations, the Gopakumar-Vafa invariants provide a powerful tool to obtain the spectrum~\cite{Paul-KonstantinOehlmann:2019jgr}.
Now being able to calculate the corresponding invariants for the singular Jacobian fibrations could prove equally useful.
However, we should point out that the genus one fibrations that lead to twisted gauge algebras generally contain multiple fibers.
This reiterates the importance of understanding their modular implications.

Our interpretation of particular large volume limits in terms of non-commutative resolutions has been based on Higgs transitions in M- and F-theory as well as transformations of restrictions of Jacobi forms.
This enabled us to demonstrate the presence of a torsional B-field in those cases.
The corresponding worldsheet theory is related by stringy K\"ahler deformations to the non-linear sigma model on a smooth genus one fibration, and this implies that it has to be smooth.
From a mathematical perspective, a non-commutative resolution is defined in terms of the associated category of branes.
Based on earlier results from~\cite{dolgachev1992elliptic,Caldararu:2002ab,Braun:2014oya,Mayrhofer:2014laa}, we have proposed two candidates for the category of branes, one in terms of a small non-K\"ahler resolution and the other using a large resolution that is not Calabi-Yau.
Ideally, it would be possible to construct the branes directly in terms of twisted sheaves on the singular fibration.
A possible starting point is the resolution $X^{(4)}_{2,\text{n.c.}1}$ that admits a concrete realization as a hybrid point in the gauged linear sigma model of the genus one fibration $X^{(4)}_1$.
This provides access to the toolbox of matrix factorizations~\cite{Herbst:2008jq} and also leads to the question how the torsional B-field manifests itself in the GLSM.
We hope to come back to this in future work.

In recent years, there has also been a renewed interest in the arithmetic properties of the periods at special attractor points in the complex structure moduli space of Calabi-Yau manifolds~\cite{Candelas:2019llw,Yang:2020sfu,Yang:2020lhd,Kachru:2020sio,Kachru:2020sio,Candelas:2021tqt,bksz}.
Modular properties of the periods arise in this context as a consequence of the factorization of the Hasse-Weil zeta function.
The factorization implies a motivic relationship between the periods on the Calabi-Yau and the periods on particular elliptic curves.
This phenomenon also occurs in the period integrals on Calabi-Yau manifolds that are related to Feynman integrals~\cite{Bonisch:2020qmm,Bonisch:2021yfw}.
Given the relation between the complex structure moduli space and the stringy K\"ahler moduli space via mirror symmetry, it is natural to ask for possible connections between the modular properties that appear on both sides of the duality.
However, the modular properties in the complex structure moduli space  do not require a torus fibration structure and investigations have therefore largely been restricted to one-parameter families.
It would be very interesting to perform a similar analysis in the complex structure moduli space of the mirror of a torus fibered Calabi-Yau.

\appendix
\section{The generic elliptic fibration $X^{(1)}_0$ over $\mathbb{P}^2$}
\label{app:x18}
We will now review the relevant properties of the generic elliptic fibration $X^{(1)}_0$ over $\mathbb{P}^2$.
This can be realized as a degree $18$ hypersurface in the weighted projective space $\mathbb{P}_{11169}$.
Starting with~\cite{Candelas:1994hw}, it has been the canonical example to study the modular properties of topological strings on elliptic fibrations~\cite{Andreas:2001ve,Klemm:2012sx,Alim:2012ss,Huang:2015sta}.
The toric data of the ambient space is given by
\begin{align}
\begin{blockarray}{crrrrrrl}
	&&&&&C_1&C_2\\
\begin{block}{c(rrrr|rr)l}
	  X& 1& 0& 0& 0& 2& 0&\\
	  Y& 0& 1& 0& 0& 3& 0&\\
	  Z&-2&-3& 0& 0& 1&-3&\leftarrow\text{holomorphic section }S_0\\
	x_1&-2&-3& 1& 0& 0& 1&\leftarrow\text{vertical divisor }D_b\\
	x_2&-2&-3& 0& 1& 0& 1&\phantom{x}\hspace{1.0cm}\text{\ditto}\\
	x_3&-2&-3&-1&-1& 0& 1&\phantom{x}\hspace{1.0cm}\text{\ditto}\\
	   & 0& 0& 0& 0&-6& 0&\\
\end{block}
\end{blockarray}\,.
\label{eqn:X18toricData}
\end{align}
The vertices of the polytope are listed in the first four columns while the linear relations that correspond to a basis of the Mori cone are contained in the last two.
Using TOPCOM~\cite{Rambau:TOPCOM-ICMS:2002} one can check that the vertices admit a unique fine regular star triangulation.

The topological invariants of the Calabi-Yau hypersurface $X^{(1)}_0$ are well known.
A basis of the K\"ahler cone is given by
\begin{align}
	J_1=S_0+3D_b\,,\quad J_2=D_b\,,
\end{align}
and the corresponding triple intersections $c_{ijk}=J_i\cdot J_j\cdot J_k$ are
\begin{align}
	c_{111}=9\,,\quad c_{112}=3\,,\quad c_{112}=1\,,\quad c_{111}=0\,.
\end{align}
Using PALP~\cite{Kreuzer:2002uu}, one can verify the Hodge numbers and Euler characteristic
\begin{align}
	h^{1,1}=2\,,\quad h^{2,1}=272\,,\quad \chi=-540\,,
\end{align}
and the intersections with the second Chern class read
\begin{align}
	\vec{b}=c_2\cdot \vec{J}=(102,\,36)^\intercal\,.
\end{align}
The Picard-Fuchs system is generated by the two operators
\begin{align}
	\begin{split}
	\mathcal{D}_1=&\theta_1 \left(\theta_1-3 \theta _2\right)-12 z_1 \left(1+6 \theta_1\right) \left(5+6 \theta_1\right)\,,\\
	\mathcal{D}_2=&\theta _2^3+z_2 \left(3 \theta _2-\theta _1+2\right) \left(3 \theta _2-\theta _1+1\right) \left(3 \theta _2-\theta _1\right)\,.
	\end{split}
	\label{eqn:x18pf}
\end{align}
This can be used to determine the discriminant polynomial, which is given by $\Delta=\Delta_1\cdot \Delta_2$ with
\begin{align}
	\Delta_1=(1-432z_1)^3-27\cdot (432z_1)^3z_2\,,\quad \Delta_2=1+27z_2\,.
\end{align}
The genus one free energy is given by~\eqref{eqn:f1ansatz} with $c_1=c_2=1$.
A modular parametrization of the K\"ahler cone is given by
\begin{align}
	\omega=\tau\cdot\left(S_0+\frac12 D\right)+t\cdot D_b\,,
\end{align}
with the height pairing of the zero section given by $D=-\pi^{-1}\pi_*(S_0\cdot S_0)=3D_b$.
In this parametrization the base degree one topological string partition function takes the form~\cite{Huang:2015sta}
\begin{align}
	Z_{d_B=1}(\tau,\lambda)=-\frac{1}{48}\frac{E_4\cdot(31 E_4^3+113E_6^2)}{\eta(\tau)^{36}\phi_{-2,1}(\tau,\lambda)}\,.
\end{align}
We provide some of the Gopakumar-Vafa invariants at genus zero, genus one and base degree one in the Tables~\ref{tab:x18gv-g0},~\ref{tab:x18gv-g1} and~\ref{tab:x18gv-d1}.
\begin{table}[h!]
	\begin{align*}
\begin{array}{c|cccccc}
	n^{(0)}_{d_1,d_2} & d_2=0 & 1 & 2 & 3 & 4 \\\hline
 d_1=0 & 0 & 3 & -6 & 27 & -192 \\
 1 & 540 & -1080 & 2700 & -17280 & 154440 \\
 2 & 540 & 143370 & -574560 & 5051970 & -57879900 \\
 3 & 540 & 204071184 & 74810520 & -913383000 & 13593850920 \\
 4 & 540 & 21772947555 & -49933059660 & 224108858700 & -2953943334360 \\
 5 & 540 & 1076518252152 & 7772494870800 & -42712135606368 & 603778002921828 \\
\end{array}
	\end{align*}
	\caption{Genus zero Gopakumar-Vafa invariants at low degrees for $X^{(1)}_0$.}
	\label{tab:x18gv-g0}
\end{table}
\begin{table}[h!]
	\begin{align*}
\begin{array}{c|cccccc}
	n^{(0)}_{d_1,d_2} & d_2=0 & 1 & 2 & 3 & 4 \\\hline
 d_1=0 & 0 & 0 & 0 & -10 & 231 \\
 1 & 3 & -6 & 15 & 4764 & -154662 \\
 2 & 3 & 2142 & -8574 & -1079298 & 48907815 \\
 3 & 3 & -280284 & 2126358 & 152278986 & -9759419622 \\
 4 & 3 & -408993990 & 521856996 & -16704086880 & 1591062421074 \\
 5 & 3 & -44771454090 & 1122213103092 & -3328467399468 & -186415241060547 \\
\end{array}
	\end{align*}
	\caption{Genus one Gopakumar-Vafa invariants at low degrees for $X^{(1)}_0$.}
	\label{tab:x18gv-g1}
\end{table}
\begin{table}[h!]
	\begin{align*}
\begin{array}{c|cccccc}
	n^{(0)}_{d_1,d_2} & d_2=0 & 1 & 2 & 3 & 4 \\\hline
 d_1=0 & 3 & 0 & 0 & 0 & 0 \\
 1 & -1080 & -6 & 0 & 0 & 0 \\
 2 & 143370 & 2142 & 9 & 0 & 0 \\
 3 & 204071184 & -280284 & -3192 & -12 & 0 \\
 4 & 21772947555 & -408993990 & 412965 & 4230 & 15 \\
 5 & 1076518252152 & -44771454090 & 614459160 & -541440 & -5256 \\
\end{array}
	\end{align*}
	\caption{Base degree one Gopakumar-Vafa invariants at low genera for $X^{(1)}_0$.}
	\label{tab:x18gv-d1}
\end{table}

Using e.g. SageMath~\cite{sagemath} one can check that the polytope~\eqref{eqn:X18toricData} contains three points
\begin{align}
	\rho_1=(-1,-1,\,0,\,0)\,,\quad \rho_2=(-1,-2,\,0,\,0)\,,\quad \rho_3=(\,0,-1,\,0,\,0)\,,
	\label{eqn:x18irrPoints}
\end{align}
in the interior of facets.
The corresponding divisors do not intersect the Calabi-Yau hypersurface and the points are therefore irrelevant and have been omitted in~\eqref{eqn:X18toricData}.
However, the irrelevant points are in one to one correspondence with toric automorphisms that arise from \textit{roots} of the dual toric variety~\cite{Cox:2000vi}.

A root $(x,m)$ consists of a homogeneous coordinate $x$ and a monomial $m$ that is not equal to the coordinate itself but has the same weight under the scaling relations.
The coordinate change $x\rightarrow x+\lambda m$ with $\lambda\in\mathbb{C}$ then defines a one-parameter family of toric automorphisms.
In general, the automorphisms that arise from irrelevant points can be used to remove the corresponding monomials from the defining equation of the mirror.
Sometimes a residual discrete automorphism group remains even after fixing this gauge and induces automorphisms of the Calabi-Yau hypersurface.
This is the case for the mirror of the degree $18$ hypersurface in $\mathbb{P}_{11169}$.

The vertices of the dual polytope and the homogeneous coordinates are given by
\begin{align}
\begin{blockarray}{crrrr}
\begin{block}{c(rrrr)}
	 x_1 &-1&-1&12&-6\\
	 x_2 &-1&-1&-6&12\\
	 x_3 &-1&-1&-6&-6\\
	 x_4 & 2&-1& 0& 0\\
	 x_5 &-1& 1& 0& 0\\
\end{block}
\end{blockarray}\,.
\end{align}
To study the automorphisms, we can ignore the remaining points. 
The irrelevant points~\eqref{eqn:x18irrPoints} correspond to the family of toric automorphisms
\begin{align}
	x_4\rightarrow x_4+\lambda_1 x_1^2x_2^2x_3^2\,,\quad x_5\rightarrow x_5+\lambda_2x_1^3x_2^3x_3^3+\lambda_3 x_1x_2x_3x_4\,.
	\label{eqn:toricam}
\end{align}
After removing the irrelevant monomials, the defining equation of the mirror is given by
\begin{align}
	\begin{split}
		P=&a_0\cdot x_1 x_2 x_3 x_4 x_5+a_1\cdot x_1^{18}+a_2\cdot x_2^{18}+ a_3\cdot x_3^{18}+a_5\cdot x_5^2+a_6\cdot x_4^3+a_4\cdot x_2^6 x_3^6 x_1^6\\
		=&x_1 x_2 x_3 x_4 x_5+z_2 x_1^{18}+x_2^{18}+x_3^{18}+x_2^6 x_3^6 x_1^6+x_5^2-\sqrt{z_1} x_4^3\,,
	\end{split}
\end{align}
where $z_1$ and $z_2$ are the algebraic complex structure coordinates
\begin{align}
	z_1=\frac{a_4a_5^3a_6^2}{a_0^6}\,,\quad z_2=\frac{a_1a_2a_3}{a_4^3}\,.
\end{align}
that are defined by the linear relations in~\eqref{eqn:X18toricData}.
By acting with~\eqref{eqn:toricam} on the polynomial $P$ we find that a $\mathbb{Z}_2$ subgroup
\begin{align}
	\lambda_1=\frac16\frac{a_0^2}{a_5a_6}\,,\quad \lambda_2=-\frac{1}{12}\frac{a_0^3}{a_5^2a_6}\,,\quad \lambda_3=-\frac{1+i}{2}\frac{a_0}{a_5}\,,
\end{align}
of the automorphism group does not reintroduce irrelevant monomials.
The action on the defining equation can be compensated by a transformation of the complex structure coordinates
\begin{align}
	z_1\rightarrow \frac{1-432z_1}{432}\,,\quad z_2\rightarrow-z_2\left(\frac{432z_1}{1-432z_1}\right)^3\,.
	\label{eqn:x18inv}
\end{align}
This defines an involution in the complex structure moduli space that identifies the large volume limit at $z_1=z_2=0$ with an equivalent large volume limit at $z_1=\frac{1}{432}\,,\quad z_2=0$.
In particular, it transforms the Picard-Fuchs system generated by the operators~\eqref{eqn:x18pf} into itself while exchanging the components of the discriminant $\Delta_1=0$ and $\Delta_2=0$.
The second large volume limit lies at a triple tangency between the principal component of the discriminant $\Delta_1=0$ and the large base complex structure limit $z_2=0$.
As discussed in~\cite{Candelas:1994hw}, the coordinate transformation~\eqref{eqn:x18inv} blows up this tangency and resolves it into normal crossing divisors.

One can show, that the mirror dual involution in the stringy K\"ahler moduli space of $X^{(1)}_0$ amounts to T-dualizing both cycles of the $T^2$ fiber~\cite{Candelas:1994hw,Andreas:2001ve,Klemm:2012sx,Alim:2012ss}.
The action on the category of topological B-branes is given by the Fourier-Mukai transform with the kernel given by the Poincar\'e sheaf on $X^{(1)}_0\times_{\mathbb{P}^2}X^{(1)}_0$~\cite{Andreas:2001ve}.
Calculating the transformation of the brane charges one can show that the complexified volume of the generic fiber transforms as
\begin{align}
	\tau\rightarrow -\frac{1}{\tau}\,.
\end{align}
This is related by large volume monodromies to the action
\begin{align}
	\tau\rightarrow \frac{\tau}{\tau+1}\,,
\end{align}
that is induced by the fiberwise conifold monodromy which in turn corresponds to the Fourier-Mukai transform with the kernel being the ideal sheaf of the relative diagonal in $X^{(1)}_0\times_{\mathbb{P}^2}X^{(1)}_0$~\cite{Schimannek:2019ijf}.
As a consequence, the involution~\eqref{eqn:x18inv} is part of the monodromy group and the two large volume limits are indeed equivalent.

\section{Modular forms for $\Gamma_1(N)$}
\label{app:modforms}
In this appendix we construct bases of $\Gamma_1(N)$ modular forms for $1\le N\le 5$ and derive their transformations under the relvant Fricke involutions and $\Gamma_0(M)$ actions.
We make use of the fact that the space of $\Gamma_1(N)$ modular forms can be decomposed into spaces of $\Gamma_0(N)$ modular forms with character
\begin{align}
	M_k(\Gamma_1(N))=\bigoplus\limits_{\epsilon\in D(N,\mathbb{C})}M_k(\Gamma_0(N),\epsilon)\,,
\end{align}
where $D(N,\mathbb{C})$ are the Dirichlet characters of modulus $N$.
We will then use general dimension formulas for $M_k(\Gamma_1(N))$ to determine subspaces that generate the $\Gamma_1(N)$ modular forms and find that in all cases relevant to us they are generated by Eisenstein series.
As a reference we use the book~\cite{stein2007modular} and the corresponding algorithms implemented in~\cite{sagemath}.
To calculate the transformation properties of the generators under the transfer matrices that relate the various genus one fibrations in the Tate-Shafarevich group we also find expressions in terms of Dedekind eta functions
for all of the relevant Eisenstein series.

\subsection{Eisenstein series and $\Gamma_0(N)$ modular forms with character}
The space of modular forms of weight $k$ for $\Gamma_1(N)$ can be decomposed as
\begin{align}
	M_k(\Gamma_1(N))=S_k(\Gamma_1(N))\oplus E_k(\Gamma_1(N))\,,
	\label{eqn:eisensteinCuspDecomposition}
\end{align}
where $S_k(\Gamma)$ is the space of cusp forms and $E_k(\Gamma)$ is generated by Eisenstein series.
To describe the generators of the Eisenstein subspaces, let us first define the series
\begin{align}
	E_{k,\chi,\psi}(q)=c_0+\sum\limits_{m\ge 1}\left(\sum\limits_{n|m}\psi(n)\cdot\chi(m/n)\cdot n^{k-1}\right) q^m\,.
	\label{eqn:eisenstein}
\end{align}
Here $k$ is a positive integer, $\psi$ and $\chi$ are primitive Dirichlet characters with respective conductors $R$ and $L$ while $c_0$ is given by
\begin{align}
	c_0=\left\{\begin{array}{cl}0&\text{ if }L>1\,,\\-\frac{B_{k,\psi}}{2k}&\text{ if }L=1\,.\end{array}\right.
\end{align}
The generalized Bernoulli numbers $B_{k,\psi}$ are defined by the identity
\begin{align}
	\sum\limits_{a=1}^R\frac{\psi(a)\cdot x\cdot e^{ax}}{e^{Rx}-1}=\sum\limits_{k=0}^\infty B_{k,\psi}\cdot \frac{x^k}{k!}\,.
\end{align}

Let $t$ be a positive integer such that $\chi(-1)\psi(-1)=(-1)^k$.
One can show that, except when $k=2$ and $\chi=\chi=1$, the Eisenstein series $E_{k,\chi,\psi}(q^t)$ is an element of $M_k(\Gamma_0(RLt),\chi\psi)$.
The collection of these Eisenstein series with $RLt\mid N$ and $\chi\psi=\epsilon$ form a basis of the Eisenstein subspace $E_k(N,\epsilon)$.
Moreover, $E_2(q)-tE_2(q^t)$ with $E_2(q)=-24\cdot E_{2,1,1}(q)$ is a $\Gamma_0(t)$ modular form of weight $k=2$.

\subsection{Modular forms for $\Gamma_1(N)$ and Dedekind eta functions}
To obtain generating elements of the spaces of modular forms for $\Gamma_1(N)$ we will use general formulas for the dimensions of those spaces as well as the subspaces of cusp forms and Eisenstein series.
Using the latter we can verify that the relevant forms can be expressed in terms of the Eisenstein series~\eqref{eqn:eisenstein}.

The dimensions of spaces of modular forms are closely related to the geometric properties of the modular curves $X(\Gamma)=\overline{\mathbb{H}/\Gamma}$.
We denote the modular curves associated to $\Gamma_0(N)$ and $\Gamma_1(N)$ respectively by $X_0(N)$ and $X_1(N)$ and for $1\le N\le 5$ both $X_0(N)$ and $X_1(N)$ have genus $g=0$.
Moreover, when $N\le 2$ the groups $\Gamma_0(N)$ and $\Gamma_1(N)$ are isomorphic and the corresponding spaces of modular forms are the same.
\begin{table}[h!]
	\begin{align*}\begin{array}{c|c|c|c|c|c}
		&X_1(1)&X_1(2)&X_1(3)&X_1(4)&X_1(5)\\\hline
		\epsilon_2&1&1&0&0&0\\
		\epsilon_3&1&0&1&0&0\\
		\epsilon_\infty^\text{reg}&1&2&2&2&4\\
		\epsilon_\infty^\text{irr}&0&0&0&1&0\\
		\epsilon_\infty&1&2&2&3&4
	\end{array}
	\end{align*}
	\caption{The numbers $\epsilon_2,\epsilon_3$ of elliptic points of order $2,3$ and the number $\epsilon_\infty$ of cusps for the modular curves $X_1(N)$ with $N=1,\ldots,5$.}
	\label{tab:specialPointsXn}
\end{table}
General formulas for the dimensions of spaces of weight $k$ modular forms and the Eisenstein subspaces can be found e.g. in~\cite{firstcourse2005}.
They depend on the genus $g$ of the modular curve, the number $\epsilon_2$, $\epsilon_3$ of elliptic points of order $2$ and $3$ and the number of cusps $\epsilon_\infty$.
For some weights it is necessary to distinguish between regular and irregular cusps with respective numbers $\epsilon_\infty^{\text{reg}}$ and $\epsilon_\infty^{\text{irr}}$ such that $\epsilon_\infty=\epsilon_\infty^{\text{reg}}+\epsilon_\infty^{\text{irr}}$.

The corresponding values for $\Gamma_1(N)$ with $1\le N\le5$ are summarized in Table~\ref{tab:specialPointsXn} and for those groups the dimensions of $M_k(\Gamma)$ take the form
\begin{align}
	\begin{split}
	\text{dim}\,M_k(\Gamma)=&\left\{\begin{array}{cl}
		0&\text{if $N\le 2$ and $2\nmid k$}\,,\\
		\left\lfloor\frac{k}{4}\right\rfloor\epsilon_2+\left\lfloor\frac{k}{3}\right\rfloor\epsilon_3+\frac{k}{2}\epsilon_\infty-k+1&\text{if $2\mid k$ and $k\ge 0$}\,,\\
		\left\lfloor\frac{k}{3}\right\rfloor\epsilon_3+\frac{k}{2}\epsilon_\infty^{\text{reg}}+\frac{k-1}{2}\epsilon_\infty^{\text{irr}}-k+1&\text{if $2\nmid k$ and $k\ge 1$}\,,\\
		0&\text{otherwise}\,.
	\end{array}\right.
	\end{split}
	\label{eqn:dimensionModularForms}
\end{align}
On the other hand, the dimensions of the subspaces of Eisenstein series are given by
\begin{align}
	\text{dim}\,E_k(\Gamma)=&\left\{\begin{array}{cl}
		0&\text{if $N\le 2$ and $2\nmid k$}\,,\\
		\epsilon_\infty&\text{if $2\mid k$ and $k\ge 4$}\,,\\
		\epsilon_\infty^{\text{reg}}&\text{if $2\nmid k$ and $k\ge 3$}\,,\\
		\epsilon_\infty-1&\text{if $k=2$}\,,\\
		\epsilon_\infty^{\text{reg}}/2&\text{if $k=1$}\,,\\
		0&\text{otherwise}\,.
	\end{array}\right.
	\label{eqn:eisensteinDim}
\end{align}
Using the decomposition~\eqref{eqn:eisensteinCuspDecomposition} one can also determine the dimensions of the subspaces of cusp forms which, however, will not be relevant for our discussion.

\subsubsection{Modular forms for $\Gamma_1(2)$}
Using the general expressions~\eqref{eqn:dimensionModularForms} as well as the data from Table~\eqref{tab:specialPointsXn}, we obtain the dimensions of the spaces of weight $k$ modular forms for $\Gamma_1(2)$,
\begin{align}
	\begin{split}
		\text{dim}\,M_k(\Gamma_1(2))=&\left\{\begin{array}{cl}
			\left\lfloor\frac{k}{4}\right\rfloor+1&\text{if $2\mid k$ and $k\ge 0$}\,,\\
			0&\text{otherwise}\,.
		\end{array}\right.\\
	\end{split}
\end{align}
This implies that the ring is generated by one form of weight 2 and another form of weight 4.
With~\eqref{eqn:eisensteinDim} we also find that the dimensions of the respective subspaces of Eisenstein series are $1$ and $2$.
As generator of the weight 2 space we take the normalized series
\begin{align}
	E_{2,2}(q)=2E_2(2q)-E_2(q)=1+24q+24q^2+96q^3+24q^4+\mathcal{O}(q^5)\,,
	\label{eqn:g12weight2}
\end{align}
while our weight 4 generator is the usual $E_4$ Eisenstein series
\begin{align}
	E_4(q)=240\cdot E_{4,1,1}(q)=1+240q+2160q^2+6720q^3+17520q^4+\mathcal{O}(q^5)\,.
\end{align}
Together they generate the ring of modular forms for $\Gamma_1(2)$.
Note that we use $E_{N,w}(\tau)$ to denote a $\Gamma_1(N)$ Eisenstein series of weight $w$ and this should not be confused with the Jacobi Eisenstein series, which are Jacobi forms, and will not be relevant for our discussion.

To calculate the action of the transfer matrix that relates the topological string partition function on genus one fibrations with $2$-sections and on the associated Jacobian fibration with discrete torsion
we express the generators in terms of Dedekind eta functions.
The relevant relations are
\begin{align}
	\begin{split}
		E_{2,2}(\tau)=&\frac{\eta(\tau)^8+32\eta(4\tau)^8}{\eta(2\tau)^4}=\sqrt{\frac{\eta(\tau)^{24}+64\eta(2\tau)^{24}}{\eta(\tau)^8\eta(2\tau)^8}}\,,\\
		E_4(\tau)=&\frac{\eta(\tau)^{16}}{\eta(2\tau)^8}+256\frac{\eta(2\tau)^{16}}{\eta(\tau)^8}\,.
	\end{split}
	\label{eqn:gamma12gens}
\end{align}
We found these relations, as well as analogous relations for the generators of modular forms for the other relevant congruence subgroups, using a combination of experimentation and the online encyclopedia of integer sequences (OEIS)~\cite{oeis}.

The special $\Gamma_1(2)$ modular forms
\begin{align}
	\Delta_4(\tau)=\frac{1}{192}\left(E_4-E_{2,2}^2\right)\,,\quad \Delta_4'(\tau)=\frac{1}{192}\left(4E_{2,2}^2-E_4\right)\,,
\end{align}
are determined by the restrictions of the generators of weak Jacobi forms $\phi_{0,1}(\tau,z),\phi_{-2,1}(\tau,z)$,
\begin{align}
	\begin{split}
		q^{\frac12}\phi_{-2,1}(2\tau,\tau)=&-(\Delta_4)^{-\frac12}\,,\quad q^{\frac12}\phi_{0,1}(2\tau,\tau)=E_{2,2}(\Delta_4)^{-\frac12}\,,\\
		q^{\frac12}\phi_{-2,1}\left(\tau,\frac12\right)=&-2(\Delta_4')^{-\frac12}\,,\quad q^{\frac12}\phi_{0,1}\left(\tau,\frac12\right)=E_{2,2}(\Delta_4')^{-\frac12}\,.
	\end{split}
\end{align}

\subsubsection{Modular forms for $\Gamma_1(3)$}
The dimensions of the spaces of $\Gamma_1(3)$ modular forms of weight $k$ are given by
\begin{align}
	\begin{split}
		\text{dim}M_k(\Gamma_1(3))=&\left\{\begin{array}{cl}
			\left\lfloor\frac{k}{3}\right\rfloor+1&\text{if $k\ge 0$}\,,\\
			0&\text{otherwise}\,.
		\end{array}\right.\\
	\end{split}
\end{align}
We thus find that the ring is generated by a form of weight $3$ and another form of weight $1$.
Again, the spaces associated to those weights are generated by Eisenstein series.
The two Dirichlet characters of modulus $3$ are
\begin{align}
	\begin{array}{c|ccc}
		n\,\text{mod}\,3&0&1&2\\\hline
		\chi^{3}_0(n)&0&1&1\\
		\chi^{3}_1(n)&0&1&-1\\
	\end{array}\,,
\end{align}
and $\chi^{3}_1$ is primitive.
We can therefore choose the normalized Eisenstein series
\begin{align}
	\begin{split}
		E_{3,1}(q)=&6E_{1,\chi^{3}_1,1}(q)=1+6q+6q^3+6q^4+\mathcal{O}(q^5)\,,\\
		E_{3,3}(q)=&E_{3,1,\chi^{3}_1}(q)=q+3q^2+9q^3+13q^4+\mathcal{O}(q^5)\,,
	\end{split}
\end{align}
as generators of the ring of modular forms for $\Gamma_1(3)$.
They can be expressed in terms of Dedekind eta functions as
\begin{align}
	\begin{split}
		E_{3,1}(\tau)=\left(\frac{\eta(\tau)^9}{\eta(3\tau)^3}+27\frac{\eta(3\tau)^9}{\eta(\tau)^3}\right)^{\frac13}\,,\quad E_{3,3}(\tau)=\frac{\eta(3\tau)^9}{\eta(\tau)^3}\,.
	\end{split}
	\label{eqn:g13dedekind}
\end{align}

The eta functions transform under the Fricke involution $\tau\mapsto -1/3\tau$ as
\begin{align}
	\begin{split}
		\eta(\tau)\mapsto\eta\left(-\frac{1}{3\tau}\right)=\sqrt{3}\sqrt{-i\tau}\eta(3\tau)\,,\quad \eta(3\tau)\mapsto\eta\left(-\frac{1}{\tau}\right)=\sqrt{-i\tau}\eta(\tau)\,,
	\end{split}
\end{align}
and we deduce the transformations of the Eisenstein series
\begin{align}
	E_{3,1}(\tau)\mapsto 3^{\frac12}(-i\tau) E_{3,1}(\tau)\,,\quad E_{3,3}(\tau)\mapsto 3^{\frac32}(-i\tau)^3 \frac{E_{3,1}(\tau)^3-27E_{3,3}(\tau)}{27}\,.
\end{align}
Note that due to the third root in~\eqref{eqn:g13dedekind}, the transformation of $E_1^{(3)}(\tau)$ only follows up to a third root of unity.
However, demanding that applying the Fricke involution twice gives back $E_1^{(3)}(\tau)$ fixes this factor to be one.

The special $\Gamma_1(3)$ modular forms
\begin{align}
	\Delta_6(\tau)=E_{3,3}(\tau)^2\,,\quad \Delta_6'(\tau)=\frac{1}{27}\left(E_{3,1}(\tau)^3-27E_{3,3}(\tau)\right)^2\,,
	\label{eqn:g13specialForms}
\end{align}
are determined by the restrictions of the generators of weak Jacobi forms
\begin{align}
	\begin{split}
		q^{\frac{k^2}{3}}\phi_{-2,1}(3\tau,k\tau)=&(\Delta_6)^{-\frac13}\,,\quad q^{\frac{k^2}{3}}\phi_{0,1}(3\tau,k\tau)=E_{3,1}^2(\Delta_6)^{-\frac13}\,,\\
		\phi_{-2,1}\left(\tau,\frac{k}{3}\right)=&-(\Delta_6')^{-\frac13}\,,\quad \phi_{0,1}\left(\tau,\frac{k}{3}\right)=3E_{3,1}^2(\Delta_6')^{-\frac13}\,,
	\end{split}
\end{align}
where $k\in\{1,2\}$.

\subsubsection{Modular forms for $\Gamma_1(4)$}
\label{sec:mfG14}
Again we use equation~\eqref{eqn:dimensionModularForms} and Table~\eqref{tab:specialPointsXn} to determine the dimensions of the spaces of modular forms
\begin{align}
	\begin{split}
		\text{dim}M_k(\Gamma_1(4))=&\left\{\begin{array}{cl}
			\left\lfloor\frac{k}{2}\right\rfloor+1&\text{if $k\ge 0$}\,,\\
			0&\text{otherwise}\,.
		\end{array}\right.\\
	\end{split}
\end{align}
This implies that the ring is generated by two forms of respective weights one and two.
The two Dirichlet characters of modulus $4$ are
\begin{align}
	\begin{array}{c|cccc}
		n\,\text{mod}\,4&0&1&2&3\\\hline
		\chi^{4}_0&0&1&0&1\\
		\chi^{4}_1&0&1&0&-1
	\end{array}\,,
\end{align}
with only $\chi^{4}_1$ being primitive.
The space of $\Gamma_1(4)$ modular forms of weight one is therefore generated by the normalized Eisenstein series
\begin{align}
	E_{4,1}(q)=&4E_{1,\chi^{4}_1,1}=1+4q+4q^2+4q^4+\mathcal{O}(q^5)\,.
\end{align}
As a generator of weight two we can take the $\Gamma_1(2)$ modular form $E_{2,2}(\tau)$ from~\eqref{eqn:g12weight2}.
The weight one generator can be expressed in terms of Dedekind eta functions as
\begin{align}
	E_{4,1}(\tau)=\frac{\eta(2\tau)^{10}}{\eta(\tau)^4\eta(4\tau)^4}=e^{-\frac{\pi i}{6}}\frac{\eta\left(\tau+\frac{1}{2}\right)^4}{\eta\left(2\tau\right)^2}\,.
\end{align}

The relevant eta functions transform under the Fricke involution $\tau\mapsto 1/4\tau$ as
\begin{align}
	\begin{split}
	\eta(\tau)\mapsto\eta\left(-\frac{1}{4\tau}\right)=&\sqrt{4}\sqrt{-i\tau}\eta(4\tau)\,,\\
	\eta(2\tau)\mapsto\eta\left(-\frac{1}{2\tau}\right)=&\sqrt{2}\sqrt{-i\tau}\eta(2\tau)\,,\\
	\eta(4\tau)\mapsto\eta\left(-\frac{1}{\tau}\right)=&\sqrt{-i\tau}\eta(\tau)\,,
	\end{split}
\end{align}
and this implies the action on the Eisenstein series
\begin{align}
	\begin{split}
		E_{2,2}(\tau)\mapsto4(-i\tau)^2\left[3E_{4,1}(\tau)^2-E_{2,2}(\tau)\right]\,,\quad E_{4,1}(\tau)\mapsto2(-i\tau)E_{4,1}(\tau)\,.
	\end{split}
\end{align}
We also need the transformation under the $\Gamma_1(2)$ action
\begin{align}
	\tau\mapsto\frac{\tau}{2\tau+1}\,.
	\label{eqn:g12utrafo}
\end{align}
To this end we use the general relation
\begin{align}
	\eta\left(\frac{a\tau+b}{c\tau+d}\right)=\epsilon(a,b,c,d)(c\tau+d)^{\frac12}\eta(\tau)\,,
\end{align}
where the phase factor is given by
\begin{align}
	\epsilon(a,b,c,d)=\left\{\begin{array}{cl}
		\exp\left(\frac{b\pi i}{12}\right)&\text{ if }c=0,d=1\,,\\
		\exp\left(\pi i\left[\frac{a+d}{12c}-s(d,c)-\frac14\right]\right)&\text{ if }c>0\,,
	\end{array}\right.
\end{align}
and $s(h,k)$ is the Dedekind sum
\begin{align}
	s(h,k)=\sum\limits_{n=1}^{k-1}\frac{n}{k}\left(\frac{hn}{k}-\left\lfloor\frac{hn}{k}\right\rfloor-\frac12\right)\,.
\end{align}
The eta functions then transform under~\eqref{eqn:g12utrafo} as
\begin{align}
	\begin{split}
		\eta(\tau)&\mapsto e^{-\frac{\pi i}{6}}\sqrt{1+2\tau}\eta(\tau)\,,\quad \eta(2\tau)\mapsto e^{-\frac{\pi i}{12}}\sqrt{1+2\tau}\eta(2\tau)\,,\\
		\eta\left(\tau+\frac12\right)&\mapsto\sqrt{2}\sqrt{1+2\tau}\eta(4\tau)\,,\quad\eta(4\tau)\mapsto e^{-\frac{\pi i}{12}}\sqrt{1+2\tau}\frac{1}{\sqrt{2}}\eta\left(\tau+\frac12\right)\,,\\
	\end{split}
\end{align}
and this implies the action on the Eisenstein series
\begin{align}
	E_{2,2}\mapsto (1+2\tau)^2E_{2,2}\,,\quad E_{4,1}\mapsto (1+2\tau)\sqrt{E_{2,2}-E_{4,1}^2}\,.
\end{align}

The special $\Gamma_1(4)$ modular forms
\begin{align}
	\begin{split}
	\Delta_8(\tau)=2^{-12}E_{4,1}^2\left(E_{2,2}-E_{4,1}^2\right)^3\,,&\quad \Delta_8'(\tau)=2^{-4}E_{4,1}^2\left(2E_{4,1}^2-E_{2,2}\right)^3\,,\\
	E_{4,2}(\tau)=\frac14\left(3E_{4,1}^2+E_{2,2}\right)\,,&\quad E_{4,2}'(\tau)=6E_{4,1}^2-E_{2,2}\,,
	\end{split}
\end{align}
are determined by the restrictions of the generators of weak Jacobi forms
\begin{align}
	\begin{split}
		q^{\frac{k^2}{4}}\phi_{-2,1}(4\tau,k\tau)=&(\Delta_8)^{-\frac14}\,,\quad q^{\frac{k^2}{4}}\phi_{0,1}(4\tau,k\tau)=E_{4,2}(\Delta_4)^{-\frac14}\,,\\
		\phi_{-2,1}\left(\tau,\frac{k}{4}\right)=&(\Delta_8')^{-\frac14}\,,\quad \phi_{0,1}\left(\tau,\frac{k}{4}\right)=E_{4,2}'(\Delta_8')^{-\frac14}\,,
	\end{split}
\end{align}
where $k\in\{1,3\}$.
On the other hand, the forms
\begin{align}
	\Delta_8''(\tau)=2^{-4}E_{4,1}^6\left(E_{2,2}-E_{4,1}^2\right)\,,\quad E_{4,2}''(\tau)=4E_{2,2}-3E_{4,1}^2\,,
\end{align}
arise from the restrictions
\begin{align}
	\begin{split}
		q^{\frac{k^2}{4}}\phi_{-2,1}\left(\tau+\frac12,\frac{k}{2}\tau\right)=&(\Delta_8'')^{-\frac14}\,,\quad q^{\frac{k^2}{4}}\phi_{0,1}\left(\tau+\frac12,\frac{k}{2}\tau\right)=E_{4,2}''(\Delta_8'')^{-\frac14}\,,
	\end{split}
\end{align}
where again $k\in\{1,3\}$.
\subsubsection{Modular forms for $\Gamma_1(5)$}
The dimensions of the spaces of $\Gamma_1(5)$ modular forms are given by
\begin{align}
	\begin{split}
		\text{dim}M_k(\Gamma_1(5))=&\left\{\begin{array}{ll}
			k+1&\text{if $k\ge 0$}\,,\\
			0&\text{otherwise}\,,
		\end{array}\right.\\
	\end{split}
\end{align}
and we find that the ring is generated by two forms of weight one.
The Dirichlet characters of modulus $5$ are
\begin{align}
	\begin{array}{c|ccccc}
		n\,\text{mod}\,5&0&1&2&3&4\\\hline
		\chi^{5}_0&0&1&1&1&1\\
		\chi^{5}_1&0&1&i&-i&-1\\
		\chi^{5}_2&0&1&-1&-1&1\\
		\chi^{5}_3&0&1&-i&i&-1\\
	\end{array}\,,
\end{align}
with $\chi^{5}_i$ for $i\in\{1,2,3\}$ being primitive. Since $\chi^{5}_2(-1)=1$, only $\chi^{5}_1$ and $\chi^{5}_3$ together with the trivial character lead to Eisenstein series that are modular forms for $\Gamma_1(5)$.
As generators we take the normalized weight one series
\begin{align}
	\begin{split}
		E_{5,1,a}(q)=&5\left(\frac{1}{3+i}E_{1,\chi^{5}_1,1}+\frac{1}{3-i}E_{1,\chi^{5}_3,1}\right)=1+3q+4q^2+2q^3+q^4+\mathcal{O}(q^5)\,,\\
		E_{5,1,b}(q)=&5\left(\frac{1}{1-3i}E_{1,\chi^{5}_3,1}-\frac{1}{1+3i}E_{1,\chi^{5}_1,1}\right)=q-2q^2+4q^3-3q^4+\mathcal{O}(q^5)\,,\\
	\end{split}
\end{align}
which are related to Dedekind eta functions via
\begin{align}
	E_{5,1,a}^2-11E_{5,1,a}E_{5,1,b}-E_{5,1,b}^2=\frac{\eta(\tau)^5}{\eta(5\tau)}\,,\quad E_{5,1,a}E_{5,1,b}=\frac{\eta(5\tau)^5}{\eta(\tau)}\,.
\end{align}

The eta functions transform under the Fricke involution $\tau\mapsto-1/5\tau$ as
\begin{align}
	\eta(\tau)\mapsto \sqrt{5}\sqrt{-i\tau}\eta(5\tau)\,,\quad \eta(5\tau)\mapsto \sqrt{-i\tau}\eta(\tau)\,,
\end{align}
such that we can deduce the action on the modular forms
\begin{align}
	\begin{split}
	E_{5,1,a}\mapsto 5^{-\frac14}\sqrt{11+z_-}\left(E_{5,1,a}+z_- \cdot E_{5,1,b}\right)\,,\\
	E_{5,1,b}\mapsto 5^{-\frac14}\sqrt{11+z_-}\left(z_-\cdot E_{5,1,a}-E_{5,1,b}\right)\,,
	\end{split}
\end{align}
where $z_\pm$ are the roots
\begin{align}
	z_\pm=-\frac12\left(11\pm5\sqrt{5}\right)\,,
\end{align}
of the polynomial $\Delta=1-11z-z^2$.
On the other hand, as discussed in~\cite{Knapp:2021vkm}, the $\Gamma_0(5)$ transformation
\begin{align}
	\tau\mapsto\frac{2\tau-1}{5\tau-2}\,,
\end{align}
acts on the modular forms as
\begin{align}
	E_{5,1,a}\mapsto E_{5,1,b}\,,\quad E_{5,1,b}\mapsto -E_{5,1,a}\,.
\end{align}

Of particular importance will be the modular forms
\begin{align}
	\begin{split}
		\Delta_{10}(\tau)=E_{5,1,a}^{\,6}E_{5,1,b}^{\,4}\,,&\quad \Delta_{10}'(\tau)=E_{5,1,a}^{\,4}E_{5,1,b}^{\,6}\,,\\
		E_{5,2}(\tau)=E_{5,1,a}^2+6E_{5,1,a}E_{5,1,b}+E_{5,1,b}^2\,,&\quad E_{5,2}'(\tau)=E_{5,1,a}^2-6E_{5,1,a}E_{5,1,b}+E_{5,1,b}^2\,,
	\end{split}
\end{align}
that arise from the restrictions of weak Jacobi forms as
\begin{align}
	q^{\frac{k^2}{5}}\phi_{-2,1}(5\tau,k\tau)=&\left\{\begin{array}{rl}\Delta_{10}^{-\frac15}&\text{ for }k\in\{1,4\}\\\Delta_{10}'^{-\frac15}&\text{ for }k\in\{2,3\}\end{array}\right.\,,\\
	q^{\frac{k^2}{5}}\phi_{0,1}(5\tau,k\tau)=&\left\{\begin{array}{rl}E_{5,2}\Delta_{10}^{-\frac15}&\text{ for }k\in\{1,4\}\\E_{5,2}'\Delta_{10}'^{-\frac15}&\text{ for }k\in\{2,3\}\end{array}\right.\,,
\end{align}
as well as the modular forms
	{\small
\begin{align}
	\begin{split}
	\Delta_{10}''(\tau)=&-\frac{z_+^5}{25\sqrt{5}}\left(z_-E_{5,1,a}-E_{5,1,b}\right)^4\left(E_{5,1,a}+z_-E_{5,1,b}\right)^6\,,\\
	\Delta_{10}'''(\tau)=&-\frac{z_+^5}{25\sqrt{5}}\left(z_-E_{5,1,a}-E_{5,1,b}\right)^6\left(E_{5,1,a}+z_-E_{5,1,b}\right)^4\,,\\
	E_{5,2}''(\tau)=&\frac{1}{25}\left(\left(191+12z_-\right)E_{5,1,a}^2-66\left(11+2z_-\right)E_{5,1,a}E_{5,1,b}+\left(59-12z_-\right)E_{5,1,b}^2\right)\,,\\
	E_{5,2}'''(\tau)=&\frac{1}{25}\left(\left(191+12z_+\right)E_{5,1,a}^2-66\left(11+2z_+\right)E_{5,1,a}E_{5,1,b}+\left(59-12z_+\right)E_{5,1,b}^2\right)\,,\\
	\end{split}
\end{align}
	}
that arise as
\begin{align}
	\phi_{-2,1}\left(\tau,\frac{k}{5}\right)=&\left\{\begin{array}{rl}\Delta_{10}''^{-\frac15}&\text{ for }k\in\{1,4\}\\\Delta_{10}'''^{-\frac15}&\text{ for }k\in\{2,3\}\end{array}\right.\,,\\
	\phi_{0,1}\left(\tau,\frac{k}{5}\right)=&\left\{\begin{array}{rl}E_{5,2}''\Delta_{10}''^{-\frac15}&\text{ for }k\in\{1,4\}\\E_{5,2}'''\Delta_{10}'''^{-\frac15}&\text{ for }k\in\{2,3\}\end{array}\right.\,.
\end{align}

\section{Gopakumar-Vafa invariants}
\label{app:gvs}
\subsection{$X^{(4)}_0$}
\label{app:gvX40}
\begin{table}[H]
	\begin{align*}
\begin{array}{c|ccccc}
	n^{(0)}_{(d_1,d_2),0} & d_2=0 & 1 & 2 & 3 & 4 \\\hline
 d_1=0 & 0 & 3 & -6 & 27 & -192 \\
 1 & 132 & -264 & 660 & -4224 & 37752 \\
 2 & 132 & 35978 & -144176 & 1268610 & -14537940 \\
 3 & 132 & 51011056 & 18693288 & -228274152 & 3397719960 \\
 4 & 132 & 5443202915 & -12483195020 & 56026973260 & -738484301224 \\
 5 & 132 & 269129066440 & 1943124647024 & -10678034962080 & 150944449677308 \\
\end{array}
	\end{align*}
	\caption{Genus 0 Gopakumar-Vafa invariants for $X^{(4)}_0$ with $\mathbb{Z}_4$ charge $0$.}
\end{table}

\begin{table}[H]
	\begin{align*}
\begin{array}{c|ccccc}
	n^{(0)}_{(d_1,d_2),\pm1} & d_2=0 & 1 & 2 & 3 & 4 \\\hline
 d_1=0 & 0 & 0 & 0 & 0 & 0 \\
 1 & 128 & -256 & 640 & -4096 & 36608 \\
 2 & 128 & 35840 & -143616 & 1261568 & -14448000 \\
 3 & 128 & 51025408 & 18707712 & -228375808 & 3398664960 \\
 4 & 128 & 5443274752 & -12483367936 & 56027811840 & -738492641536 \\
 5 & 128 & 269130074368 & 1943122546176 & -10678030021632 & 150944505966464 \\
\end{array}
	\end{align*}
	\caption{Genus 0 Gopakumar-Vafa invariants for $X^{(4)}_0$ with $\mathbb{Z}_4$ charge $\pm1$.}
\end{table}

\begin{table}[H]
	\begin{align*}
\begin{array}{c|ccccc}
	n^{(0)}_{(d_1,d_2),2} & d_2=0 & 1 & 2 & 3 & 4 \\\hline
 d_1=0 & 0 & 0 & 0 & 0 & 0 \\
 1 & 152 & -304 & 760 & -4864 & 43472 \\
 2 & 152 & 35712 & -143152 & 1260224 & -14445960 \\
 3 & 152 & 51009312 & 18701808 & -228357232 & 3398801040 \\
 4 & 152 & 5443195136 & -12483128768 & 56026261760 & -738473750064 \\
 5 & 152 & 269129036976 & 1943125131424 & -10678040601024 & 150944541311592 \\
\end{array}
	\end{align*}
	\caption{Genus 0 Gopakumar-Vafa invariants for $X^{(4)}_0$ with $\mathbb{Z}_4$ charge $2$.}
\end{table}

\begin{table}[H]
	\begin{align*}
\begin{array}{c|ccccc}
	n^{(1)}_{(d_1,d_2),0} & d_2=0 & 1 & 2 & 3 & 4 \\\hline
 d_1=0 & 0 & 0 & 0 & -10 & 231 \\
 1 & 3 & -6 & 15 & 1092 & -37158 \\
 2 & 3 & 510 & -2046 & -271570 & 12288647 \\
 3 & 3 & -70396 & 534294 & 38019450 & -2438717094 \\
 4 & 3 & -102235910 & 130472356 & -4176344608 & 397769305554 \\
 5 & 3 & -11192756330 & 280552809188 & -832115693476 & -46603783112291 \\
\end{array}
	\end{align*}
	\caption{Genus 1 Gopakumar-Vafa invariants for $X^{(4)}_0$ with $\mathbb{Z}_4$ charge $0$.}
\end{table}

\begin{table}[H]
	\begin{align*}
\begin{array}{c|ccccc}
	n^{(1)}_{(d_1,d_2),\pm1} & d_2=0 & 1 & 2 & 3 & 4 \\\hline
 d_1=0 & 0 & 0 & 0 & 0 & 0 \\
 1 & 0 & 0 & 0 & 1152 & -36864 \\
 2 & 0 & 512 & -2048 & -270592 & 12221952 \\
 3 & 0 & -70144 & 531456 & 38086400 & -2440249344 \\
 4 & 0 & -102263808 & 130483200 & -4176063488 & 397770053632 \\
 5 & 0 & -11192985088 & 280553999104 & -832121521536 & -46603758311936 \\
\end{array}
	\end{align*}
	\caption{Genus 1 Gopakumar-Vafa invariants for $X^{(4)}_0$ with $\mathbb{Z}_4$ charge $\pm1$.}
\end{table}

\begin{table}[H]
	\begin{align*}
\begin{array}{c|ccccc}
	n^{(1)}_{(d_1,d_2),2} & d_2=0 & 1 & 2 & 3 & 4 \\\hline
 d_1=0 & 0 & 0 & 0 & 0 & 0 \\
 1 & 0 & 0 & 0 & 1368 & -43776 \\
 2 & 0 & 608 & -2432 & -266544 & 12175264 \\
 3 & 0 & -69600 & 529152 & 38086736 & -2440203840 \\
 4 & 0 & -102230464 & 130418240 & -4175615296 & 397753008256 \\
 5 & 0 & -11192727584 & 280552295696 & -832108662920 & -46603941324384 \\
\end{array}
	\end{align*}
	\caption{Genus 1 Gopakumar-Vafa invariants for $X^{(4)}_0$ with $\mathbb{Z}_4$ charge $2$.}
\end{table}

\begin{table}[H]
	\begin{align*}
\begin{array}{c|ccccc}
	n^{(g)}_{(d_1,1),0} & g=0 & 1 & 2 & 3 & 4 \\\hline
 d_1=0 & 3 & 0 & 0 & 0 & 0 \\
 1 & -264 & -6 & 0 & 0 & 0 \\
 2 & 35978 & 510 & 9 & 0 & 0 \\
 3 & 51011056 & -70396 & -744 & -12 & 0 \\
 4 & 5443202915 & -102235910 & 103845 & 966 & 15 \\
 5 & 269129066440 & -11192756330 & 153597224 & -136352 & -1176 \\
\end{array}
	\end{align*}
	\caption{Base degree 1 Gopakumar-Vafa invariants for $X^{(4)}_0$ with $\mathbb{Z}_4$ charge $0$.}
\end{table}

\begin{table}[H]
	\begin{align*}
\begin{array}{c|ccccc}
	n^{(g)}_{(d_1,1),\pm1} & g=0 & 1 & 2 & 3 & 4 \\\hline
 d_1=0 & 0 & 0 & 0 & 0 & 0 \\
 1 & -256 & 0 & 0 & 0 & 0 \\
 2 & 35840 & 512 & 0 & 0 & 0 \\
 3 & 51025408 & -70144 & -768 & 0 & 0 \\
 4 & 5443274752 & -102263808 & 103424 & 1024 & 0 \\
 5 & 269130074368 & -11192985088 & 153638144 & -135680 & -1280 \\
\end{array}
	\end{align*}
	\caption{Base degree 1 Gopakumar-Vafa invariants for $X^{(4)}_0$ with $\mathbb{Z}_4$ charge $\pm1$.}
\end{table}

\begin{table}[H]
	\begin{align*}
\begin{array}{c|ccccc}
	n^{(g)}_{(d_1,1),2} & g=0 & 1 & 2 & 3 & 4 \\\hline
 d_1=0 & 0 & 0 & 0 & 0 & 0 \\
 1 & -304 & 0 & 0 & 0 & 0 \\
 2 & 35712 & 608 & 0 & 0 & 0 \\
 3 & 51009312 & -69600 & -912 & 0 & 0 \\
 4 & 5443195136 & -102230464 & 102272 & 1216 & 0 \\
 5 & 269129036976 & -11192727584 & 153585648 & -133728 & -1520 \\
\end{array}
	\end{align*}
	\caption{Base degree 1 Gopakumar-Vafa invariants for $X^{(4)}_0$ with $\mathbb{Z}_4$ charge $2$.}
\end{table}

\subsection{$X^{(4)}_2$}
\label{app:gvX42}
\begin{table}[H]
	\begin{align*}
\begin{array}{c|ccccc}
	n^{(0)}_{(d_1,d_2),0} & d_2=0 & 1 & 2 & 3 & 4 \\\hline
 d_1=0 & 0 & 0 & -4 & 0 & -48 \\
 1 & 128 & 256 & 384 & 1280 & 6272 \\
 2 & 152 & 99664 & -30992 & -126848 & -759480 \\
 3 & 128 & 3155968 & 7549440 & 16893184 & 88847360 \\
 4 & 132 & 52323072 & 5509090960 & -1370469376 & -7723924760 \\
 5 & 128 & 604121088 & 465504784128 & 988194620416 & 1811907204864 \\
\end{array}
	\end{align*}
	\caption{Genus 0 Gopakumar-Vafa invariants for $X^{(4)}_2$ with $\mathbb{Z}_2$ charge $0$.}
\end{table}

\begin{table}[H]
	\begin{align*}
\begin{array}{c|ccccc}
	n^{(0)}_{(d_1,d_2),1} & d_2=0 & 1 & 2 & 3 & 4 \\\hline
 d_1=0 & 0 & -4 & 0 & -12 & 0 \\
 1 & 128 & 256 & 384 & 1280 & 6272 \\
 2 & 132 & 100032 & -31448 & -125184 & -767028 \\
 3 & 128 & 3155968 & 7549440 & 16893184 & 88847360 \\
 4 & 152 & 52333784 & 5509072256 & -1370447536 & -7723742224 \\
 5 & 128 & 604121088 & 465504784128 & 988194620416 & 1811907204864 \\
\end{array}
	\end{align*}
	\caption{Genus 0 Gopakumar-Vafa invariants for $X^{(4)}_2$ with $\mathbb{Z}_2$ charge $1$.}
\end{table}

\begin{table}[H]
	\begin{align*}
\begin{array}{c|ccccc}
	n^{(1)}_{(d_1,d_2),0} & d_2=0 & 1 & 2 & 3 & 4 \\\hline
 d_1=0 & 0 & 0 & 0 & 0 & 9 \\
 1 & 0 & 0 & 0 & 0 & -1024 \\
 2 & 0 & 8 & -304 & -960 & 104064 \\
 3 & 0 & -512 & 33024 & 202752 & -6234624 \\
 4 & 3 & -200064 & 50883252 & -122761856 & -209484538 \\
 5 & 0 & -6313472 & 5670980352 & 22729640448 & 78534126336 \\
\end{array}
	\end{align*}
	\caption{Genus 1 Gopakumar-Vafa invariants for $X^{(4)}_2$ with $\mathbb{Z}_2$ charge $0$.}
\end{table}

\begin{table}[H]
	\begin{align*}
\begin{array}{c|ccccc}
	n^{(1)}_{(d_1,d_2),1} & d_2=0 & 1 & 2 & 3 & 4 \\\hline
 d_1=0 & 0 & 0 & 0 & 0 & 0 \\
 1 & 0 & 0 & 0 & 0 & -1024 \\
 2 & 3 & 0 & -234 & -1216 & 106681 \\
 3 & 0 & -512 & 33024 & 202752 & -6234624 \\
 4 & 0 & -199304 & 50877472 & -122749440 & -209690688 \\
 5 & 0 & -6313472 & 5670980352 & 22729640448 & 78534126336 \\
\end{array}
	\end{align*}
	\caption{Genus 1 Gopakumar-Vafa invariants for $X^{(4)}_2$ with $\mathbb{Z}_2$ charge $1$.}
\end{table}

\begin{table}[H]
	\begin{align*}
\begin{array}{c|ccccc}
	n^{(g)}_{(d_1,1),0} & g=0 & 1 & 2 & 3 & 4 \\\hline
 d_1=0 & 0 & 0 & 0 & 0 & 0 \\
 1 & 256 & 0 & 0 & 0 & 0 \\
 2 & 99664 & 8 & 0 & 0 & 0 \\
 3 & 3155968 & -512 & 0 & 0 & 0 \\
 4 & 52323072 & -200064 & 0 & 0 & 0 \\
 5 & 604121088 & -6313472 & 768 & 0 & 0 \\
 6 & 5457155040 & -105265520 & 298928 & 16 & 0 \\
 7 & 41134792704 & -1227180032 & 9472000 & -1024 & 0 \\
 8 & 269236532736 & -11229163776 & 158569728 & -400128 & 0 \\
\end{array}
	\end{align*}
	\caption{Base degree 1 Gopakumar-Vafa invariants for $X^{(4)}_2$ with $\mathbb{Z}_2$ charge $0$.}
\end{table}

\begin{table}[H]
	\begin{align*}
\begin{array}{c|ccccc}
	n^{(g)}_{(d_1,1),1} & g=0 & 1 & 2 & 3 & 4 \\\hline
 d_1=0 & -4 & 0 & 0 & 0 & 0 \\
 1 & 256 & 0 & 0 & 0 & 0 \\
 2 & 100032 & 0 & 0 & 0 & 0 \\
 3 & 3155968 & -512 & 0 & 0 & 0 \\
 4 & 52333784 & -199304 & -12 & 0 & 0 \\
 5 & 604121088 & -6313472 & 768 & 0 & 0 \\
 6 & 5457212544 & -105246336 & 300096 & 0 & 0 \\
 7 & 41134792704 & -1227180032 & 9472000 & -1024 & 0 \\
 8 & 269236922660 & -11229110040 & 158595796 & -398536 & -20 \\
\end{array}
	\end{align*}
	\caption{Base degree 1 Gopakumar-Vafa invariants for $X^{(4)}_2$ with $\mathbb{Z}_2$ charge $1$.}
\end{table}

\subsection{$X^{(5)}_0$}
\label{sec:gvX50}
\begin{table}[H]
	\begin{align*}
\begin{array}{c|ccccc}
	n^{(0)}_{(d_1,d_2),0} & d_2=0 & 1 & 2 & 3 & 4 \\\hline
 d_1=0 & 0 & 3 & -6 & 27 & -192 \\
 1 & 90 & -180 & 450 & -2880 & 25740 \\
 2 & 90 & 29270 & -117260 & 1030120 & -11796650 \\
 3 & 90 & 40818484 & 14923920 & -182269500 & 2713157820 \\
 4 & 90 & 4354611955 & -9987106460 & 44828505450 & -590900946660 \\
 5 & 90 & 215303747352 & 1554493980250 & -8542333784568 & 120753741368778 \\
\end{array}
	\end{align*}
	\caption{Genus 0 Gopakumar-Vafa invariants for $X^{(5)}_0$ with $\mathbb{Z}_5$ charge $0$.}
\end{table}

\begin{table}[H]
	\begin{align*}
\begin{array}{c|ccccc}
	n^{(0)}_{(d_1,d_2),\pm1} & d_2=0 & 1 & 2 & 3 & 4 \\\hline
 d_1=0 & 0 & 0 & 0 & 0 & 0 \\
 1 & 100 & -200 & 500 & -3200 & 28600 \\
 2 & 100 & 28850 & -115600 & 1016225 & -11641250 \\
 3 & 100 & 40815550 & 14950200 & -182549900 & 2717025075 \\
 4 & 100 & 4354596400 & -9986764325 & 44823849500 & -590823324900 \\
 5 & 100 & 215303680450 & 1554497429950 & -8542398268200 & 120755025918450 \\
\end{array}
	\end{align*}
	\caption{Genus 0 Gopakumar-Vafa invariants for $X^{(5)}_0$ with $\mathbb{Z}_5$ charge $\pm1$.}
\end{table}

\begin{table}[H]
	\begin{align*}
\begin{array}{c|ccccc}
	n^{(0)}_{(d_1,d_2),\pm2} & d_2=0 & 1 & 2 & 3 & 4 \\\hline
 d_1=0 & 0 & 0 & 0 & 0 & 0 \\
 1 & 125 & -250 & 625 & -4000 & 35750 \\
 2 & 125 & 28200 & -113050 & 994700 & -11400375 \\
 3 & 125 & 40810800 & 14993100 & -183006850 & 2723321475 \\
 4 & 125 & 4354571400 & -9986212275 & 44816327125 & -590697868950 \\
 5 & 125 & 215303571950 & 1554503015325 & -8542502642700 & 120757104858075 \\
\end{array}
	\end{align*}
	\caption{Genus 0 Gopakumar-Vafa invariants for $X^{(5)}_0$ with $\mathbb{Z}_5$ charge $\pm2$.}
\end{table}

\begin{table}[H]
	\begin{align*}
\begin{array}{c|ccccc}
	n^{(g)}_{(d_1,1),0} & g=0 & 1 & 2 & 3 & 4 \\\hline
 d_1=0 & 3 & 0 & 0 & 0 & 0 \\
 1 & -180 & -6 & 0 & 0 & 0 \\
 2 & 29270 & 342 & 9 & 0 & 0 \\
 3 & 40818484 & -57484 & -492 & -12 & 0 \\
 4 & 4354611955 & -81811190 & 85065 & 630 & 15 \\
 5 & 215303747352 & -8954366490 & 122915960 & -112040 & -756 \\
\end{array}
	\end{align*}
	\caption{Base degree 1 Gopakumar-Vafa invariants for $X^{(5)}_0$ with $\mathbb{Z}_5$ charge $0$.}
\end{table}

\begin{table}[H]
	\begin{align*}
\begin{array}{c|ccccc}
	n^{(g)}_{(d_1,1),\pm1} & g=0 & 1 & 2 & 3 & 4 \\\hline
 d_1=0 & 0 & 0 & 0 & 0 & 0 \\
 1 & -200 & 0 & 0 & 0 & 0 \\
 2 & 28850 & 400 & 0 & 0 & 0 \\
 3 & 40815550 & -56500 & -600 & 0 & 0 \\
 4 & 4354596400 & -81802600 & 83350 & 800 & 0 \\
 5 & 215303680450 & -8954314100 & 122899250 & -109400 & -1000 \\
\end{array}
	\end{align*}
	\caption{Base degree 1 Gopakumar-Vafa invariants for $X^{(5)}_0$ with $\mathbb{Z}_5$ charge $\pm1$.}
\end{table}

\begin{table}[H]
	\begin{align*}
\begin{array}{c|ccccc}
	n^{(g)}_{(d_1,1),\pm2} & g=0 & 1 & 2 & 3 & 4 \\\hline
 d_1=0 & 0 & 0 & 0 & 0 & 0 \\
 1 & -250 & 0 & 0 & 0 & 0 \\
 2 & 28200 & 500 & 0 & 0 & 0 \\
 3 & 40810800 & -54900 & -750 & 0 & 0 \\
 4 & 4354571400 & -81788800 & 80600 & 1000 & 0 \\
 5 & 215303571950 & -8954229700 & 122872350 & -105300 & -1250 \\
\end{array}
	\end{align*}
	\caption{Base degree 1 Gopakumar-Vafa invariants for $X^{(5)}_0$ with $\mathbb{Z}_5$ charge $\pm2$.}
\end{table}

\addcontentsline{toc}{section}{References}
\bibliographystyle{utphys}
\bibliography{names}
\end{document}